\documentclass[12pt]{report}
\usepackage{epsfig,manthesis}
\newcommand{\half}{\frac{1}{2}}                                 
\newcommand{\Tr}{{\rm Tr}}                                    
\newcommand{\be}{\begin{equation}}                     
\newcommand{\ee}{\end{equation}}                              
\newcommand{\beq}{\begin{eqnarray}}
\newcommand{\eeq}{\end{eqnarray}}
\newcommand{\rar}{\rightarrow}
\newcommand{\bra}{\langle}
\newcommand{\ket}{\rangle}
\newcommand{\mpi}{m_{\pi}}

\def\Lcal{{\cal{L}}}

\def\Lcali{{\cal{L}}_{\pi N}^{(1)}}
\def\Lcalii{{\cal{L}}_{\pi N}^{(2)}}
\title{Chiral Symmetry and the Nucleon-Nucleon Interaction.}
\author{Keith George Richardson}
\supervisor{Dr M.C. Birse}
\department{Department of Physics}
\submissiondate{September 1999}
\degree{Doctor of Philosophy}
\faculty{Faculty of Science}
\institution{University of Manchester}
\begin{document}
\maketitle{lof}{lot}{Various aspects of the application of Effective Field Theory (EFT) 
to the Nucleon-Nucleon (NN) interaction are considered.  We look for 
contributions beyond One Pion Exchange which are predicted by Chiral 
Symmetry.
\\ \indent
Using the
formalism of the Wilson Renormalisation Group (RG) we review
power counting in a simplified EFT containing only nucleons.  In the
case of weak scattering at low energy, we find a natural expansion of the 
scattering matrix around the unique trivial fixed point of the RG.  
\\ \indent
We show 
that when scattering is strong at low energy, the calculation can be
organised in a useful and systematic way by expanding the potential
around a non-trivial fixed point corresponding to a bound state of
two nucleons at threshold.  The resulting
expansion of the inverse of the scattering matrix reproduces the effective 
range expansion order by order.
\\ \indent
The extension of this EFT to include pions in a manner consistent with
chiral symmetry is discussed.  By considering a modified effective
range expansion, we find that the small 
momentum expansion in S-wave scattering converges slowly, if at all.  
\\ \indent
The NN potential is written down to third order in small momenta.  This 
potential contains the leading order and next to leading order two-pion 
exchange contributions.
Using the potential with a cut-off in coordinate space, 
we calculate phase shifts in peripheral partial waves for which the 
EFT predictions are parameter-free, and for which we may use the 
expansion around the trivial fixed point.
\\ \indent
We find several partial waves in which the effects of two-pion exchange 
can be isolated, although no strong evidence is found for the convergence of
the small momentum expansion at this order.
}
\lfoot{}
\preface{Declaration and Copyright Notice}

\indent
No portion of the work referred to in this  \expandafter{\@thesis} has
been submitted in support of an application for another degree or
qualification of this or any other university or other institution of
learning.

\vspace{2cm}

Copyright in text of this thesis rests with the Author.  Copies (by any
process) either in full, or of extracts, may be made {\bf only} in
accordance with instructions given by the Author and lodged in the John
Rylands University Library of Manchester.  Details may be obtained from
the Librarian. This page must form part of any such copies made. Further
copies (by any process) of copies made in accordance with such
instructions may not be made without permission (in writing) of the
Author.

The ownership of any intellectual property rights which may be described
in this thesis is vested in the University of Manchester,  and may not
be made available for use by third parties without the written
permission of the University, which will prescribe the terms and
conditions of any such agreement.

Further information on the conditions under which disclosures and
exploitation may take place is available from the Head of Department of
Physics.

\preface{The Author}
The author was born in Ayrshire in 1973.  His secondary school education
was undertaken in Belmont Academy, Ayr and Charleston Academy, Inverness.
From 1992-1995 he attended Edinburgh University and graduated with
a B.Sc. Hons. in Mathematical Physics.  He is keenly interested
in music, both as a listener and a participant, and at weekends may 
often be found exploring the Lakeland Fells with his wife Sandra.
\preface{Acknowledgements}
Special thanks are due to my supervisor, Dr. Mike Birse, for
patient and thoughtful teaching, constant support and guidance, 
and also for invaluable editorial comment during the preparation of this
thesis.
\newline\newline
\noindent
The work presented in Chapter \ref{calc} is the result of a collaborative
effort involving Mike Birse, Judith McGovern and myself.\newline
\newline
\noindent
Grateful thanks are due to S. Richardson, A. Munro and E. Munro for
final proof reading of this thesis.  I have also benefited greatly
from many informal and enjoyable discussions with 
N. Evanson,
S. Jones, R. Davidson, N. Petropoulos, G. Kerley and 
A. Sabio Vera.
D. Thompson has provided much appreciated computer support. \newline\newline 
This work was funded by PPARC to whom I am grateful.
\newline\newline
\noindent
My wife, Sandra Richardson, has provided limitless
moral and almost limitless financial support over 
the last three years.  Without her encouragement, and that
of my parents, none of
the work which went into the writing of this thesis
would have been possible.
\newline\newline
\noindent
This thesis is dedicated to Mum, Dad and Sandra. 

\chapter{Introduction}
\label{intro1}

The history of theoretical attempts to calculate the Nucleon-Nucleon (NN)
interaction  
from field theory is long and painful.  Yukawa's early success,
the prediction of the pion in 1935, led to the investigation of meson
exchange potentials, and has culminated in the many  
phenomenological models avaliable today, such as One Boson Exchange
(OBE) potentials \cite{polemics}.

Of course, for a long time, the aim of meson theory was to emulate the 
description of electromagnetic interactions allowed by QED.  
Potentials such as those mentioned above fall a long way short of 
this sort of success.   While they have in common
the model independent tail contributed by One Pion Exchange (OPE), they 
contain many free parameters which must be fitted to experimental data. 
OBE potentials contain the fictional $\sigma$ meson and 
form factors which are not known theoretically.  Furthermore, such 
models tend to ignore irreducible contributions to the potential from 
the exchange of two or more mesons.  Most importantly, they 
do not provide a systematic approach to NN interactions.

QCD has long supplanted meson theory as the theory of the strong 
interaction. 
Quantum Chromodynamics (QCD) is an SU(3) gauge theory describing
the interactions of coloured spin-half quarks as mediated by 
spin-one gluons.

Gluons themselves carry colour charge, and the resulting interactions
among gluons cause the quark-gluon coupling constant to become 
small for large energies, and zero in the high energy limit.  This
phenomenon is known as asymptotic freedom.  In this regime, the 
coupling constant provides a natural small expansion parameter, allowing
use of perturbation theory in, for example,  studies of
deep inelastic scattering.   

The situation at low energies is very different.  Here the coupling
constant is large, and causes quarks to become bound into the
observed colour-neutral hadrons such as nucleons and pions.
Despite our improved understanding of the strong interaction,
the non-perturbative physics involved in the confinement of
quarks makes the calculation of the low energy interactions
of these hadrons directly from QCD extremely difficult.   

Many QCD motivated models of the nucleon exist \cite{mikesol,modofnuc}, 
but where these have been used to calculate NN interactions,
the long range contributions usually still come from pion 
and $\sigma$ exchange.
In a model with both QCD and hadronic degrees of freedom, it is 
difficult to avoid double counting.  

Fortunately, we can exploit QCD without trying to solve it
directly.  The QCD Lagrangian has certain symmetries in the limit
of vanishing quark masses.  Although these symmetries 
are explicitly broken or disguised by a variety of mechanisms,
they have important implications for the spectrum of particles
and their interactions.  

Chiral Perturbation Theory (CHPT),
which is introduced in Chapter \ref{chiral}, is an Effective Field 
Theory (EFT) which makes use of the fact that the up and down (and 
to a lesser extent strange) quark masses are small compared to typical 
hadronic scales.  CHPT has been very 
succesful in the pionic and single-nucleon sectors.
(For references and a recent introduction to EFT and CHPT 
techniques, see \cite{pich98}).

Recently, much attention has been focussed on the extension of these ideas 
to the many-nucleon systems of interest in nuclear physics, and the
two-nucleon interaction in particular.  Chiral dynamics predicts
in a natural way that the long distance, or low-energy, part of the 
two-nucleon interaction should be given by OPE but, more interestingly, it
also
predicts that there should be pieces with a two-pion range.

Various complications      arise
which are related to the large attraction in S-wave NN scattering which
gives rise to the deuteron and the unnaturally large spin singlet scattering 
length.  To take account of strong, low-energy scattering, it is 
necessary to modify 
the structure of the effective theory.  These problems and some solutions 
which have been suggested will be discussed in Chapter \ref{calc}.   
 
The resulting expansion of the NN interaction involves the exchange 
of pions and contributions from an infinite number of contact 
interactions arranged according
to the number of powers of momentum and pion masses which they contain.
In Chapter \ref{pions}, the NN potential is calculated up to third
order in this expansion.  This potential includes irreducible
two-pion exchange which introduces three low-energy constants.  We
use values of these constants which have been obtained from $\pi N$ 
scattering.

In Chapter \ref{results}, we examine the resulting phase shifts 
in peripheral partial waves ($l>1$).  These phase shifts are the
best place to look for the effects of two-pion exchange because
they receive no contributions from undetermined low energy 
constants to the order at which we shall be working.  For 
high angular momenta, the centrifugal barrier also tends
to obscure the effects of any regularisation procedure 
used to remove unphysical singularities.

The conclusions are presented in Chapter \ref{conc} where we discuss
evidence for two-pion exchange.

\chapter{QCD and Chiral Symmetry} \label{chiral}
\section{Introduction}
Detailed reviews of chiral symmetry in nuclei are given by Birse 
\cite{mikerev} and by Bernard, Kaiser and Mei{\ss}ner \cite{chpt}.  
The aim of this chapter is to give an overview of the ideas which will
be most relevant in the context of NN scattering.  We shall consider the 
implications of approximate chiral symmetry for the interactions of
pions and up to one nucleon.

Extra complications arise when two (or more) nucleons are included.
The discussion of these is deferred to Chapters \ref{calc} 
and \ref{pions}.

\section{Chiral Symmetry}

Consider a simplified two flavour theory of massless 
non-interacting quarks given by the Lagrangian
\be \label{simple}
\Lcal=\bar{\psi}i\!\!\not\!{\partial}\psi,
\ee
where $\bf{\psi}$ is a two component isospinor
containing the four component up and down quark spinors.
It is easy to check that this Lagrangian is invariant under the SU(2) 
isospin and axial isospin rotations
\beq \label{trans}
\psi'&=&e^{-i\vec\beta.\vec\tau/2}\psi
\;=\;(1-\half i{{\vec\beta.\vec\tau}})\psi, \\ \nonumber
\psi'&=&e^{-i\vec\alpha.\vec\tau\gamma_5/2}\psi
\;=\;(1-\half i{{\vec\alpha.\vec\tau}}\gamma_5)\psi,
\eeq
with isospin Pauli matrices $\vec{\tau}$ and infinitesimal vectors
$\vec{\alpha}$ and $\vec{\beta}$. Associated
with this symmetry are the conserved vector and axial-vector Noether
currents:
\be \label{curr}
J^\mu= \frac{1}{2}\bar{\psi}\gamma^\mu\psi\qquad \mbox{and}\qquad
A^\mu=\half\bar{\psi}\gamma^\mu\gamma_5\psi.
\ee
This invariance is referred to as chiral symmetry because the Lagrangian 
(\ref{simple}) can be decomposed into left and right-handed parts,
\be \label{split}
\Lcal=\bar{\psi}_Li\!\!\not\!{\partial}\psi_L
+\bar{\psi}_Ri\!\!\not\!{\partial}\psi_R,
\ee
containing fields which are helicity eigenstates,
\be
\psi_L=\half(1+\gamma_5)\psi\qquad\mbox{and}\qquad
\psi_R=\half(1-\gamma_5)\psi,
\ee
whose isospin can be 
rotated independently using combinations of the transformations
(\ref{trans}).  The symmetry group can therefore be written as
$SU(2)_L{\times}SU(2)_R$, and the associated conserved left-handed and 
right-handed currents are ${J^\mu_L}$ and ${J^\mu_R}$.
These are related to the currents (\ref{curr}) by,  
\be \label{eq:axvec}
V^\mu={J^\mu_L}+{J^\mu_R}, \qquad\mbox{and}\qquad A^\mu={J^\mu_L}-{J^\mu_R}.
\ee
The corresponding conserved charges $Q$ and $Q_5$ 
are obtained by integrating the zero components of the currents over
all space.

The Lagrangian (\ref{simple}), which has the same structure as the 
quark kinetic energy term in QCD, remains chirally symmetric if interactions 
with vector fields $\bar{\psi}\gamma_{\mu}\psi$ are included, but structures such as 
$\bar{\psi}\psi$ and therefore fermion mass terms and couplings to scalar 
fields explicitly break the symmetry.  

The relevance of chiral symmetry to the real 
world follows from the fact that typical upper values for the current quark 
masses\footnote{The masses which appear in the QCD Lagrangian.} 
are~\cite{pdg} $5$MeV for the up quark and $9$MeV for the down quark at 
a renormalisation scale of $2$GeV. 
These masses carry large errors since 
they must be extracted non-perturbatively, but they 
are certainly small compared to typical hadronic scales such as the $\rho$ mass ($770$MeV).  
The chiral limit ($m_q\rar 0$) is therefore a useful place to start when 
constructing effective field theories to describe low energy QCD.  

In the real world, the small quark masses cause the axial and vector 
currents (\ref{eq:axvec}) to have a small but non-zero divergence
This is the starting point for PCAC (Partial Conservation of the 
Axial Current) current algebra calculations ~\cite{pcac}, 
an application of chiral symmetry which predates QCD.  As we shall see
in the case of $\pi N$ scattering, 
chiral effective field theories reproduce current algebra predictions
of matrix elements  
at tree level but chiral perturbation theory allows systematic progress to 
one loop order and beyond.

The relevant part of the QCD Lagrangian can be written as follows:  
\be \label{qcdlag}
\Lcal_{QCD}=-{\frac{1}{4}}{F^{\mu\nu}_a}{F^a_{\mu\nu}}
+\bar{\bf{\psi}}
(i{\not\!\!{D}}-{\cal{M}})\bf{\psi},
\ee
where ${\cal{M}}$ is a matrix containing the quark masses, $\not\!\!{D}$
is the gauge covariant derivative and the first term contains only 
gluonic degrees of freedom.

In the chiral limit, ${\cal{M}}\rightarrow 0$, $\Lcal_{QCD}$ is 
invariant under global $SU(N_f)_L \times SU(N_f)_R\times U(1)_V \times U(1)_A$ 
transformations, where $N_f$ is the number of quark flavours.  The charge
associated with $U(1)_V$ invariance (quark number) is conserved,
but $U(1)_A$ is broken by an anomoly.

If exact chiral symmetry was realised in the particle spectrum 
(the linear or Wigner-Weyl mode) then, experimentally,  we should observe 
degenerate multiplets of hadrons with opposite parity and also some 
massless baryons~\cite{weinbook2}.  No such pattern is observed.

\begin{figure} 
\begin{center}
\epsfxsize=10cm
\epsfig{file=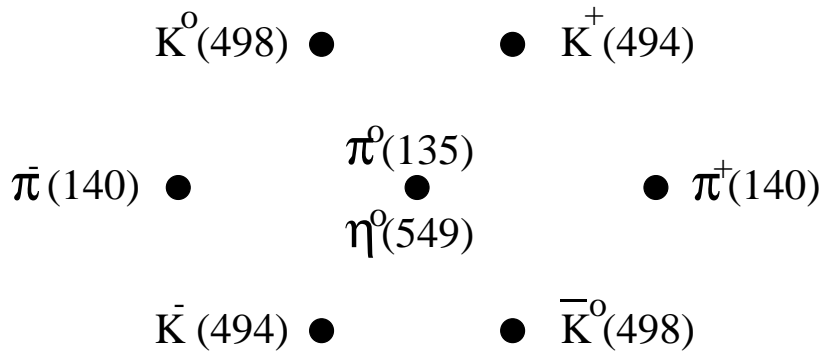}
\caption[The Pseudoscalar Meson Octet.]
{\label{fig:octet} The $0^-$ Pseudoscalar Meson Octet with masses in MeV.}
\label{fig:pin}
\end{center}
\end{figure}

Instead, a striking feature of the hadron spectrum is the 
hierarchy of meson masses.  
The pions, which have approximately equal masses 
and are abnormally light in hadronic terms, belong to a pseudoscalar 
meson octet whose members are the eight lightest physically observed hadrons.  These are shown in Figure \ref{fig:octet}. 

This pattern of masses is a strong indication that, in addition 
to the explicit symmetry breaking caused by the quark masses, 
the large classical symmetry group is somehow reduced 
dynamically.  

From now on we shall focus on the case of two flavour chiral symmetry
in which only the masses of the 
up and down quarks are treated as small quantities.   
The next lightest (strange) quark ($m_s\sim90-170$MeV) is 
considerably more massive and this leads to much stronger explicit symmetry 
breaking which is reflected in the Kaon masses.
We
shall also ignore effects produced by the mass difference
$m_u-m_d$ which means that we ignore the divergence
of the vector current.  

Examination of the particle spectrum suggests that 
only the $SU(2)_V \sim SO(3)$ part of the 
$SU(2)_L{\times}SU(2)_R \sim SO(4)$ symmetry of the massless Lagrangian 
is preserved by the vacuum.  This is known as the non-linear or 
Nambu-Goldstone mode.
Goldstone's Theorem \cite{Gold} states that when a continuous 
symmetry of the Lagrangian is not respected by the vacuum,
then the particle spectrum should contain one massless boson
corresponding to each broken generator of the original symmetry group. 
In the case of massless two flavour QCD, there would be a triplet of 
massless pseudoscalar 
bosons belonging to the three-sphere $SO(4)/SO(3)$.    
Attributing their small masses to the explicit symmetry breaking associated 
with the current quark masses, and observing that they have the 
correct quantum numbers, we identify the pions with these Goldstone bosons.

The relationship between the small masses associated with the 
pseudo-Goldstone bosons and the non-zero current quark masses 
is made explicit by the Gell-Mann-Oakes-Renner relation \cite{gor}:
\begin{equation} \label{eq:gor}
\bar{m}\langle0|\bar{\psi}{\psi}|0\rangle\simeq-f_\pi^2{m_\pi^2},
\end{equation}
where $\frac{1}{2}\langle0|\bar{\psi}{\psi}|0\rangle$ is the value of the
quark condensate in the QCD vacuum and $\bar{m}=\frac{1}{2}(m_u+m_d)$
is the average light quark mass.

The current matrix element which occurs in pion decay is,
\be \label{me}
\bra0|A^{\mu}_i(x)|\pi_{j}(q)\ket = if_{\pi}q^{\mu}e^{-iq.x}\delta_{ij},
\ee
which defines the pion decay constant $f_{\pi}=92.5\pm 0.2 \mbox{MeV}$. 
This quantity  can be measured in the charged pion decay 
process $\pi^+\rightarrow\mu^++\nu_\mu$.  Taking
the divergence of (\ref{me}) we find,
\be
\bra0|\partial_{\mu}A^{\mu}_i(x)|\pi_{j}(q)\ket = f_{\pi}m_{\pi}^2e^{-iq.x}
\delta_{ij}.
\ee
The field $\partial_\mu A^\mu/f_\pi \mpi^2$ clearly has the same behaviour
as a canonically normalised pion field close to a single pion pole, 
$q^2\sim \mpi^2$.  
In PCAC calculations, it is assumed that matrix elements of the field 
are smooth functions of $q^2$ between $q^2=0$ and $q^2=m_\pi^2$.

In the effective theory which is introduced in the next section,
the fact that pions interact weakly at low energies is crucial.
In fact, if pions were true Goldstone Bosons, their interactions would
vanish at zero energy.  This is because, in the chiral limit, 
global chiral rotations are equivalent to the creation of zero
energy pions.  Since the vacuum would be invariant under such rotations,
observable consequences of the interactions of pions with each other,
and with other particles, would vanish at threshold.

Since pions are only approximate Goldstone bosons, such 
amplitudes have residues proportional to powers of the 
pion mass at zero energy.

\section{Chiral Perturbation Theory: $\pi\pi$ scattering.}
Although the mechanism which causes chiral symmetry to be hidden is not
yet understood, using the ideas outlined in the previous section, 
we can still deduce much about the interactions of the
particles which constitute the asymptotic spectrum of QCD.  
To make this approach to hadronic interactions systematic, we 
first appeal to an unproved but plausible `theorem' of 
Weinberg \cite{wein79} which states that:

\begin{quote}
For a given set of asymptotic states, perturbation theory with the
most general Lagrangian containing all terms allowed by the assumed
symmetries will yield the most general S-matrix elements consistent 
with analyticity, perturbative unitarity, cluster decomposition and
the assumed symmetries.
\end{quote}

In QCD, the asymptotic states are the hadrons, and at lowest 
energies we
may simply consider the pions.  The above theorem indicates 
that if we
build a Lagrangian involving pion fields, and we carefully include
in it all terms consistent with unitarity, charge conjugation 
symmetry, 
parity and approximate chiral symmetry, then we will have an 
effective 
theory describing their interactions equivalent to that given by 
the low energy limit of QCD.  It will become clear however, that the effective 
Lagrangian contains many terms whose coefficients are not constrained
by the imposed symmetry.  These `low energy constants' must 
be determined experimentally.

It was 
shown some time ago that, at tree level, this
approach was equivalent to current algebra methods at lowest 
order, 
although an unambiguous continuation to one loop order and 
beyond had to 
wait for the work of Weinberg \cite{wein79}.

Going beyond tree level involves deciding which of the Feynman diagrams 
generated by the infinite number of terms in the Lagrangian should be kept.  
It might seem at first that four momentum integrals could mix 
up terms with any number of loops.  However, Weinberg 
showed that there is a consistent way to order the loop contributions.  His
scheme consists of a power counting rule according to which one assigns 
an order to 
each diagram.  This {\it{chiral power}} depends on the number of derivative 
interactions, vertices of various types,  and loops contained in the diagram.

Since the addition of baryons to Weinberg's scheme complicates the discussion
somewhat, it is instructive to see how the theory works in their absence. 
The lowest order chiral perturbation theory (CHPT) Lagrangian is presented
below.  To this order CHPT coincides exactly with the 
non-linear sigma model ~\cite{dotsm}.  In the absence of external currents, 
the Lagrangian can be constructed using the following choice of 
chiral covariant derivative,
\be
\partial_\mu U\qquad\mbox{where}\qquad U=e^{i\vec{\tau}.\vec{\pi}/f_\pi}
\ee
and takes the form 
\be \label{l0pipi}
\Lcal_{\pi\pi}^{(2)}=\frac{f_\pi^2}{4}
\;\Tr\left[\partial_\mu U\partial^\mu U^\dagger\right]
+\frac{f_\pi^2m_\pi^2}{4}
\;\Tr\left[U+U^\dagger\right].
\ee
The first term is chirally invariant, whereas the second term, whose 
coefficient has been fixed to give the correct pion mass, 
provides explicit breaking of chiral symmetry.
This choice of $U$ is known as the exponential paramaterisation. 
The Lagrangian is used by expanding in powers of the pion fields.
Keeping only terms with up to four pion fields,
\be
\Lcal_{\pi\pi}^{(2)}=\frac{1}{2}\partial_\mu\vec{\pi}.\partial^\mu\vec{\pi}
+\frac{1}{6f_\pi^2}\left[
(\vec{\pi}.\partial_\mu\vec{\pi})^2-\vec{\pi}^2
(\partial_\mu\vec{\pi}.\partial^\mu\vec{\pi})
\right]-\frac{\mpi^2}{2}\vec{\pi}^2
+\frac{\mpi^2}{48f_\pi^2}\vec{\pi}^4.
\ee
It is important to note
that chiral Lagrangians are not unique, but they will always give the
same predictions for on-shell observables.  Differing off-shell predictions
are a consequence of using different interpolating pion fields.  

From the Lagrangian, we infer the following Feynman rules.
The relativistic propagator for a pion with four-momentum $l$ is given by
\be
\frac{i}{l^2-\mpi^2+i\epsilon},
\ee
and the four pion vertex with incoming pion 
four-momenta $q_{1,2}$ and outgoing four-momenta $q_{3,4}$ by
\be \label{fourpi}
\frac{i}{f_\pi^2}\left[(q_1-q_3)^2-\mpi^2\right] + \mbox{permutations},
\ee
where isospin factors have been omitted for simplicity.  

Dimensional analysis indicates that factors of $\mpi$  or $q$ in the 
Lagrangian must always be paired with inverse 
powers of $f_\pi$.  Furthermore, it is found in actual calculations
that $f_\pi$ typically appears multiplied by factors of $4\pi$, which
are generated by loop integrals.  The scale $4\pi f_\pi$
is also roughly the scale at which the theory is expected to break down
due to the exclusion of heavy particles.

This suggests an expansion of the amplitude in powers 
of $(q^2\;\sim\;\mpi^2)/(4\pi f_\pi)^2$.   
Counting
powers of small momenta $Q\sim q,\mpi$, a propagator is clearly $O(Q^{-2})$ 
and the vertex $O(Q^2)$.
Loops, involving an integral over one four-momentum variable, count
as $O(Q^4)$.  The total chiral power $\nu$ is then given by
\be
\nu=\sum_{i}V_id_i-2I_\pi+4L.
\ee
In the above expression, 
$V_i$ is the number of vertices of type $i$, 
$d_i$ is the number of derivatives or pion masses at a type $i$ vertex, 
$I_\pi$ is the number of internal pion lines and $L$ is the number of loops.  
For example, the four pion vertex (\ref{fourpi}) has $d_i=2$.
Using Euler's 
topological identity relating the number of vertices, edges and faces
on a planar graph, $I_\pi$ can be eliminated to give:
\be \label{pc}
\nu=\sum_{i}(d_i-2)V_i+2L+2.
\ee
This result is known as Weinberg's power counting theorem.
  
\begin{figure}
\begin{center}
\epsfig{file=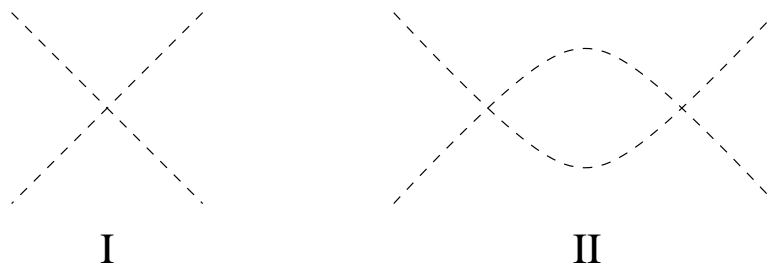}
\caption[The Tree Level and One Loop contributions to the $\pi\pi$ 
scattering amplitude.]{The (I) Tree Level and (II) One Loop contributions 
to the $\pi\pi$ scattering amplitude.  The dashed lines represent
pion fields.}
\label{fig:pipi}
\end{center}
\end{figure}

Consider the $\pi\pi$ scattering diagrams in Figure \ref{fig:pipi}.   
The four pion vertex from (\ref{l0pipi}) contains two derivatives.
In Figure \ref{fig:pipi} (I), there is one vertex with two derivatives.
This gives a total chiral power of $\nu=2$.  Now
consider Figure \ref{fig:pipi} (II).  This diagram includes two vertices 
with two derivatives each and one loop.  
The corresponding chiral power is $\nu=4$ which means that 
the one-loop 
contribution to $\pi\pi$ scattering is suppressed by two chiral powers relative
to the tree level amplitude.  

Calculation of this loop diagram leads to a 
divergence multiplying a $\nu=4$ term.  
Infinities of this type are removed by renormalising the 
relevant counterterm.  By construction, 
all appropriate counterterms with four derivatives are 
already included in the Lagrangian.  In the present
example,  the necessary term is part of $\Lcal^{(4)}_{\pi\pi}$.  
Contact terms from $\Lcal^{(4)}_{\pi\pi}$ will contribute at tree 
level when working to $O(Q^4)$.

Using these techniques to calculate $\pi\pi$ scattering lengths 
\cite{weinbook2},
one obtains values consistent with those extracted from experiment.  
The procedure of renormalising the theory order by order is characteristic 
of effective field theories, and is perfectly systematic despite the
non-renormalisability (in the strict sense) of the non-linear sigma 
model Lagrangian (\ref{l0pipi}).

The price to be paid is a large number of undetermined constants and 
new interactions every time we go to higher order in the calculation of 
a given quantity.  Once these constants are determined however, they may 
be used in the calculation of more processes giving the theory predictive 
power.  

Ultimately, it may even be possible to calculate these numbers 
directly using, for example, lattice QCD, but estimates of their 
sizes can be obtained using  
resonance saturation, the  large $N_c$ limit and extended NJL models.

\section{Heavy Baryons}

A problem arises when one tries to include relativistic heavy baryons, 
such as nucleons, in the theory.  Since these are not massless in the 
chiral limit, they introduce a new energy scale, and upset the neat counting 
scheme mentioned above.  The solution, borrowed from heavy quark
effective field theory \cite{hqet}, is to consider perturbations
around the limit in which the nucleon is considered as a static
source.  It is then 
possible to recover a consistent power counting scheme, although we now
have a dual expansion in $Q$ and $1/M$.  This technique is known
as heavy baryon chiral perturbation theory (HBCHPT).

In HBCHPT, the fully relativistic $\pi N$ Lagrangian is reduced 
by splitting the nucleon spinors into upper (light) and lower (heavy)
components, and integrating out the latter \cite{chpt}.  The resulting
nucleon propagator is independent of the nucleon mass $M$.  The 
terms involving $M$ which would arise in the expansion of the 
denominator of a relativistic propagator are now generated by
terms in the reduced Lagrangian.

In the present
context, we are only interested in contributions involving two
nucleons and up to two pions.  These will allow us to calculate
$\pi N$ scattering lengths up to order $Q^2$ in section \ref{pin}.
Using a common notation, the corresponding lowest order ($\nu=1$) chirally 
invariant two flavour, two nucleon Lagrangian is written:
\be \label{l0i}
\Lcal_{\pi N}^{(1)}=\bar{N}\left(iD_0-\frac{g_A}{2}\vec{\sigma}.\vec{u}\right)N,
\ee
defining the vector and axial vector quantities:
\be \label{def1}
D_\mu=\partial_\mu+\half\left[u^\dagger,\partial_\mu u\right] 
\qquad \mbox{and} \qquad
u_\mu=i\{u^\dagger,\partial_\mu u\}.
\ee
$D_\mu$ is a chirally covariant derivative 
and $u$, which contains the pion fields, is defined in the
following way:
\be \label{def2}
u^2=U=e^{i\vec{\tau}.\vec{\pi}/f_\pi}.
\ee

This Lagrangian is used by expanding the above matrices in
powers of the pion fields.  The part we require is obtained 
by keeping terms with up to two pion fields,
\be \label{l0}
\Lcal_{\pi N}^{(1)}=\bar{N}\left[i\partial_0-\frac{1}{4f_\pi^2}
\vec{\tau}.(\vec{\pi}\times\partial_0\vec{\pi})-\frac{g_A}{2f_\pi}
\vec{\tau}.(\vec{\sigma}.\vec{\nabla})\vec{\pi}\right]N +\ldots.
\ee 
As usual, $\vec{\sigma}$ and $\vec{\tau}$ are the Pauli spin and 
isospin matrices.
 
Chiral symmetry can be realised in the Lagrangian formalism in 
many ways, and there are several popular conventions.  Weinberg
\cite{wein2} and Ord\'o\~nez {\it{et al}}. \cite{orvk} prefer to
parameterise the pionic three sphere using stereographic
coordinates. This leads to a Lagrangian which looks quite different 
from the expression (\ref{l0i}).  As may be checked by expanding in 
powers of the
pion field, it contains the same structures as (\ref{l0}) to the order 
considered here.  Any differences which occur at higher orders
will, of course, disappear in the calculation of physical 
observables.
  
An important feature of (\ref{l0}) is
the appearance, in the final term, of pseudovector $\pi N$ coupling.
In this representation of chiral symmetry, all pion nucleon
interactions contain at least one derivative.

In time ordered perturbation theory calculations of $\pi N$ scattering
using Lagrangians with pseudoscalar $\pi N$ coupling, there are
large contributions from nucleon pair terms.  Before the implications
of chiral symmetry for the pion nucleon interaction were realised,
these unnaturally large contributions had to be removed by hand. 

In contrast, when all terms consistent with chiral symmetry are included
in the Lagrangian, as demanded by the philosophy of EFT, the
resulting amplitude is naturally small. For example,
in the linear sigma model a neat cancellation 
occurs between the pair terms and the contribution where the
two pions couple first to a $\sigma$ field which then couples 
to the nucleon.  The advantage of using 
the Lagrangian (\ref{l0}) is that all amplitudes 
involving Goldstone bosons vanish term by term at threshold in the 
chiral limit.

The last term in this Lagrangian is independently chirally invariant,
and its coefficient must therefore be fixed independently of the effective
theory.  It can
be shown that $g_A=1.26$ is the axial-vector beta decay constant,
and the pion nucleon coupling then obeys the Goldberger-Treiman
relation for on shell nucleons:
\be
g_{\pi N}=\frac{Mg_A}{f_\pi}.
\ee

The propagator for a nucleon with a four velocity $v$ and small off-shell 
momentum $l$ satisfying $v.l\sim Q$ may be written in the following way:
\be
\frac{i}{v.l+i\epsilon},
\ee
so that it is clearly $O(1/Q)$.
We are now in a position to extend the
power counting rule (\ref{pc}) to include nucleons.  Again denoting
the total chiral power by $\nu$,
\be
\nu=\sum_{i}V_id_i-2I_\pi-I_N+4L,
\ee
where $I_N$ is the number of internal nucleon lines.
This expression may be recast in a more useful form using Euler's
identity as before, and another topological identity
\be
\sum_{i}V_in_i=2I_N+E_N,
\ee
with $n_i$ equal to the number of nucleon fields 
appearing in a type $i$ vertex, and $E_N$ the number of external 
nucleon lines.  This leads to
\be
\nu=\sum_{i}V_i\left(d_i+\frac{n_i}{2}-2\right)+2L-E_N+2.
\ee
This form is particularly useful since the expression in brackets is
always greater than or equal to zero.

We shall also require the two nucleon part of the 
second order Lagrangian which
contains four terms whose coefficients are not determined by
chiral symmetry:
\beq \label{orig}
\Lcal_{\pi N}^{(2)}=\bar{N}\left[
\frac{1}{2M}\vec{D}.\vec{D}
+i\frac{g_A}{4M}\{\vec{\sigma}.\vec{D},v.u\} +2c_1\mpi^2(U+U^\dagger)\right. 
\\ \nonumber \left.
+\left(c_2-\frac{g_A^2}{8M}\right)(v.u)^2+c_3u.u+i\half\left(c_4+\frac{1}{4M}\right)
\vec{\sigma}.(\vec{u}\times\vec{u})\right]N,
\eeq
using the definitions (\ref{def1},\ref{def2}).  
Notice the appearance of terms in this Lagrangian which are 
proportional to the inverse nucleon mass.  The coefficients of these terms 
are fixed to ensure that
Lorentz invariance is recovered order by order in the energy expansion.

If the quark mass splitting
is taken into account, the Lagrangian contains an extra isospin violating
term whose coefficient is $c_5$.
After expanding in powers of the pion field, (\ref{orig}) becomes:

\beq \label{pinlag2}
\Lcal_{\pi N}^{(2)}&=&\bar{N}\frac{\nabla^2}{2M}N
-\frac{2c_1}{f_\pi^2}\mpi^2\;\bar{N}N\;\vec{\pi}^2 \\ \nonumber
&+&\left(c_2-\frac{g_A^2}{8M}\right)\frac{1}{f_\pi^2}
\bar{N}N\;\partial_0\vec{\pi}.\partial_0\vec{\pi}
-\frac{c_3}{f_\pi^2}\bar{N}N
\left[(\nabla\vec{\pi})^2-(\partial_0\vec{\pi})^2\right] \\ \nonumber
&+&\left(4c_4+\frac{1}{M}\right)
\frac{1}{8f_\pi^2}\;\epsilon_{ijk}\;\epsilon_{abc}\;\bar{N}
\sigma_k\tau_c N\;\partial_i\pi_a\partial_j\pi_b+\ldots.
\eeq

As already mentioned, Ord\'o\~nez {\it et al.} \cite{orvk} employ a different 
representation of chiral symmetry.  In this language, the low
energy constants corresponding to the $c_i$ are $B_{1,2,3}$.
Since their primary interest is $NN$ scattering, they only consider
the three terms which give rise to non-zero contributions to that
amplitude.  Comparing their Lagrangian with (\ref{pinlag2}),
we can read off the relationships
\be \label{bici}
B_1=4c_3, \qquad B_2=-\left(4c_4+\frac{1}{M}\right)\qquad\mbox{and}\qquad
B_3=8c_1.
\ee  
 
\section{$\pi N$ Scattering at tree level.} \label{pin}
To demonstrate the usefulness of the field theory defined by
(\ref{l0}) and (\ref{pinlag2}), we shall compare
the calculation of the $\pi N$ S-wave scattering
lengths up to order $\nu=2$ with their experimental values.  
The three possible diagrams containing vertices from $\Lcal_{\pi N}^{(1)}$
and $\Lcal_{\pi N}^{(2)}$ are shown in Figure \ref{fig:pin}.
\begin{figure}
\begin{center}
\epsfxsize=14.1cm
\epsfig{file=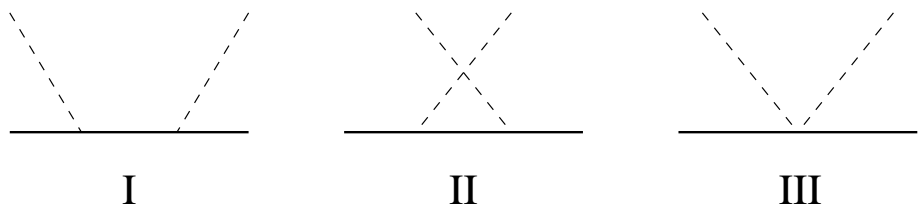}
\caption[The Feynman Diagrams for $\pi N$ scattering at tree level.]
{The Feynman Diagrams for $\pi N$ scattering at tree level.  
Dashed and solid lines represent pion and nucleon fields respectively.}
\label{fig:pin}
\end{center}
\end{figure}

The amplitudes corresponding to the two diagrams with an
intermediate nucleon vanish at threshold because the 
$\pi N$ vertex in (\ref{l0}) contains a spatial derivative 
acting on a pion field:
\be \label{nu1a}
T^{ba}_I=T^{ba}_{II}=0.
\ee
$a$ and $b$ label the isospin of incoming and outgoing pions. 
The seagull diagram, Figure \ref{fig:pin} III,
with a vertex from $\Lcal_{\pi N}^{(1)}$ produces:
\be \label{nu1b}
T^{ba}_{III}=\frac{\mpi}{2f_\pi^2}i\epsilon_{bac}\tau_c.
\ee
Together, (\ref{nu1a}) and (\ref{nu1b}) constitute the entire
$\nu=1$ calculation:
\be
T^{ba}_1=\frac{\mpi}{2f_\pi^2}i\epsilon_{bac}\tau_c.
\ee

The $\nu=2$ contributions also come from the seagull diagram, 
but now with insertions from $\Lcal_{\pi N}^{(2)}$.  The sum of these is
given by
\be
T^{ba}_2=\frac{2\mpi^2}{f_\pi^2}\left(c_2+c_3-2c_1-\frac{g_A^2}{8M}
\right)\delta_{ba}.
\ee
Notice that there is no contribution from $c_4$ at threshold since the
corresponding vertex contains two spatial derivatives. 
Decomposing the total amplitude into isospin even and odd parts,
\begin{equation} \label{piNamp}
T^{ab} = T_+\;\delta_{ab}+T_-\;i\epsilon_{abc}\tau_{c},
\end{equation}
where
\be
T_+=\frac{2\mpi^2}{f_\pi^2}\left(c_2+c_3-2c_1-\frac{g_A^2}{8M}
\right)\qquad\mbox{and}\qquad
T_-=\frac{\mpi}{2f_\pi^2}.
\ee
$T_+$ contains one term with a fixed coefficient proportional to
$1/M$ which would arise in the expansion of the intermediate
nucleon propagators in Figure \ref{fig:pin} I, II in a fully relativistic
calculation.

The even and odd scattering lengths are defined by
\begin{equation} \label{eq:scatlength1}
a_{\pm}=\frac{1}{4\pi}(1+\frac{m_\pi}{M})^{-1}T_{\pm}.
\end{equation}
The S-wave scattering lengths which are normally quoted are labelled by
total isospin and are related to $a^{\pm}$ as follows:
\begin{equation} \label{eq:Sscatt}
a_{\frac{1}{2}}=a_++2a_- \qquad a_{\frac{3}{2}}=a_+-a_-.
\end{equation}
To order $m_\pi^2$ these are given by
\be
a_{\frac{1}{2}}=\frac{\mpi}{4\pi f_\pi^2}+\frac{\mpi^2}{2\pi f_\pi^2}
\left(c_2+c_3-2c_1-\frac{4+g_A^2}{8M}\right),
\ee

\be
a_{\frac{3}{2}}=\frac{-\mpi}{8\pi f_\pi^2}+\frac{m_\pi^2}{2\pi f_\pi^2}
\left(c_2+c_3-2c_1+\frac{2-g_A^2}{8M}\right).
\ee
The order $\mpi$ results were originally obtained by Weinberg 
using current algebra \cite{curpin}:
\begin{equation} \label{sl}
a_{\frac{1}{2}}=-2a_{\frac{3}{2}}=\frac{1}{4\pi}\frac{m_\pi}{f_\pi^2}
=0.175m_\pi^{-1},
\end{equation}
and compare very favourably with the empirically determined values 
\cite{pionsnuclei}, 
\begin{equation}
a_{\frac{1}{2}}=(0.173\pm0.003)m_\pi^{-1} \qquad \mbox{and} \qquad  
a_{\frac{3}{2}}=(-0.101\pm0.004)m_\pi^{-1}.
\end{equation}
Notice that (\ref{sl}) provides a natural explanation for the smallness
of the experimentally measured isospin averaged amplitude 
$a_\half+2a_\frac{3}{2}$.

The determination of the $c_i$ from $\pi N$ scattering observables
is discussed by 
B\"uttiker and Mei\ss ner \cite{meisstriangle}.  In Chapter \ref{results} 
we shall 
use the central values from the work of Bernard {\it et al.} \cite{cidet}.  
These are listed below in units of GeV$^{-1}$
along with the quoted uncertainties:
\beq
c_1=-0.93\pm 0.10 \qquad c_2=3.34\pm 0.20 \\ \nonumber
c_3=-5.29\pm 0.25 \qquad c_4=3.63\pm 0.10.
\eeq

The $O(m_\pi^2)$ contributions to $\pi N$ scattering at threshold
using these values do not lead to a marked improvement 
of the results (\ref{sl}).  This can
be explained by the relative importance of the one loop contribution
which enters alongside vertices from $\Lcal_{\pi N}^{(3)}$ 
at order $\mpi^3$.  

The $O(m_\pi^3)$ calculation of $a_-$ and $a_+$ 
is given by Bernard {\it et al.} \cite{chpt}.  
It is interesting to note that the $O(\mpi^4)$ contribution 
to the dominant scattering length, $a_-$, is exactly zero \cite{mpi4}.

\chapter{Low Energy S-wave NN Scattering}
\label{calc}
\section{Introduction}
The success of effective chiral Lagrangians in the pion and 
pion-nucleon sectors does not carry over easily to the two-nucleon case.  
There are several interrelated reasons for this which will
be addressed in due course.  The most
obvious problem is that $NN$ scattering is, empirically, not weak 
at low energies in $l=0$ partial waves.  In the $^1S_0$ channel, the 
scattering length $a\simeq-24$ fm is unnaturally large indicating a 
nearly bound state, and there is a bound state in the coupled 
$^3S_1- {^3D_1}$ channels.
These features indicate that a naive perturbative expansion of the 
corresponding partial wave amplitudes will fail. 

Weinberg~\cite{wein2} argued that the techniques of effective field theory 
should still be applicable, so long as power counting rules are applied
to the irreducible potential rather than the amplitude.  
Solving the Lippmann-Schwinger or Schr\"odinger equation would sum up 
large contributions from intermediate, nearly on-shell nucleons.

According to the rules of effective field theory,  the Lagrangian  
must now include appropriate
four-nucleon contact interactions which have the form ~\cite{wein2};
\be \label{eq:nnlag}
\Lcal_{NN}=-\frac{1}{2}C_S(\bar{N}N)(\bar{N}N)
-\frac{1}{2}C_T(\bar{N}\vec{\sigma}N)(\bar{N}\vec{\sigma}N)+\ldots
\ee
Omitted terms contain derivatives acting on nucleon fields which lead
to a momentum and energy dependent potential.  In the approach to power 
counting explained in the previous chapter, dimensional counting was 
used to assign a chiral order to each term in the Lagrangian.
Such an ordering is clearly possible here, although it is not obvious that
it is useful when an infinite number of diagrams must be summed.

There is a more immediate, but not unrelated, complication; the 
potential obtained from (\ref{eq:nnlag}) is highly singular.  Singular 
potentials are an unavoidable consequence of using local meson-nucleon 
and nucleon-nucleon couplings 
in the Lagrangian, and one must therefore specify a regularisation and 
renormalisation scheme in order to obtain finite physical quantities.  
Many such schemes have been proposed 
\cite{afg,ksw,md1,md2,lep,rbm,bvk,pkmr,ksw2,geg,uvk,ch}, 
but the inconsistencies in the results obtained when the potential 
must describe strong scattering have led to the conclusion that Weinberg's 
power counting fails in such systems. 
(For more details and references see \cite{drp,mcb} and Section \ref{regsch}).

A different approach to power counting has recently been suggested by Kaplan, 
Savage and Wise (KSW) \cite{ksw2} in the framework of the 
Power Divergence Subtraction scheme (PDS).  It has also been 
obtained by van Kolck \cite{uvk} in a 
scheme independent approach.  Although this power counting seems
to avoid the problems just mentioned, its basis and relation to 
more familiar schemes is not immediately obvious.

In this chapter, to understand the failure of Weinberg's 
power counting and the apparent success of modified power counting, 
we shall use the method of the Wilson (or `exact') renormalisation 
group \cite{wrg} applied to a potential which is allowed to have a 
cut-off dependence \cite{bmr}.  
This is implemented by imposing a momentum cut-off on the integrals 
encountered when iterating the potential, and demanding that 
observables calculated in this way are cut-off independent.

For simplicity, we restrict our attention to the field theory containing
only nucleons with interactions given
by (\ref{eq:nnlag}).  This theory can be obtained by integrating pions 
out of the full HBCHPT which was described in the previous chapter.  We
therefore expect 
that it should be valid up to momenta of the order of the 
pion mass.  

From (\ref{eq:nnlag}) we obtain a potential with an infinite number 
of terms.
The potential is rewritten in units of the cut-off $\Lambda$.  If $\Lambda$ 
is allowed to vary, the potential must change to ensure that the theory
still has the correct long-range behaviour.  The variation of the 
potential with
$\Lambda$ is given by a renormalisation group (RG) equation.
As the cut-off is lowered to zero, all finite ranged physics 
is integrated out of the theory, and 
physical masses and energy scales no longer appear explicitly.  
In the limit, the (dimensionless) rescaled potential must therefore become 
independent of $\Lambda$, the only remaining scale.  The many 
possible limiting values of the potential are fixed points of 
the renormalisation group equation.

We shall find two kinds of fixed point: the first is trivial, corresponding to 
an identically zero potential (no scattering), and the second corresponds 
to potentials producing a bound state exactly at threshold.  

By looking at perturbations of the potential around the single 
fixed point of the first type, 
we find that it is stable and the scaling behaviour of the perturbations 
leads to Weinberg's power counting.  
This organisation of the potential is systematic in a natural situation
where perturbation theory is applicable, but not in the case of interest
(low energy S-wave NN scattering).

The simplest fixed point of the second type is unstable, and 
the power counting found here closely matches the power counting
which was found by KSW using the framework of PDS.  Perturbations 
around this fixed 
point systematically generate terms in the effective range expansion, 
which has been used to parameterise low energy S-wave NN data
for many years. 

\begin{section} {The Scattering Equation}

The irreducible potential obtained from (\ref{eq:nnlag})
only receives contributions from tree level graphs, and has the 
following simple expansion in S-wave scattering:
\begin{equation} \label{eq:pot1}
V(k',k,p)=C_{00}+C_{20}(k^2+k'^2)+C_{02}p^2\cdots,
\end{equation}
where $k$ and $k'$ are relative momenta and $p=\sqrt{ME}$ is the 
corresponding on-shell value at centre of mass energy $E$.  The coefficients
in this potential are combinations of those in the 
Lagrangian (\ref{eq:nnlag}): $C_{00}=C_S-3C_T$ etc.

We shall work with the reactance matrix $K$ since it has a simpler
relationship to the effective range expansion than the scattering matrix $T$.
The Lippmann-Schwinger (LS) equation for the off-shell $K$ matrix 
corresponding to the potential (\ref{eq:pot1}) is given by
\begin{equation} \label{eq:lse}
K(k',k,p)=V(k',k,p)+\frac{M}{2\pi^2}{\cal{P}}
\int_{0}^{\infty}q^2dq\,\frac{V(k',q,p)K(q,k,p)}
{{p^2}-{q^2}},\nonumber
\end{equation}
where ${\cal{P}}$ denotes a principal value integral.  The reactance matrix
consequently satisfies standing wave boundary conditions.   

The inverse of the on-shell $K$-matrix differs from that of the on-shell
$T$-matrix by a term $iM p/4\pi$, which ensures that $T$ is unitary if $K$ is
Hermitian.   Observables are obtained from $K$ by expanding the inverse
of its on-shell value in powers of energy:

\begin{equation} \label{eq:ere}
\frac{1}{K(p,p,p)}
=-\frac{M}{4\pi}\left(-\frac{1}{a}+\frac{1}{2}r_{e}\;p^2
+\cdots\right).
\end{equation}
This is just the familiar effective range expansion
where $a$ is the scattering length and $r_e$ is the effective range.
For weak scattering at threshold, $a<0$ corresponds to attraction and $a>0$ 
to repulsion.
$a\rightarrow\pm\;\infty$ signals a zero energy bound state
approached from above or below.

\section{Regularisation Schemes} \label{regsch}
A natural approach to regularising this theory is to impose a momentum
space cut-off at $\Lambda<\Lambda_0$ where $\Lambda_0$ is a scale
corresponding to underlying physics which has been integrated out of the
effective theory \cite{md1,md2}. 
This can be done either by introducing a form factor 
in the potential reflecting the non-zero range of the interaction, 
or a cut-off on the 
momenta of intermediate virtual nucleon states.  In the case of a sharp 
momentum cut-off, the divergent integrals which arise in solving the 
scattering equation (\ref{eq:lse}) are of the form
\begin{equation} \label{eq:int1}
I_n=\frac{M}{2\pi^2}{\cal{P}}\int_{0}^{\Lambda}dq\,\frac{q^{2n+2}}
{{p^2}-{q^2}},\nonumber
\end{equation}
where extra factors of $q^{2}$ in the numerator of the integrand occur
in loop integrals with insertions of momentum dependent factors from the 
potential.

Expanding this integral in powers of energy, we see that it has $n+1$
power law divergences,
\begin{equation} \label{eq:int2}
I_n=\frac{M}{2\pi^2}\left[
-\frac{\Lambda^{2n+1}}{2n+1}
-\frac{\Lambda^{2n-1}}{2n-1}p^2
+\ldots
-\Lambda p^{2n}
+p^{2n}\:I(p)
\right],
\end{equation}
\end{section}
where $I(p)$ is a function of $p$ which is finite as 
$\Lambda\rightarrow\infty$;
\begin{equation} \label{eq:ip}
I(p)=\frac{p}{2}\ln{\frac{\Lambda+p}{\Lambda-p}}.
\end{equation} 
A sharp cut-off is used here to avoid unnecessary complications.
If a different choice was made, the divergences in the $I_n$
would appear with different numerical coefficients, and the structure of 
$I(p)$ would change. Neither of these modifications would affect the 
conclusions below.

If the potential (\ref{eq:pot1}) is truncated at a given `order' in the
energy/momentum expansion  it has an $n$-term separable form, and we can
obtain an explicit expression for the $K$-matrix and thence the observables
in the effective range expansion.  The undetermined coefficients in the
potential may then be fixed by demanding that they reproduce the observed 
effective range expansion up to the same order. 

In this way, by including more and more terms in the potential, it is 
possible to determine whether the calculation is systematic.  By systematic 
we mean that the inclusion of extra terms in the potential should not result
in large changes to the coefficients which have already been fitted.  
Satisfying this condition ensures that the resulting Lagrangian is 
meaningful outside the process in which it is determined.

Two distinct cases have been identified.  When the coefficients in the
effective range expansion are natural
the cut-off $\Lambda$ can be chosen
in such a way that the calculation is systematic.  As long as 
\begin{equation} 
\Lambda<\!<\frac{1}{a}\sim\frac{1}{r_e},
\end{equation}
the requirements given above are satisfied.  In this weak scattering
regime, perturbation theory is valid for $p<\Lambda$, and Weinberg's power 
counting rules apply.  

The scattering length $a$ is unnaturally large in $S$-wave NN scattering
however:
\begin{equation}
a\simeq-24\:\hbox{fm}>\!> \frac{1}{\Lambda_0},
\end{equation}
where the scale of excluded physics $\Lambda_0\sim\mpi$.
In this case choosing $\Lambda<\!<1/a$ would lead to a 
(systematic) theory with an extremely limited range of validity.  
One might hope that choosing the cut-off between the energy scales set by
the inverse scattering length and effective range
could be useful:
\begin{equation}
\frac{1}{a}<\!<\Lambda<\!<\frac{1}{r_e},
\end{equation}
but now corrections to the coefficients in the potential resulting from 
the addition of extra terms will contain powers of both $1/\Lambda a$ and 
$\Lambda r_e$.  This choice does not lead to a systematic expansion in 
either $\Lambda$ or $1/\Lambda$.
 
An alternative approach is to use Dimensional Regularisation (DR) in 
which the loop integrals, $I_n$, are continued to $D$ dimensions.  In the 
commonly employed minimal subtraction scheme ($\overline{MS}$) \cite{ksw}, 
any logarithmic divergence (pole 
at $D=4$) is subtracted and power law divergences do not appear.  Since 
no logarithmic divergences appear in (\ref{eq:int2}), the loop integrals 
$I_n$ are set to zero and the $K$-matrix is simply given by the first Born
approximation,
\begin{equation}
K(k',k;p)=V(k',k,p).
\end{equation}
From the on-shell version of this identity we can obtain the 
effective range expansion which is found to converge only in the 
region $p<\sqrt{2/ar_e}$. As in the case of a cut-off $\Lambda<\!<1/a$, 
this scheme is always systematic, but useless when $a$ is large.

In the PDS scheme, linear divergences (corresponding to poles at $D=3$)
are subtracted so that the $I_n$ do not vanish.
The resulting $K$-matrix now has a 
dependence on the subtraction scale $\mu$, and choosing
\begin{equation}
\frac{1}{a}<\mu\sim p<\!<\frac{1}{r_e},
\end{equation} 
where $p$ is the momentum scale of interest, a systematic power
counting scheme emerges which appears to be useful.  The coefficient
of the leading order counterterm scales like $1/\mu$, and this 
term must therefore be treated non-perturbatively.

Note that the subtracted
term, which is linear in $\Lambda$, is the only divergence which occurs 
when the potential is momentum independent ($C_{2i,0}=0$ in (\ref{eq:pot1})).
PDS must therefore be interpreted as a momentum independent scheme.
Gegelia \cite{geg} 
has obtained a similar result by performing a momentum subtraction at 
the unphysical point $p=i\mu$.  

The results obtained using these schemes 
agree with those of van Kolck (in a scheme independent approach) \cite{uvk} 
and more recently Cohen and Hansen \cite{ch}
(in coordinate space).

The relationship between cut-off regularisation in coordinate space
and the many possible DR schemes has been clarified by
Phillips, Beane and Birse \cite{bbp}.  They start from the general
expression for the $I_n$ continued to the entire complex dimension
and energy planes.  These functions have poles at odd values of $D$ 
corresponding to infrared divergences for $D\le 1-2n$
and ultraviolet divergences for $D\ge 3-2n$.

Subtracting the $n+1$ UV divergences between $D=3-2n$ and $D=4$ recovers 
the power law divergences in (\ref{eq:int2}) with $\Lambda$ replaced
by the DR scale $\mu$.  Subtracting the infinite number of
remaining UV poles ($D>4$) reproduces the function $I(p)$ which
is defined in 
(\ref{eq:ip}).  Cut-off regularisation is, therefore, exactly 
equivalent to DR in the `uvPDS' scheme where all
UV divergences are subtracted.

In the original form of PDS \cite{ksw2}, only the $D=3$ divergence
is subtracted. This is therefore the simplest of an infinite number of
equivalent schemes.

In the next section, we shall see that PDS power counting can be
understood in terms of an expansion around a fixed point of the  
renormalisation group (RG) equation which we derive for the 
potential.  We shall also make clear that it is, at least to all
orders considered so far, equivalent to an effective range 
expansion (as suggested by van Kolck \cite{uvk}).

\section {The Renormalisation Group}

The starting point for the derivation of the RG equation
is the LS equation (\ref{eq:lse})
where all quantities are now allowed to depend on $\Lambda$.  $V$ is
the potential required to reproduce the observed $K$-matrix to 
all orders, and it will 
therefore be $\Lambda$ dependent after renormalisation  
in a cut-off scheme.  The $\Lambda$ dependence of the free Green's function
$G_0$ regulates divergent loop integrals.  As above, we use a sharp
cut-off $\Lambda$ on loop momenta.  This choice will simplify the
discussion but, as before, our results apply equally well to
any reasonable choice of momentum space cut-off.

To proceed, it is helpful to 
demand that the entire off-shell $K$-matrix is independent of 
$\Lambda$.  Note that this is a stronger requirement than is necessary 
simply to ensure cut-off independence of the resulting observables.
After differentiating the LS-equation with respect to $\Lambda$ and setting 
$\partial K/\partial\Lambda=0$, the RG-equation for 
$V(k',k,p,\Lambda)$ can be obtained by operating on the resulting expression
from the right with $(1+G_0K)^{-1}$ to produce
\begin{equation}\label{eq:rge}
{\partial V\over\partial\Lambda} 
={M\over2\pi^2}V(k',\Lambda,p,\Lambda){\Lambda^2\over\Lambda^2-p^2}
V(\Lambda,k,p,\Lambda).
\end{equation}
In the derivations which follow, we shall impose the boundary condition
that $V$ should have an expansion in powers of $p^2$, $k^2$ and $k'^2$.
This means that it can be obtained from the type of Lagrangian proposed by 
Weinberg \cite{wein2} and written in the form (\ref{eq:pot1}).

To bring out the interesting features of this approach, it is useful 
to introduce the dimensionless momentum variables, $\hat{k}=k/\Lambda$ etc.,
and the scaled potential
\begin{equation}
\hat V(\hat k',\hat k,\hat p,\Lambda)={M\Lambda\over 2\pi^2}
V(\Lambda\hat k',\Lambda\hat k,\Lambda\hat p,\Lambda),
\end{equation}
where the overall factor of $M/2\pi^2$ comes from the LS-equation 
(\ref{eq:lse}).  
Finally, we rewrite the RG-equation in terms of these new variables:
\begin{equation}\label{eq:scrge}
\Lambda{\partial\hat V\over\partial\Lambda}=
\hat k'{\partial\hat V\over\partial\hat k'}
+\hat k{\partial\hat V\over\partial\hat k}
+\hat p{\partial\hat V\over\partial\hat p}
+\hat V+\hat V(\hat k',1,\hat p,\Lambda){1\over 1-\hat p^2}
\hat V(1,\hat k,\hat p,\Lambda).
\end{equation}

In the following, the idea of a fixed point will
be important.  As $\Lambda$ varies, the RG-equation (\ref{eq:scrge})
describes how the rescaled potential must flow to ensure that the
long range behaviour of the theory does not change.  As the cut-off
is taken to zero, more and more  physics is integrated out of the 
theory.  In the limit, the cut-off is the only remaining energy scale and 
the rescaled potential, being dimensionless,  must become independent of 
$\Lambda$.  The many possible limiting values of the potential are fixed 
points of the renormalisation group equation.
To find fixed points, we look for solutions of (\ref{eq:scrge}) which satisfy
\begin{equation}
\Lambda\frac{\partial\hat V}{\partial\Lambda}=0.
\end{equation}
Near a fixed point, all physical energies and masses are large compared to 
the cut-off and power counting becomes possible.  

The simplest example is the trivial
fixed point $\hat V(\hat k',\hat k,\hat p,\Lambda)=0$. 
It is easy to see that the $K$-matrix calculated from this
potential is also zero, corresponding to no scattering.  
It is necessary to know how perturbations around the fixed point potential 
scale if we are to use it as the basis of a power counting
scheme.  This can be done by linearising the RG-equation by looking 
for eigenfunctions which scale as an integer power of $\Lambda$,
\begin{equation}
\hat V(\hat k',\hat k,\hat p,\Lambda)=
C\Lambda^\nu \phi(\hat k',\hat k,\hat p).
\end{equation}
Substituting this back into (\ref{eq:scrge}), the linear eigenvalue equation
for $\phi$ is found to be
\begin{equation}\label{eq:linrge.tr}
\hat k'{\partial\phi\over\partial\hat k'}
+\hat k{\partial\phi\over\partial\hat k}
+\hat p{\partial\phi\over\partial\hat p}
+\phi=\nu\phi.
\end{equation}
The solutions of this equation which satisfy the boundary conditions 
specified above are easily found to be
\begin{equation}
\phi(\hat k',\hat k,\hat p)=\hat k^{\prime l}\hat k^m \hat p^n,
\end{equation}
with RG eigenvalues $\nu=l+m+n+1$, where $l$, $m$ and $n$ are non-negative 
even integers.  The momentum expansion of the rescaled potential around the 
trivial fixed point is therefore given by 
\begin{equation}
\hat V(\hat k',\hat k,\hat p,\Lambda)=\sum_{l,n,m}\widehat C_{lmn}\left(
{\Lambda\over\Lambda_0}\right)^\nu \hat k^{\prime l}\hat k^m \hat p^n.
\end{equation}
(For a Hermitian potential one must take 
$\widehat C_{lmn}=\widehat C_{mln}$.)  The coefficients
$\widehat C_{lmn}$ have been made dimensionless by taking out a factor 
$1/\Lambda_0^\nu$ where, as before, $\Lambda_0$ is a scale corresponding
to underlying (integrated out) physics.

Since all of
the allowed eigenvalues are positive, this potential vanishes as 
$\Lambda/\Lambda_0\rightarrow 0$, and the trivial fixed point is stable.  
The power counting scheme associated with this fixed point can be 
made explicit by considering the unscaled potential: 
\begin{equation}\label{eq:potexp.tr}
V(k',k,p,\Lambda)={2\pi^2\over M\Lambda_0}\sum_{l,n,m}\widehat C_{lmn}
{k^{\prime l}k^m p^n\over\Lambda_0^{l+m+n}}.
\end{equation}
As long as the $\widehat C_{lmn}$ are natural, the contributions
of energy and momentum dependent terms to the potential are suppressed 
by powers of $1/\Lambda_0$.  Assigning an order $d=\nu-1$ to these
perturbations leads to a scheme which is exactly equivalent to that
originally suggested by Weinberg \cite{wein2}.

In Section \ref{regsch}, we saw that power counting 
in this way is useful in the case where scattering is weak at
low energies, and either dimensional regularisation with minimal 
subtraction or a cut-off may be used.   When the scattering length 
is unnaturally large however, we are forced to choose $\Lambda$ 
to be so small that the theory becomes essentially useless.  
To find a 
useful expansion of a potential which describes
low-energy S-wave scattering, it is necessary to 
look for a non-trivial fixed point which describes strong scattering
at low energies.

The simplest fixed point of this kind can be found by looking for
a momentum independent potential, $\widehat V=\widehat V(\hat p)$, 
which satisfies the full RG-equation.  This becomes  
\begin{equation}\label{eq:fprge.ere}
\hat p{\partial\hat V_0\over\partial\hat p}
+\hat V_0(\hat p)+{\hat V_0(\hat p)^2\over 1-\hat p^2}=0.
\end{equation}
Solving this equation subject to the boundary condition that the potential be 
analytic in $\hat p^2$ as $\hat p^2\rightarrow 0$ we obtain
\begin{equation}
\widehat V_0(\hat p)=-\left[1-{\hat p\over 2}\ln{1+\hat p\over 1-\hat p}
\right]^{-1}.
\end{equation}
The corresponding unscaled potential is 
\begin{equation} \label{eq:potfp.ere}
V_0(p,\Lambda)=-{2\pi^2\over M}\left[\Lambda-{p\over 2}
\ln{\Lambda+p\over\Lambda-p}\right]^{-1}=\;-{2\pi^2\over M}\left[\Lambda-I(p)
\right]^{-1},
\end{equation}
where $I(p)$ is the same function that was defined in (\ref{eq:ip}).
The RG-equation (\ref{eq:scrge}) was derived using a sharp cut-off, and
this form for the fixed point potential is a consequence of that 
choice.  Choosing another type of cut-off would change the second 
($p$ dependent) term in the square brackets, but the $1/\Lambda$ 
behaviour as $p\rightarrow 0$ is independent of the form of cut-off used.

Inserting the potential (\ref{eq:potfp.ere}) 
in the LS-equation, we find $1/K=0$.  This
fixed point therefore corresponds to a zero energy bound state (i.e.
the scattering length $a$ is infinite).  
Not surprisingly, this fixed point has properties which are quite different
from those we found in the trivial case.
As before, we look at perturbations around the fixed point which 
scale as an integer power of $\Lambda$:
\begin{equation} 
\hat V(\hat k',\hat
k,\hat p,\Lambda)=\hat V_0(\hat p)+C\Lambda^\nu \phi(\hat k',\hat k,\hat p),
\end{equation}
where $\phi$ is a function which is well behaved for small momenta and 
energy and satisfies the linearised RG-equation:
\begin{equation}\label{eq:linrge.ere}
\hat k'{\partial\phi\over\partial\hat k'}
+\hat k{\partial\phi\over\partial\hat k}
+\hat p{\partial\phi\over\partial\hat p}+\phi\\
+{\hat V_0(\hat p)\over 1-\hat p^2}\left[\
\phi(\hat k',1,\hat p)+\phi(1,\hat k,\hat p)\right]
=\nu\phi.
\end{equation}

The general case $\phi=\phi(\hat k',\hat k,\hat p)$ can be tackled by
breaking it down into simpler types of perturbations.  Perturbations 
which depend only on energy $\phi=\phi(\hat p)$ turn out to be particularly 
important, so we shall consider these first.  Equation (\ref{eq:linrge.ere}) 
can be integrated subject to the same boundary conditions as before, 
and the result is simply 
\begin{equation}\label{enper.ere}
\phi(\hat p)=\hat p^{\nu+1} \hat V_0(\hat p)^2,
\end{equation}
with eigenvalues $\nu=-1,1,3,5,...$.  

The  first of these eigenvalues is negative, and is associated
with an unstable perturbation since it gives rise to a term in the 
potential which scales like $\Lambda_0/\Lambda$ as $\Lambda\rightarrow0$.
  
\begin{figure} \label{fig:rgflow}
\begin{center}
\includegraphics[width=7cm]{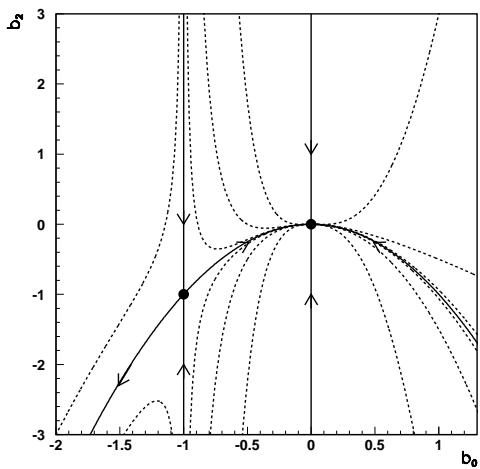}
\caption[The RG flow of the first two terms of the rescaled potential
in powers of energy.]
{The RG flow of the first two terms of the rescaled potential
in powers of energy.  The fixed points are indicated by the black dots.
The solid lines are flow lines that approach one of the fixed points
along a direction corresponding to an RG eigenfunction; the dashed
lines are more general flow lines.  The arrows indicate the direction 
of the flow as $\Lambda\rightarrow 0$.
}
\end{center}
\end{figure}

Figure \ref{fig:rgflow} shows the  
renormalisation group flow of the first two coefficients in the 
expansion of the potential in powers of energy:
\be
\hat{V}_0(\hat{p})=b_0+b_2 p^2+\ldots.
\ee
The trivial fixed point is at the origin, and the non-trivial one is at
$b_0=b_2=-1$.  Although the position of the fixed point is determined
by our choice of regulator, the pattern of RG flow is general.
The arrows show the direction of flow as $\Lambda\rightarrow 0$.

Potentials which start on the critical surface $b_0=-1$ flow 
towards the trivial fixed point in this limit.  A potential
that starts with a small perturbation away from this surface, however, 
initially flows towards the unstable fixed point for large $\Lambda$,
but approaches either the trivial fixed point or infinity as 
$\Lambda\rightarrow 0$.

We must also consider perturbations which are momentum dependent.  This time 
we look for a function $\phi$ of the form
\begin{equation}
\phi(\hat
k',\hat k,\hat p)=\hat k^n\phi_1(\hat p)+\phi_2(\hat p)
\end{equation}
which satisfies (\ref{eq:linrge.ere}).  The solutions are given by
\begin{equation}\label{eq:mompert.ere}
\phi(\hat k',\hat k,\hat p)=\left[\hat k^n-\hat p^n+\left({1\over n+1}
\right.\right.+{\hat p^2\over n-1}+\cdots\\
+\left.\left.{\hat p^{n-2}\over 3}\right)\hat V_0(\hat p)
\right]\hat V_0(\hat p).
\end{equation}
If a Hermitian potential is required, a matching perturbation of this form 
should be added with $k$ replaced by $k'$.  As usual, the
eigenvalues are given by the condition that the potential be 
a well behaved function of energy and squared momenta.  In this 
case they are $\nu=n=2,4,6\dots$.

In general, new solutions to the linearised RG-equation (\ref{eq:linrge.ere})
can be obtained by multiplying existing ones by $p^m$ where $m$ is a positive
even integer.  Applying this to the functions in equation (\ref{eq:mompert.ere})
results in a new set of eigenfunctions with corresponding eigenvalues 
$\nu=n+m$. 	

An important point to note is that the momentum dependent
eigenfunctions have different eigenvalues from the corresponding purely
energy dependent ones and so they scale differently with $\Lambda$.
This is quite unlike the more familiar case of perturbations around the trivial
fixed point where, for example, the $\hat p^2$ and $\hat k^2$ terms in the 
potential are both of the same order, $\nu=3$. It means that, 
in the vicinity of the non-trivial fixed point, one cannot use the 
usual equations of motion to eliminate energy dependence from the 
potential in favour of momentum dependence.

One more kind of perturbation is possible: a product of two factors 
of the type given in equation (\ref{eq:mompert.ere}), where one is a 
function of $\hat k$ and the other is a function of $\hat k'$, is a
solution of the linearised RG-equation with eigenvalues $\nu=5,7,9\dots$.  
These solutions can also be multiplied by powers of $p^2$ to give 
eigenfunctions with higher eigenvalues.

Any perturbation which solves equation (\ref{eq:linrge.ere}) subject
to the given boundary conditions can be expressed as a sum of those
 given above,
so we are in a position to expand a general potential around this
fixed point.  Near the fixed point we need only  
consider the perturbations with the lowest eigenvalues.  Including
the unstable perturbation, the first two give the following unscaled
potential: 
\begin{eqnarray} \label{eq:pot2}
V(k',k,p,\Lambda)=V_0(p,\Lambda)
&+&{M\Lambda_0\over 2\pi^2}\sum_{\nu=-1}^1\widehat C_\nu\left({p\over\Lambda_0}
\right)^{\nu+1}V_0(p,\Lambda)^2\\ \nonumber
&+&\widehat C_2\left[k'^2+k^2-2p^2+{2\over 3}{M\Lambda^3\over 2\pi^2}V_0\right]
{V_0(p,\Lambda)\over\Lambda_0^2}.
\end{eqnarray}
Although this potential looks complicated, it has a two-term separable
structure, and we can obtain an exact expression for the $K$-matrix
\cite{md2,rbm}.  Expanding the inverse of the on-shell $K$-matrix, 
we find the following effective range expansion
\begin{equation}\label{eq:enere.ere}
\frac{1}{K(p,p,p)}=-{M\Lambda_0\over 2\pi^2}
\sum_{\nu=-1}^1\widehat C_\nu\left({p\over\Lambda_0}\right)^{\nu+1}+\;\cdots
\end{equation}
to first order in the $\widehat C_\nu$.  

To this order, the potential is determined uniquely by comparing equation
(\ref{eq:enere.ere}) with the effective range expansion (\ref{eq:ere}):
\begin{equation} \label{eq:coefts}
\widehat C_{-1}=-{\pi\over 2\Lambda_0 a},\qquad 
\widehat C_1={\pi\Lambda_0 r_e\over 4}\qquad \ldots.
\end{equation}
The identification of the terms in the potential and the effective 
range expansion is straightforward at this order because only energy dependent 
perturbations contribute to the scattering.  It is not clear 
that this equivalence persists to higher order in the $\widehat C_\nu$.  
To illustrate the situation, we have calculated corrections to the 
unscaled potential 
(\ref{eq:pot2}) up to order $\widehat C_{-1}\widehat C_{2}$.  We find the 
following extra contributions:
\begin{equation}
\frac{M}{2\pi^2}\widehat C_{-1}\widehat C_{2}\left(
k'^2+k^2+A\;p^2+\frac{4}{3}\frac{M\Lambda^3}{2\pi^2}V_0\right)
\frac{V_0(p,\Lambda)^2}{\Lambda_0},
\end{equation}
where $A$ is a constant of integration which is not fixed
by the boundary conditions.  
This unfixed term arises from the solution of the homogeneous part of the
linearised RG equations, and has exactly the same structure as the 
$\nu=1$ term in (\ref{eq:pot2}). 

Both of these perturbations contribute to the effective range.
To avoid spoiling the one to one correspondence
between observables and terms in the potential, we are free to 
choose $A=-2$ which ensures that the contribution of 
$\widehat C_{-1}\widehat C_{2}$ to the effective range vanishes.

So long as analogous procedures can be carried out to all orders, the 
effective theory defined by an expansion around this fixed point is 
systematic and completely equivalent to the effective range expansion.
This corresponds to the fact that the parts of our potential which now
contribute to observables act like a quasipotential: an energy-dependent 
boundary condition on the logarithmic derivative of the wave function at 
the origin.  Such an equivalence has previously been suggested by 
van Kolck \cite{uvk}.

As has already been pointed out, the form of the fixed point potential 
(\ref{eq:potfp.ere}) depends on the type of cut-off used, but for small 
$p$ it always behaves like $1/\Lambda$.  We can therefore compare 
the terms that appear in the expansion of our potential (\ref{eq:pot2}) 
with those of KSW \cite{ksw2}. The subtraction scale $\mu$
in PDS acts as a resolution scale, and plays a similar role to 
$\Lambda$ in a cut-off scheme.
To first order in $1/a$ and $p^2$, their potential can be written in the form
\be
V(p,\mu)=\frac{4\pi}{M\mu}\left[-1-\frac{1}{\mu a}+\frac{r_e}{2\mu}p^2
+\cdots \right].
\ee
The leading order term in this expansion scales like $1/\mu$ which 
agrees with the $1/\Lambda$ dependence of the fixed point potential
(\ref{eq:potfp.ere}), and the factors of $1/\mu^2$ in the second and 
third terms match the $1/\Lambda^2$ factors in the energy dependent
perturbations of (\ref{eq:pot2}).
The power counting for perturbations around the non-trivial fixed point 
agrees with that of KSW if, as before, we assign 
them an order $d=\nu-1$. 

The fixed point is unstable and, since $1/a\neq 0$, as 
$\Lambda\rightarrow0$ the potential will either flow to the non-trivial
fixed point or diverge to infinity.  As long as it is possible to
choose the cut-off so that $1/a<\!<\Lambda<\!<\Lambda_0$ however, the
behaviour of the potential is dominated by the flow towards the non-trivial
fixed point, and the eigenfunctions found above still define a systematic 
expansion of the potential, as noted in \cite{ksw2}.

The approach of van Kolck \cite{uvk} is equivalent to starting at the trivial
fixed point, and following the flow line corresponding to 
the physical $K$-matrix back to the region $1/a<\!<\Lambda$ where
the power counting around the non-trivial fixed point is again useful.
This procedure involves solving the RG equation to all orders 
in the scattering length, resumming terms involving powers of
$\Lambda a$.

\section {Summary} 
In the case of weak scattering, the effective
field theory originally suggested by Weinberg \cite{wein2} to
describe NN scattering is systematic, 
and gives predictions which are independent of the regularisation scheme
used (coordinate space or momentum space cut-off, dimensional
regularisation ($\overline{MS}$), etc.).  This can be understood
in terms of an expansion of the potential around the trivial fixed point 
of the RG equation.  This is the expansion used in the next chapter
to describe NN scattering in higher partial waves.

In the presence of a resonance or bound state close to threshold, 
conventional regularisation schemes do not lead 
to power counting schemes which are both systematic and useful.  Recently,
several equivalent power counting schemes based on new 
regularisation or subtraction schemes have been proposed which appear to avoid 
the problems mentioned above.  We have demonstrated that these 
correspond to an expansion of the
potential around the simplest non-trivial fixed point of the
RG.  Terms in this potential have a one to one 
correspondence with on-shell scattering observables in the effective
range expansion to all orders considered so far.
The success of the effective range
expansion \cite{ere,bl} can therefore be understood in terms of an 
effective field theory based around this non-trivial fixed 
point \cite{bvk,pkmr,ksw2}.  
\chapter{Including Pions}
\label{pions}
\section{Introduction}
The low energy effective field theory without pions, which was considered
in the previous chapter,  is not constrained by chiral symmetry - 
all information about the interaction is contained in the values of 
coefficients of contact interactions. 

While it is important to understand how power counting works
in the restricted EFT, the inclusion of pions should significantly 
add to its usefulness since it is well known, and understood from chiral
dynamics, that they provide the longest range part of the interaction 
between nucleons. The first success of HBCHPT, which is constrained 
by chiral dynamics,  is that the leading order, finite range contribution
is the crucial One Pion Exchange (OPE) potential.

In this chapter, we consider the predictions of the effective field theory
up to third order in small momenta. This involves calculating the
one loop, two pion exchange (TPE) graphs. Adding higher order terms to the 
interaction potential allows us to test the convergence of the chiral 
expansion in the two-nucleon sector.  

As in the nucleon-only theory, naive power counting rules must be modified 
for reducible loop diagrams; those loops which involve pure two-nucleon
intermediate states are enhanced by a factor of $M/Q$.  
To a given order in the chiral expansion, this 
effect increases the number of diagrams which must be summed.  
To avoid calculating these extra diagrams explicitly, we follow Weinberg's
prescription \cite{wein2} and define an effective potential to be used in 
a Schr\"odinger equation.

As well as testing for convergence of predictions for observables
as more terms are added to the potential, we shall compare the 
results obtained when corrections to OPE are treated 
both perturbatively and non-perturbatively.  If the expansion is 
under control, then iterating the TPE potential should generate only
higher order contributions and should not result in a qualitatively 
different fit to experimental measurements.

The potentials which are presented in this chapter will be of most interest
in those partial waves for which four-nucleon contact interactions do not
contribute at the order to which we will be working ($\nu=3$).  In partial
waves which exhibit weak scattering at low energies, Weinberg power 
counting is valid, and contact interactions contribute at orders 
$\nu=0,2,4,\ldots$.  The $\nu=2$ counterterms contribute to $P$-wave 
scattering, and so the $D$-waves are the first in which we do not 
have the freedom to tune short distance physics.

Since the low energy constants required for a complete calculation to order 
$\nu=3$ have already been determined from $\pi N$ scattering, the EFT 
makes parameter-free predictions in partial waves for which $l>1$.
Of course, our decision to stop calculating at order $\nu=3$ is
purely arbitrary.  A $\nu=4$ counterterm would allow
a free parameter in $D$-waves, but counterterm contributions
to $F$-wave scattering do not enter until $\nu=6$.  For this reason
we expect that $F$-waves and above will provide the best opportunity
to test the EFT expansion.  

Furthermore, potentials derived from low energy effective theories are 
not expected to be reliable at short distances.  In fact, as we shall see, 
parts of the chiral TPE potential are substantially more singular
at the origin than the centrifugal barrier term in the Schr\"odinger
equation.  In a useful EFT, it is expected that the range of 
these singularities will 
be set by the \emph{a priori} unknown `high' energy scale $\Lambda_0$.
Provided the separation between low and high energy scales is 
large enough, the insensitivity of observables to the core of the 
potential as $l$ becomes large, and at low energies,  should 
obscure the effects of whatever procedure is used to regularise 
these singularities.

On the other hand, as $l$ becomes large the longest range interaction,
OPE, becomes dominant.  To find evidence for chiral TPE, it is necessary to
search for a window in energy and angular momentum in which NN observables
are not sensitive to short distance behaviour of the potential, but are still
not completely explained by OPE.

In $l>1$ partial waves, scattering is weak at low energies and 
Weinberg's simple power counting in $m_\pi$ and $p$ ought to be valid.  
In the language of Chapter~\ref{calc} we can expand around the 
trivial fixed point of the potential.  

The new EFT is expected 
to break down at the high energy scale which is set by the mass of 
the lightest relevant degree of freedom not included.  
For the purpose of comparison, One Boson Exchange (OBE) potentials 
provide a good description of the on-shell NN interaction.  The 
most important contributions to OBE potentials are 
OPE, the scalar $\sigma(\sim 550)$ exchange and vector $\rho(770)$ 
and $\omega(783)$ exchange.  Other mesons such as the relatively 
light pseudoscalar $\eta (549)$ are sometimes included, but 
they couple weakly to nucleons and can be safely ignored in 
the present discussion. Although the $\sigma$ does not 
correspond to an observed narrow resonance in $\pi\pi$ scattering, 
it is included to obtain the correct amount of intermediate range attraction.  

It is a well established fact that $\sigma$  exchange is a
parameterisation of the exchange of two correlated pions in 
a relative S-wave state.
In particular, the Paris~\cite{paris}
and Stony Brook~\cite{stony}
potentials 
successfully replace $\sigma$ exchange by TPE derived by 
dispersion relations from $\pi\pi$ and $\pi N$ scattering.  The relationship
between this $\sigma$ and the chiral partner of the pion is discussed by
Birse~\cite{mikerev}.

One of the aims of this investigation is to decide whether the TPE to 
be derived from the effective field theory is able to reproduce 
this important contribution.  If it is, then one might expect the 
breakdown scale to be set by the $\rho$ mass
($770$MeV).  

In practice, at least to the order in the chiral expansion considered here, 
the breakdown occurs at a much lower energy except in the partial waves 
which are well described by OPE alone. 
  
The potentials which will be presented in this chapter constitute
a complete calculation up to third order in small masses and momenta.  
This requires vertices taken from 
$\Lcali$ and $\Lcalii$. 
Contributions from the pion only Lagrangian $\Lcal_{\pi\pi}$ 
involve two loops and 
enter at order $\nu=4$, so correlated TPE diagrams with two loops are 
not included explicitly here.
A complete calculation to order $\nu=4$ introduces many other
two loop graphs which have not yet been evaluated.

The most general coordinate space potential that we will obtain 
from the effective Lagrangian may be decomposed into four radial
functions multiplying different combinations of spin factors:
\begin{equation} \label{decomp}
U(r)=U_C(r)+U_S(r)\;\vec{\sigma_1}.\vec{\sigma_2}+U_T(r)\;S_{12}+U_{LS}(r)
\;\vec{L}.\vec{S}, 
\end{equation}
with the further isospin decompositions 
$U_C=V_C(r)+W_C(r)\;\vec{\tau_1}.\vec{\tau_2}$ and so on.  
$\vec{\sigma}$ and $\vec{\tau}$ are the usual spin and
isospin Pauli matrices and the subscripts $C,S,T$ and $LS$ label the central, 
spin, tensor and spin orbit contributions respectively. The tensor
operator $S_{12}$ is given by
\be
S_{12}=3\frac{\vec{\sigma_1}.\vec{r}\vec{\sigma_2}.\vec{r}}{r^2}
-\vec{\sigma_1}.\vec{\sigma_2}.
\ee
For given values 
of total spin $S$ and orbital angular momentum $l$, the spin and isospin 
dot products evaluate to numbers: $\vec{\sigma}_1.\vec{\sigma}_2
=2\left[S(S+1)-\frac{3}{2}\right]$, and for isospin, 
$\vec{\tau}_1.\vec{\tau}_2=2\left[I(I+1)-\frac{3}{2}\right]$.  
$I=(l+S+1)$ (mod $2$) for neutron-proton scattering.

In singlet
spin states ($S=0$), the matrix elements of the tensor $S_{12}$ and 
spin-orbit $\vec{L}.\vec{S}$ operators vanish. In triplet states ($S=1$)  
$S_{12}$ can mix together partial waves which differ by two units of orbital 
angular momentum $l$, but parity conservation implies that states with
$l=j$ are decoupled.  Partial waves will be labelled using the 
universal Russell-Saunders notation $^{2S+1}l_j$.  

The potentials are to be used in non-relativistic Schr\"odinger 
equations which may be reduced by the usual methods to linear one 
dimensional differential equations in the radial variable $r$.  In uncoupled 
channels ($l=j$) the radial wavefunctions $u_j(p,r)$ satisfy 
\be \label{se1}
\left(\frac{d^2}{dr^2}+p^2-\frac{j(j+1)}{r^2}-MU(r)\right)u_j(p,r)=0.
\ee

In coupled channels we must consider two wavefunctions, which we 
shall label $u_j(p,r)$ and $w_j(p,r)$, together.  In the limit of vanishing 
tensor potential these correspond to the $l=j-1$ and $l=j+1$
wavefunctions and they satisfy the coupled differential equations
\begin{eqnarray} \label{se2}
\left(\frac{d^2}{dr^2}+p^2-\frac{j(j-1)}{r^2}-MU_{-}(r)
+\frac{2(j-1)}{2j+1}MU_T(r)\right)u_j(p,r)\\ \nonumber
=\frac{6\sqrt{j(j+1)}}{2j+1}MU_T(r)w_j(p,r)
\end{eqnarray}
and
\begin{eqnarray} \label{se3}
\left(\frac{d^2}{dr^2}+p^2-\frac{(j+1)(j+2)}{r^2}-MU_{+}(r)
+\frac{2(j+2)}{2j+1}MU_T(r)\right)w_j(p,r)\\ \nonumber
=\frac{6\sqrt{j(j+1)}}{2j+1}MU_T(r)u_j(p,r), 
\end{eqnarray}
with $U_{\pm}(r)$ given by the appropriate spin and isospin 
matrix elements of (\ref{decomp}) excluding the tensor piece.
In these expressions $M$ is twice the reduced mass of the nucleons.  
In $np$ scattering, $M=938.9$MeV and this quantity will hereafter be 
referred to simply as the nucleon mass. 

Since the potentials will be iterated to ensure that 
all $\nu=3$ contributions are included, it is important that they 
are derived from irreducible Feynman diagrams.  This 
issue arises in relation to iterated
OPE and the TPE box diagrams.
   
\section{Regularisation} \label{reg}
In general, the potentials obtained from the effective theory are
more singular than $1/r^2$ for small $r$.  Consequently, 
it is not possible to integrate (\ref{se1}),(\ref{se2}) and (\ref{se3})
without some form of regularisation.  This is a natural state of 
affairs in a field theory; singular potentials in coordinate space
are analogous to bad ultraviolet behaviour and 
divergent loop integrals in momentum space.

The procedure for handling such momentum space divergences within a natural
power counting 
regime was outlined in Chapter~\ref{chiral}.  The integrals are performed 
using a regularisation scheme which introduces a new scale 
$\Lambda$ (cut-off) or $\mu$ (DR):
\be \label{rn}
 p\mbox{,}\mpi < \Lambda\mbox{,}\mu <\!< \Lambda_0.
\ee
It is evident that the physical system must have 
a clear separation of energy scales to allow a choice of 
$\Lambda$ satisfying (\ref{rn}).   
  In cut-off regularisation, the ratio
$\Lambda/\Lambda_0$ may be left finite or taken to zero.  In the former
case, cut-off dependent contributions to observables are suppressed 
by powers of $\Lambda/\Lambda_0$.  The assumption that the 
separation of scales is sufficient to justify using the EFT may 
be checked by testing the sensitivity of observables to $\Lambda$.    
If the cut-off is to be removed, the divergent part of the 
integral is subtracted by renormalising the coefficient of the relevant 
counterterm.  

Another way of obtaining finite observables is to perform 
a perturbative expansion of the amplitude in momentum space 
counting chiral powers.  This approach 
is used by Kaiser {\it et al.} \cite{kbw}
who perform a partial calculation of the amplitude 
to third order in small momenta.  In this calculation, 
no infinities occur for finite momenta.

In the present work, the potential is calculated in momentum space
and Fourier transformed into coordinate space before any
loop integrals are performed.  The resulting potentials are as 
singular as $1/r^6$ for small $r$.
Many approaches to regularisation are possible in coordinate space.  
The connection to power counting is, in general, more difficult 
to expose than in momentum space.

For example, Ord\'o\~nez {\it et al.} \cite{orvk} 
use a Gaussian form factor in the momentum space potential which 
leads to unwieldy expressions in  coordinate space, especially
when TPE is included.  

Here we shall choose to avoid the singular region by 
solving the Schr\"odinger equation for $r>R$ and imposing
a boundary condition on the logarithmic derivative of the wave function
at $r=R$.  Coordinate space cut-offs have also been used 
by Cohen and Hansen \cite{ch} in an EFT context.
Interestingly, they were used by Feshbach and Lomon \cite{fl}
to investigate TPE potentials in the 1960's.

We have already stated however, that the effective field theory 
is parameter-free in $l>1$ partial waves.
Ideally, $R$ should be chosen to be small enough that the 
desired long distance physics predicted by the effective theory 
is included, but large enough that any unphysical singularities 
do not dominate the centrifugal barrier term in the Schr\"odinger equation.  
This is analogous to the choice (\ref{rn}) of cut-off in momentum 
space with the intermediate scale $\Lambda$ replaced by $1/R$.  

If this condition is met, then the value of the logarithmic derivative
may be expanded in powers of energy, and it is possible to determine how 
its value should scale with $R$ in order for observables to remain unchanged.
This type of analysis is closely related to the renormalisation group
treatment of the previous chapter.  In what follows we shall use an energy independent boundary condition.

Because, as will be explained, the lowest order pionic (static OPE) 
contribution is ubiquitous and uniquely defined, it will be included in 
all of our potentials.  We wish to examine corrections to the 
corresponding phase shifts. 
 
To begin, it is possible to define two solutions, $f_+$ 
and $f_-$, to the uncoupled Schr\"odinger equation (\ref{se1}) with $U(r)$
equal to the OPE potential such that their asymptotic values 
satisfy~\cite{newton}
\be 
f_{\pm}(p,r)\longrightarrow e^{\pm ipr}\;\mbox{ as }\; r\rightarrow{\infty},
\ee
suppressing angular momentum labels.
In general, both of these functions will be irregular at the
origin:
\be
f_{\pm}(p,r)\sim (pr)^{-l}\;\mbox{ as }\; pr \rightarrow 0.
\ee

The `physical wavefunction' in the $l$'th partial wave, $\phi_l(r)$, 
is obtained by choosing a linear 
combination of these functions which is regular at the origin:
\begin{eqnarray} \label{fred}
\phi_l(p,r)&\sim&(pr)^{l+1}\;\mbox{ as }\; pr\rightarrow 0, \\ \nonumber
\phi_l(p,r)&\rightarrow&\sin(pr+\delta_\pi)\;\mbox{ as }\; r\rightarrow{\infty}
\end{eqnarray}
defining $\delta_\pi$ as the phase shift due the OPE potential alone.  
We also define a corresponding irregular function
\begin{eqnarray} \label{bob}
f_l(p,r)&\sim&(pr)^{-l}\;\mbox{ as }\; pr\rightarrow 0, \\ \nonumber
f_l(p,r)&\rightarrow&\cos(pr+\delta_\pi)\;\mbox{ as }\; r\rightarrow{\infty}.
\end{eqnarray}    

It is then straightforward to demonstrate that the physical wavefunction
$\chi_l(r)$, corresponding to OPE modified by the addition 
of an extra short-range interaction, will satisfy
\be
\chi_l(r)=\sin(\delta-\delta_\pi)f_l(p,r)+\cos(\delta-\delta_\pi)\phi_l(p,r),
\ee
where $\delta\equiv\delta(p)$ is the total phase shift.

Now consider the case of interest - imposing a boundary condition on 
the wavefunction's logarithmic derivative at $r=R$:
\be
R\frac{d}{dr}\ln{\chi}_l(p,r)\;\Big|_{r=R}\;=l+1+\alpha.
\ee
Notice that, in the case $\alpha=0$, the small $pr$ behaviour
of the wavefunction is not modified by the boundary condition. 

Using the limiting behaviour given in (\ref{fred}) and (\ref{bob}),
we obtain the following expression for the tangent of the 
change in the phase shift due to the boundary condition
as $pR\rightarrow 0$:
\be
\tan{(\delta-\delta_\pi)} \sim (pR)^{2l+1} \frac{\alpha}{2l+1+\alpha}.
\ee

The case $\alpha=0$ corresponds to the EFT prediction in partial
waves with $l>1$; for small enough $pR$, phase shifts are not
affected by the boundary condition.  
Note that, if we fix $\delta-\delta_\pi$ at a particular 
(sufficiently low) energy, on dimensional grounds, $\alpha$ 
must scale like:
\be
\alpha\sim\widehat{C}(2l+1)\left(\frac{1}{R\Lambda_0}\right)^{2l+1}.
\ee
where $\widehat{C}$ is a dimensionless number.
The leading order perturbation around the trivial RG fixed
point which contributes to scattering in the $l$'th partial
wave scales in the same way, but with the inverse 
cut-off radius $R$ replaced by $1/\Lambda$.  
 
\section{Static One Pion Exchange}

At next to leading order ($\nu=0$), we must introduce OPE.
Since it is well known that OPE accounts for the 
longest range part of the NN interaction, it is not surprising
that it dominates the predictions for phase shifts in higher 
partial waves ($l>4$).

\begin{figure}
\begin{center}
\includegraphics[width=5cm]{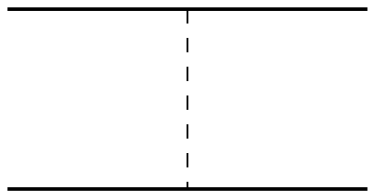}
\caption[The Feynman Diagram for OPE.]{\label{fig:ope} The Feynman Diagram for OPE.  
The solid lines represent nucleons and the dotted line represents 
a pion propagator.}
\end{center}
\end{figure}

The static OPE potential may be obtained in momentum space 
by calculating  the diagram in Figure \ref{fig:ope}. 
It is of order $\nu=0$ and requires two single derivative 
vertices taken from $\Lcali$:
\be
-\left(\frac{g_A}{2f_\pi}\right)^2\frac{\vec{\sigma_1}.\vec{q}
\;\;\vec{\sigma_2}.\vec{q}}{q^2+m_\pi^2}
\;\vec{\tau_1}.\vec{\tau_{2}}, 
\ee 
with centre of mass three momentum transfer $\vec{q}=\vec{k}-\vec{k'}$. 
Non-static corrections to this, proportional to $1/M$,
do not enter until order $\nu=3$ in the momentum expansion.
Fourier transforming this potential leads to
the familiar Yukawa potential in coordinate space plus a delta function piece 
which may be absorbed by a redefinition of the coupling constants $C_S$ and
$C_T$ multiplying the four nucleon contact terms considered in 
Chapter~\ref{calc}.  The remainder produces the following contributions:
\begin{eqnarray} \label{opepot}
W^{(0)}_{S}(r)&=&\frac{\mpi^2 g_A^2}{48\pi f_\pi^2}\;
\frac{e^{-x}}{r} \\ \nonumber
W^{(0)}_{T}(r)&=&\frac{\mpi^2 g_A^2}{48\pi f_\pi^2}\;
\frac{3+3x+x^2}{x^2}\;\frac{e^{-x}}{r}
\end{eqnarray}
where, from now on, $x$ will represent $\mpi r$ and the superscript
indicates that this potential is of order $\nu=0$.

The central part of the OPE potential behaves like $1/r$ for
small $r$, and it is therefore possible, for $S=0$, to obtain regular
and irregular solutions of the corresponding Schr\"odinger equation
in the usual way without regularisation. However, the singular behaviour 
leads to a term proportional to 
$r\ln{\mpi r}$ in the expansion of the irregular $l=0$ partial wave
solution. This term becomes important when the Yukawa potential is used 
alongside zero range potentials as in Section \ref{dere}.
    
The spin and isospin structure of the central OPE potential means that 
it is alternately attractive and repulsive for spin singlet even and odd 
partial waves respectively.  Taken on its own, the OPE potential
is too weak to produce the strong scattering observed in S-waves.
The strength of the central potential is about a third of the size
necessary to produce a bound state~\cite{pionsnuclei}.

An impressive feature of
OPE is the tensor contribution proportional to $S_{12}$.  This 
causes mixing between triplet partial waves. Without mixing of this
form the deuteron would not be bound, as the pure $^3S_1$ interaction
is not strong enough to support a bound state.  In fact, OPE produces 
the bulk of the mixing required in most coupled channels as will
be obvious from the phase shifts plotted in Chapter~\ref{results}.   
   
The $1/r^3$ behaviour of the tensor term in (\ref{opepot}) for small $r$
means that the
potential must be regularised if it is to be iterated to obtain 
the corresponding spin triplet phase shifts. 
Although the central part of OPE is
relatively well behaved for small $r$, from an EFT point of view it
is no more reliable here than the highly singular TPE contributions.

When comparing the predictions of these potentials to partial
wave analyses, it is necessary to remember that the latter are
essentially phenomenological potential models.  The good agreement
we shall find between OPE and the higher Nijmegen partial waves at 
low energies is simply
due to the fact that the potentials have the same content for large $r$.
In fact, to facilitate comparison at higher energies, we choose to
use their form for the OPE potential.  This form of the potential includes
effects due to the small pion mass splitting which is usually
ignored in chiral effective field theory treatments since the
scale of the mass differences is set by the fine structure constant.  
This OPE is given in the Appendix, and is equivalent to (\ref{opepot})
when the pion masses are made equal.

Isospin violation in NN scattering is discussed by Epelbaum and Mei\ss ner 
~\cite{nniso} in the context of PDS.  
Although we use the Nijmegen expression for OPE, we shall ignore all other 
contributions which are $O(\alpha)$.

The deviation of phase shifts from static OPE, at energies where 
phenomenological potentials begin to differ, is therefore 
of most interest to us since we hope to look for the effect of
model independent non-static OPE and TPE contributions
constrained by chiral symmetry.

\section{Non-Static OPE and TPE}

\begin{figure} 
\begin{center}
\includegraphics[width=15cm]{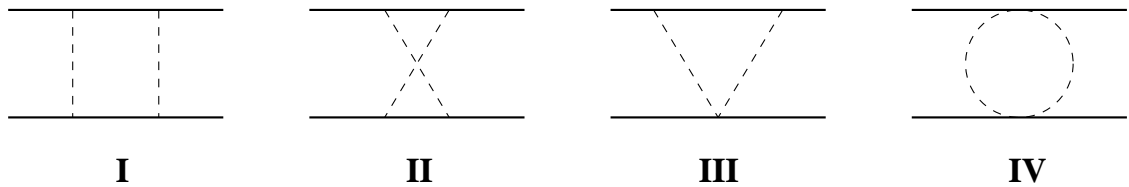}
\caption[The Two Pion Exchange Feynman Diagrams which contribute 
to NN scattering at order $\nu=3$.]
{\label{fig:tpe} The Two Pion Exchange Feynman Diagrams which 
contribute to NN scattering at order $\nu=3$.}
\end{center}
\end{figure}

The non-static corrections to OPE, and the leading order 
TPE contributions both enter at order $\nu=2$.  Non-static 
corrections to TPE start at order $\nu=3$ along with 
the leading seagull counterterms from $\Lcalii$.

The Feynman diagrams which contribute to two pion exchange up to order
$\nu=3$ are shown in Figure \ref{fig:tpe}.  Vertex corrections to OPE 
and contact interactions only lead to coupling constant and wave
function renormalisation \cite{egm}.

Non-static contributions proportional to
$1/M$ arise in a calculation of the Feynman diagram 
in Figure \ref{fig:ope} to order $\nu=2$.
This leads to an energy dependent 
potential. 
Ord\'o\~nez 
{\it et al.} \cite{orvk} use the energy dependent form of OPE directly, but
to keep the present calculation as simple as possible, we choose
to transform away this energy dependence analytically using a 
redefinition of the wavefunction.  For this reason, our TPE potential
is not the same as that of Ord\'o\~nez {\it et al.} (in the point
coupling limit).

Iterating the OPE potential once produces a 
reducible TPE interaction with the same structure as that obtained from
Figure \ref{fig:tpe} I and II involving four vertices from  
$\Lcali$ and two nucleon propagators.  Changing the definition of OPE 
alters which parts of the interaction derived from these diagrams must
be subtracted to make it irreducible.
This ambiguity, along with off-shell ambiguities, has historically been 
the source of a great deal 
of confusion regarding the correct form of the two pion exchange 
potential \cite{tmo,bw}. 

To obtain the correct potentials, it is useful in practice to
use time-ordered (or old-fashioned) perturbation theory to calculate the   
diagrams of Figures \ref{fig:ope} and \ref{fig:tpe} \cite{wein2,orvk}.  
In this approach, it is straightforward to separate reducible TPE graphs 
(which contain pure two-nucleon intermediate states) from genuine
TPE contributions to the potential. 

The correct form of the static TPE potential to be used with an energy 
independent static OPE interaction has been obtained by Coon and
Friar \cite{coonfriar,friarcoon}.  They show how to remove iterated 
static OPE
from their TPE potential by adding and subtracting it from the non-pionic 
part of the two nucleon Hamiltonian.  The former piece is implicitly included 
in the two nucleon Green's function, and the latter part is treated as a 
counterterm.  Energy dependence in the resulting potential is then removed by 
a wave function redefinition.

A different method of removing energy dependence is considered
by Epelbaoum {\it et. al.} \cite{egm},
but they use equations of motion which involve trading off energy
dependence and momentum dependence.  As we found in the previous
chapter, energy and momentum dependent perturbations scale differently 
around the non-trivial RG fixed point.  This method must therefore
be applied to S-wave scattering with care.

There are two remaining TPE diagrams which contribute at order $\nu=2$.
The double seagull or `bubble' diagram (Figure \ref{fig:tpe} IV) 
has one loop and two vertices from $\Lcali$, and the `triangle' 
diagram (Figure \ref{fig:tpe} III)
has three vertices from $\Lcali$ and one nucleon propagator.

The contributions from Figure \ref{fig:tpe} III and IV 
are spin independent because both single pion vertices 
connect to the same nucleon leg.  Since these diagrams are obviously 
irreducible, they may be calculated without further ado.  A few intermediate 
steps are given to illustrate the techniques used.  More details 
may be found in ~\cite{friarcoon}.  

In momentum space, discarding $O(1/M)$ corrections for the moment, 
the contribution from 
the bubble diagram, Figure \ref{fig:tpe} IV, can be written as
\be
\frac{-1}{128f_\pi^4}\vec{\tau_1}.\vec{\tau_2}\int\frac{d^3l}{(2\pi)^3}
\frac{1}{\omega_+\omega_-}\frac{(\omega_+-\omega_-)^2}{\omega_++\omega_+},
\ee
with $\omega_{\pm}=\sqrt{(\;\vec{q}\pm \vec{l}\;)^2+4\mpi^2}$.

This expression can be obtained either by calculating the Feynman diagram 
and integrating over $l_0$, or by summing contributions from the 
corresponding two time ordered diagrams.  Making several changes of 
variable and performing a Fourier tranform it is possible to obtain the 
following symmetric form for the potential in coordinate space:
\be
V_{IV}(r)=\frac{-1}{8f_\pi^4}\vec{\tau_1}.\vec{\tau_2}
\int\frac{d^3q}{(2\pi)^3}\int\frac{d^3l}{(2\pi)^3}
\;e^{i(\vec{q}+\vec{l}).\vec{r}}\;
\frac{(\omega_q-\omega_l)^2}{\omega_q\omega_l(\omega_q+\omega_l)},
\ee
using the notation $\omega_q=\sqrt{q^2+\mpi^2}$.  The integrand may
be simplified by discarding two divergent (zero range) contributions 
and made separable at the cost of introducing an integral over a new 
variable $\lambda$:
\begin{eqnarray}
V_{IV}(r)&=&\frac{1}{4f_\pi^4(2\pi)^6}\;\vec{\tau_1}.\vec{\tau_2}
\;\int_0^\infty d\lambda \;
\lambda^2\left[\int d^3 q \;e^{i\vec{q}.\vec{r}}\frac{1}{q^2+m^2+\lambda^2}\right]^2 \\
\nonumber
&=&\frac{\mpi}{32f_\pi^4(2\pi)^3r^2}\;\vec{\tau_1}.\vec{\tau_2}\;
\left[\left(\frac{d^2}{dr^2}-4\mpi^2\right) K_1(2\:\mpi r)\right].
\end{eqnarray}
Using the properties of the modified Bessel functions $K_n$, we arrive at the
final form of the potential:
\be
V_{IV}(r)=\frac{\mpi}{128\pi^3f_\pi^4r^4}
\;\vec{\tau_1}.\vec{\tau_2}\;\left[xK_0(2x)+K_1(2x)\right].
\ee
Note that this spin independent isovector contribution has no important counterpart
in OBE potentials. 

The single seagull or `triangle' diagram leads to
\be
\frac{-g_A^2}{16f_\pi^4}\vec{\tau_1}.\vec{\tau_2}\int\frac{d^3l}{(2\pi)^3}
\frac{\vec{q}^2-\vec{l}^2}{\omega_+\omega_-(\omega_++\omega_-)},
\ee
which becomes, in coordinate space,
\be
V_{III}(r)=\frac{-g_A^2}{f_\pi^4}\vec{\tau_1}.\vec{\tau_2}
\int\frac{d^3q}{(2\pi)^3}\int\frac{d^3l}{(2\pi)^3}
\;e^{i(\vec{q}+\vec{l}).\vec{r}}\;
\frac{\vec{q}.\vec{l}}{\omega_q\omega_l(\omega_q+\omega_l)}.
\ee
To remove the dot product we differentiate under the integrals 
and separate as before:
\be
V_{III}(r)=\frac{2g_A^2}{\pi f_\pi^4}\;\vec{\tau_1}.\vec{\tau_2}\;
\lim_{r_1,r_2\rightarrow r}
\vec{\nabla_1}.\vec{\nabla_2}
\int_0^\infty d\lambda \int d^3q\;
\frac{e^{i\vec{q}.\vec{r_1}}}{\omega_q^2+\lambda^2}
\int d^3l\;
\frac{e^{i\vec{l}.\vec{r_2}}}{\omega_l^2+\lambda^2}
\ee
which can be evaluated and simplified to give 
\be
V_{III}(r)=\frac{\mpi g_A^2}{64\pi^3 f_\pi^4 r^4}\;\vec{\tau_1}.\vec{\tau_2}\;
\left[
(2x^2+5)K_1(2x)+5xK_0(2x)
\right],
\ee
after performing the derivatives and taking the limit.

The total of the contributions mentioned above constitutes 
the static part of the $\nu=2$ TPE potential.  This is presented 
below in coordinate 
space decomposed according to (\ref{decomp}).  Only the 
non-zero contributions are shown.
\begin{eqnarray} \label{nu=2}
W^{(2)}_C&=&\frac{\mpi}{128\pi^3f_\pi^4r^4}
\left[1+2g_A^2(5+2x^2)-g_A^4(23+12x^2)\right]
K_1(2x) \\ \nonumber
&+&x\left[1+10g_A^2-g_A^4(23+4x^2)\right]K_0(2x) \\ \nonumber
V^{(2)}_S&=&\frac{g_A^4\mpi}{32\pi^3f_\pi^4r^4}
\left[3xK_0(2x)+(3+2x^2)K_1(2x)\right] \\ \nonumber
V^{(2)}_T&=&\frac{g_A^4\mpi}{128\pi^3f_\pi^4r^4}
\left[-12xK_0(2x)-(15+4x^2)K_1(2x)\right] \\ \nonumber
\end{eqnarray}
For small $r$, all of the above potentials behave like
$r^{-5}$.  For comparison, the longest
range part of each of these potentials is listed below: 
\be \label{asym1}
W_C^{(2)}\sim-\frac{\mpi^{7/2}g_A^4}{64\pi^{5/2}f_\pi^4}\;\frac{e^{-2x}}{r^{3/2}},\qquad
V_S^{(2)}\sim \frac{\mpi^{5/2}g_A^4}{32\pi^{5/2}f_\pi^4}\;\frac{e^{-2x}}{r^{5/2}},
\ee
\beq \nonumber
V_T^{(2)}\sim-\frac{\mpi^{5/2}g_A^4}{32\pi^{5/2}f_\pi^4}\;\frac{e^{-2x}}{r^{5/2}}.
\eeq
In each case the long-range contribution is from a box diagram.

The $\nu=3$ contributions to the TPE potential are now
introduced. These potentials arise from diagrams of the 
form Figure \ref{fig:tpe} III with the seagull vertex taken 
from $\Lcalii$ and may be calculated straightforwardly
using methods similar to those described above.  
There is no contribution from the term in the Lagrangian
proportional to the low energy constant $c_2$.  This 
was checked by calculating the diagram explicitly,
but it is related to the fact that the momentum transfer 
vanishes in the centre of mass \cite{kbw}.   In contrast, 
the vanishing contribution in $\pi N$ scattering was proportional 
to $c_4$.
\begin{eqnarray} \label{nu=3}
V^{(3)}_C&=&\frac{3g_A^2}{32\pi^2f_\pi^4}\frac{e^{-2x}}{r^6}
\left[2c_1x^2(1+x)^2+c_3(6+12x+10x^2+4x^3+x^4)\right] \\ \nonumber
W^{(3)}_S&=&\frac{g_A^2}{48\pi^2f_\pi^4}\frac{e^{-2x}}{r^6}
c_4(1+x)(3+3x+x^2) \\ \nonumber
W^{(3)}_T&=&-W^{(3)}_S
\end{eqnarray}
These potentials are more singular than the $\nu=2$ potentials;
they all behave like $r^{-6}$ near $r=0$.   At large distances,
the dominant terms are:
\be \label{asym2}
V_C^{(3)}\sim\frac{3g_A^2\mpi^4}{32\pi^2f_\pi^4}(2c_1+c_3)\frac{e^{-2x}}{r^2}
\qquad \mbox{and} \qquad
W_S^{(3)}\sim\frac{g_A^2\mpi^3}{48\pi^2f_\pi^4}c_4\frac{e^{-2x}}{r^3}.
\ee
Notice that these central, spin and tensor potentials have shorter
ranges than their $\nu=2$ counterparts.

The $1/M$ corrections must be calculated with care.  In a relativistic
calculation these arise when retardation in intermediate
propagators is taken into account.  In the calculation of 
$\pi N$ scattering we saw that, in HBCHPT, they 
come from terms in the $\nu=2$ Lagrangian with fixed coefficients.
The correct form of these contributions for the
energy independent potential used here is given by Friar \cite{friar97}.
For completeness, they are listed below.
\begin{eqnarray} \label{1/M}
V^{(M)}_C&=&\frac{1}{32M}\frac{3g_A^4}{32\pi^2f_\pi^4}\frac{e^{-2x}}{r^6}
\left[6+12x+16x^2+16x^3+7x^4+x^5\right] \\ \nonumber
W^{(M)}_C&=&\frac{1}{32M}\frac{-g_A^2}{4\pi^2f_\pi^4}\frac{e^{-2x}}{r^6}
\left[6+12x+10x^2+4x^3+x^4\right]\\ \nonumber
&+&
\frac{1}{32M}\frac{g_A^4}{16\pi^2f_\pi^4}\frac{e^{-2x}}{r^6}
\left[36+72x+62x^2+28x^3+8x^4+3x^5\right] \\ \nonumber
V^{(M)}_S&=&\frac{1}{32M}\frac{-g_A^4}{16\pi^2f_\pi^4}\frac{e^{-2x}}{r^6}
\left[6+12x+11x^2+6x^3+2x^4\right] \\ \nonumber
W^{(M)}_S&=&\frac{1}{32M}\frac{g_A^2}{6\pi^2f_\pi^4}\frac{e^{-2x}}{r^6}
(1+x)(3+3x+2x^2) \\ \nonumber
&+&\frac{1}{32M}\frac{-g_A^4}{24\pi^2f_\pi^4}\frac{e^{-2x}}{r^6}
\left[18+36x+31x^2+14x^3+2x^4\right] \\ \nonumber
V^{(M)}_T&=&\frac{1}{32M}\frac{3g_A^4}{32\pi^2f_\pi^4}\frac{e^{-2x}}{r^6}
\left[12+24x+20x^2+9x^3+2x^4\right] \\ \nonumber
W^{(M)}_T&=&\frac{1}{32M}\frac{-g_A^2}{6\pi^2f_\pi^4}\frac{e^{-2x}}{r^6}
(1+x)(3+3x+x^2) \\ \nonumber
&+&\frac{1}{32M}\frac{g_A^4}{48\pi^2f_\pi^4}\frac{e^{-2x}}{r^6}
\left[36+72x+52x^2+17x^3+2x^4\right] \\ \nonumber
V^{(M)}_{SO}&=&\frac{1}{32M}\frac{-3g_A^4}{2\pi^2f_\pi^4}\frac{e^{-2x}}{r^6}
(1+x)(2+2x+x^2) \\ \nonumber
W^{(M)}_{SO}&=&\frac{1}{32M}\frac{g_A^2(g_A^2-1)}{\pi^2f_\pi^4}
\frac{e^{-2x}}{r^6}(1+x)^2  
\end{eqnarray}
Taken together, (\ref{opepot}),(\ref{nu=2}),(\ref{nu=3}) and (\ref{1/M})
constitute the entire potential to order $\nu=3$ in the chiral expansion and 
next to leading order in the expansion of the inverse nucleon mass.  

This potential has been used in several other calculations. 
Friar \cite{friar97} has demonstrated that it is one of a class of 
infinitely many equivalent potentials.  The Nijmegen group
\cite{nijfriar} have recently used it as part of a new proton-proton 
partial wave analysis, and Kaiser {\it et al.} \cite{kbw} 
originally wrote down this form after taking the Fourier transform 
of their perturbatively derived on-shell scattering amplitude,
but not in the context of a particular iteration scheme.

\section{Distorted Wave Born Approximation} \label{dwba}

The Distorted Wave Born Approximation (DWBA) will be used in two ways.  
In the next section
we shall use it to examine the effect of perturbatively removing 
OPE from the phase shifts produced by a strong, short-range potential.  
 
Secondly,  it will allow us to consider treating
TPE perturbatively and non-perturbatively while keeping
OPE to all orders in Chapter \ref{results}.

The DWBA can be derived from the
well known and exact two-potential formula,
\be
\tan{\delta_l^{(1)}}-\tan{\delta_l^{(2)}}=-p\int_{0}^{\infty}u_l^{(1)}(r)
\left[U^{(1)}(r)-U^{(2)}(r)\right]u_l^{(2)}(r)dr
\ee
by writing $U^{(1)}(r)= U^{S}(r)$
and $U^{(2)}(r)= U^{S}(r)+U^{W}(r)$.  $U^{S}$ is a strong potential
to be treated to all orders, while $U^W$ is an additional
weak potential for which we expect first order perturbation theory 
to be accurate.  The substitution results in an approximate expression
relating the change in the scattering function to the (suitably normalised)
wave function due to the potential $U^S$ acting alone and the 
weak potential $U^W$:
\be \label{eqn:dwba}
\tan{\delta_l}\simeq\tan{\delta_l^{S}}-p\int_{0}^{\infty}u_l^{S}(r)
U^W u_l^{S}(r)dr.
\ee
Corrections to this expression are proportional to $(u_l-u^S_l)U^W$.
This approximation is so-called because (\ref{eqn:dwba}) is essentially 
the Born approximation with 
the plane waves replaced by waves `distorted' by the strong
potential.  Notice that it reduces to the Born approximation 
when $U^S$ and the resulting phase shifts are zero.

\section{S-wave Scattering With Pions.} \label{dere}

The subtleties discussed in Chapter \ref{calc} were
a consequence of the strength of the S-wave interaction
at low energies.  In this partial wave, convergence
of the $Q$ expansion is less obvious than in other partial waves
where interactions are weak and chiral power counting indicates 
that pions can be included perturbatively.  
The question of whether or not pions can be treated perturbatively
in this partial wave using a suitably modified power counting scheme
has been debated recently \cite{steefur,lands,ch}.

One way to investigate whether including pions perturbatively 
is valid in the presence of strong scattering 
is to consider what happens to the effective range expansion (ERE) 
(\ref{eq:ere}) when 
pions are removed.  So long as the short distance potential has
a simple structure, the effective range $r_e$ can be interpreted, 
very roughly, as half the range of the potential \cite{newton}.  
We notice immediately that in np scattering, $r_e$ is rather
large ($r_e\sim 2.7\;$fm) compared with typical nuclear scales.
Dimensional analysis suggests that the size of the 
effective range due to short distance physics alone
should be set by the inverse of the high energy scale
$1/\Lambda_0$.  Even if  $\Lambda_0$ is as low as $m_\rho/2$, 
then we would expect a modified effective range of the order of
 $0.6\;$fm.  

It is natural to ask whether removing pions, which give rise to the
long-range OPE potential, will reduce $r_e$ to a more 
natural size.  This gives us an opportunity to test the momentum
expansion in a low partial wave.  In this section we shall investigate 
removing OPE perturbatively and non-perturbatively and compare the results.

Consider a potential 
which reproduces the low energy $^1 S_0$ np phase shifts.
The subscript $l=0$ is dropped for convenience.
The potential consists of an unknown short-range piece which we
take to be zero for $r>R$, plus the relevant part of the OPE potential:
\be
U^{1\pi}(r)=-\alpha_\pi \frac{e^{-\mpi r}}{r}, \qquad \mbox{where}
\qquad \alpha_\pi=\frac{g_A^2 \mpi^2}{16\pi f_\pi^2}.
\ee 
$\alpha_\pi$ is of order $Q^{2}$.
We are implicitly making the assumption that there is sufficient
separation between long and short distance scales to allow a 
suitable choice of $R$.  The wave function due to the short range 
potential acting alone is
\be
u^S(r)=\frac{\sin{(pr+\delta_S)}}{p\cos{\delta_S}}
\ee
for $r>R$.  $u^S(r)$ has been normalised so 
that it is suitable for use in (\ref{eqn:dwba}) which now reads,
\be \label{eqn:opedwba}
\tan{\delta}\simeq\tan{\delta_{S}}+\frac{M\alpha_\pi}{p\cos^2{\delta_S}}
\int_{R}^{\infty} \sin^2{(pr+\delta_S)}\frac{e^{-\mpi r}}{r}dr.
\ee
The short distance potential cannot be made truly zero-range in 
$S$-wave scattering since the presence at the origin of the irregular 
solution, along with the $1/r$ singularity in the OPE potential, 
causes the integral in (\ref{eqn:opedwba}) to diverge 
logarithmically as $R\rightarrow 0$.

As has already been discussed in Chapter~\ref{calc}, low energy S-wave 
scattering is well described by the effective range expansion truncated
at order $p^2$.  If the power counting derived earlier is valid, then
it should be possible to obtain a `modified effective range expansion'
with the effects of one pion exchange removed perturbatively.  Inverting
both sides of (\ref{eqn:opedwba}), multiplying by $p$ and expanding
to first order in $\alpha_\pi$ we find

\begin{equation} \label{eqn:modere}
p\cot{\delta}\simeq p\cot{\delta_S}-
M\alpha_\pi\frac{(p\cot{\delta_S})^2}{p^2\cos^2{\delta_S}}\Sigma(p)
+O(\alpha_\pi^2),
\end{equation}
where $\Sigma(p)$ represents the integral in (\ref{eqn:opedwba}).
Integrating, expanding in powers of $\mpi R$ and discarding 
$O(\mpi R)$, $\Sigma(p)$ may be written,
\begin{eqnarray}
\Sigma(p)=&-&\frac{1}{2}(\gamma+\ln{(\mpi R)})\\ \nonumber
&+&\frac{1}{2}\cos{2\delta_S}\;\left[\gamma+\ln{(m_\pi R)}
+\frac{1}{2}\ln{(1+\frac{4p^2}{m_\pi^2})}\right]\\ \nonumber
&+&\frac{1}{2}\sin{2\delta_S}\;\tan^{-1}{\frac{2p}{\mpi}} + O(\mpi R).
\end{eqnarray}
$\gamma$ is Euler's constant.
Notice that it is necessary to keep terms up to order $p^4$ in $\Sigma(p)$
to obtain the modified effective range expansion to order $p^2$:
\begin{eqnarray}
-\frac{1}{a}+\frac{r_e}{2}p^2=-\frac{1}{a_S}+\frac{r_S}{2}p^2&+&
M\alpha_\pi \left[
(\gamma+\ln{\mpi R})-\frac{1}{a_S^2\mpi^2}+\frac{2}{a_S\mpi}\right] \\ \nonumber
&+&M\alpha_\pi \left[
\frac{1}{\mpi^2}
+\frac{2}{a_S^2\mpi^4} 
-\frac{8}{3a_S \mpi^3}
+\frac{r_S}{a_S\mpi^2}
-\frac{r_S}{\mpi}
\right]p^2 .
\end{eqnarray}
The above expression gives the physical quantities $a,r_e$ in 
terms of the modified quantities $a_S,r_S$.  Taking coefficients of 
$p^0$ and $p^2$, and solving for the modified coefficients,
\be \label{as}
-\frac{1}{a_S}=-\frac{1}{a}+M\alpha_\pi(\gamma+\ln{\mpi R})+\frac{2M\alpha_\pi}{\mpi a}-\frac{M\alpha_\pi}{\mpi^2 a^2}
\ee
\be \label{modre}
r_S=r_e+2M\alpha_\pi\left[\frac{-1}{\mpi^2}
+\frac{8}{3a\mpi^3}
-\frac{2}{a^2\mpi^4}+\frac{1}{\mpi}r_e
-\frac{1}{a\mpi^2}r_e\right],
\ee
ignoring corrections $O(\alpha_\pi^2)$.  Notice that 
$a_S$ is not uniquely defined due to the logarithm in (\ref{as}).
Cohen and Hansen \cite{ch} find equivalent expressions which
are also obtained in coordinate space, but using a rather different 
approach.  

As well as expanding in powers of $\alpha_\pi$, to make the 
approach systematic we must count powers of small momenta in 
other quantities.
The power counting scheme obtained around the non-trivial fixed point 
in the previous chapter implies that we should treat $1/a$ as order $Q$.

We assume that other coefficients in the modified effective range 
expansion scale according to their dimension.  The modified effective
range should therefore scale like $Q^0/\Lambda_0$.

Using these assignments, we find that the last two terms in the 
square brackets are suppressed relative to the first three by 
a factor $Q/\Lambda_0$, and should be ignored to leading order 
in small momenta.  Omitting these terms, we find $r_S=1.0\;$fm,
while including them, $r_S=4.0\;$fm (the dominant correction
comes from the fourth term in the square brackets in (\ref{modre})).

Clearly
the expansion in $Q$ is converging at best slowly.  

We investigate this further by considering a distorted wave modified 
ERE \cite{dwere}, which is obtained without expanding
in powers of $\alpha_\pi$. 
This is obtained using techniques similar
to those used to define an ERE in the presence of the Coulomb potential
where the usual expansion fails due to the potential's infinite range
(see for example \cite{stony}).

The calculation requires the solutions to the Schr\"odinger equation
in the presence of the OPE potential, 
$\phi_0$ and $f_0$, which were defined 
in (\ref{fred}) and (\ref{bob}).  Solving the
Schr\"odinger equation using a series expansion, they are found to 
have the following behaviour for small $r$:
\beq
\phi_0(r)&\sim& p\;C_0(p)\left[r-\frac{M\alpha_\pi}{2}r^2+O(r^3)\right] \\
\nonumber
f_0(r)&\sim&\frac{1}{C_0(p)}\left[1-M\alpha_\pi r\ln{\lambda_\pi(p) r}
+O(r^2\ln{r})\right].
\eeq
$C_0(p)$ and $\lambda_\pi(p)$ are functions which must be determined 
numerically.  

From these, we construct a comparison function $\phi(r)$ which is a 
solution to the same Schr\"odinger equation, but satisfying 
$\phi(0)=1$ and
\be
\phi(r)\sim \sin{(pr+\delta)}
\ee
for large $r$. $\delta$ is the phase shift obtained 
from the Schr\"odinger equation for a potential containing 
OPE and an unknown short range piece as before.  
Writing $\delta=\delta_\pi+\hat{\delta}$,
\be
\phi(r)=C_0(p)\left[f_0(r)+\phi_0(r)\cot{\hat{\delta}}\right].
\ee
The calculation now proceeds by analogy with the usual derivation
of the ERE in which $\phi(r)$ is a free wave solution.
The resulting modified ERE is given by
\be
p\cot{\delta_S}=-M\alpha_\pi\ln{\beta\lambda_\pi(p)}
+C_0^2(p)\;p\cot{\hat{\delta}}.
\ee 
$\beta$ is an arbitrary constant which is present due to the
presence of the logarithmic term in the expansion of $f_0$.  This,
in turn, is due to the $1/r$ singularity in the OPE potential.
Because of this arbitrariness in the zero energy part of the
modified ERE, it is again not possible to uniquely define a modified
scattering length.  By expanding $p\cot{\hat{\delta}}$, and
using numerically obtained values for the scattering length 
and effective range due to OPE alone,
we find an increased effective range $r_S=4.4\;$fm.  
Again, this disagrees with the result obtained in the momentum
expansion.  The similarity to the result including 
`higher order' terms is presumably accidental.  

Steele and
Furnstahl \cite{steefur} have also found an increased
effective range ($r_S=3.1\;$fm) in a modified effective range 
expansion, but using a cut-off pion Yukawa.

The fact that the effective range remains large when OPE
is removed might indicate that there is 
other important long-range physics.  The modified
effective range is certainly more natural expressed
in units of $1/(2\mpi)$ than units of $1/m_\rho$.

These problems have also been discussed by Kaplan and Steele \cite{lands}
who apply effective field theory techniques to a range of toy 
models.  The EFT expansion succeeds for simple short range
potentials where a small modified effective range is found.
The expansion fails, however, when the model of the short distance 
physics is given by two Yukawa functions with different ranges and opposite
signs.  In the latter case, the coupling constants were adjusted to 
reproduce the physical scattering length and effective range.

This suggests that the problems with the momentum expansion
found in the above calculations might be due to complexities 
in the real short distance NN interaction rather than the 
perturbative treatment of OPE.

\chapter{Results}
\label{results}
\section{Introduction.}

In Chapter \ref{pions}, the two nucleon potential as calculated
to third order ($\nu=3$) in small momenta and masses was presented.
There are many questions which we can ask about these potentials.
Some can be answered quite generally, while others will be addressed
in the context of individual partial waves.

The first calculation of NN scattering to this order 
was performed by Ord\'o\~nez {\it et al.}, although it differed
in several respects from the calculation described here.  Most importantly,
Delta resonances were incuded and $g_A$, $f_\pi$ and 
the low energy constants $B_i$
were treated as free parameters in a fit to the VPI \cite{vpi} partial
wave analysis (PWA).  
Since the determination of the
low energy constants from $\pi N$ scattering does not include 
Delta's, the $B_i$ found by these authors cannot be compared
with empirically determined values\footnote{The relationship between the $B_i$
and the $c_i$ in the absence of Delta's is given by Equation (\ref{bici})
on page \pageref{bici}.}.

Ideally, the potentials should be used to calculate cross-sections
which can be compared directly with experimental data.  Such an analysis
has been performed by Rentmeester {\it et al.} \cite{nijfriar} 
who use the same potential as 
model-independent input to a new multi-energy PWA of the
1998 Nijmegen proton-proton database \cite{nnonline}.  As in the present 
work, the potential is regularised by excluding the region $r<R$ and
imposing a boundary condition at the cut-off radius in each partial wave
considered.

In the PWA  the boundary condition is allowed to be an analytic function 
of energy and this energy dependence is controlled by a variable number of 
parameters which are adjusted to minimise $\chi^2$ in a fit to the chosen
data set for $R=1.4\;$fm and $R=1.8\;$fm.  The $c_i$ are treated as free
parameters, and the values found are reasonably close to those obtained
from $\pi N$ scattering data.  At both radii
a systematic reduction in $\chi^2_{min}$ and the number of necessary
boundary condition parameters is found as the leading order TPE and then next
to leading order TPE potentials are added.

Nevertheless, their optimum fit including the $\nu=3$ potential requires
23 boundary conditions.  Although this procedure seems to provide
evidence for the importance of the long
range part of the TPE potential,  adjusting boundary conditions in $l>1$ 
partial waves is not in keeping with a consistent EFT treatment to order 
$\nu=3$ for the reasons discussed in Chapter \ref{pions}; the first 
contact interaction which contributes to $D$-wave scattering ($l=2$) enters at
order $\nu=4$, and contributions to higher partial waves are suppressed
by larger chiral powers. 

Since the EFT predictions are parameter-free in D-waves and higher,
we restrict our attention to $l>1$ partial waves.
We shall investigate the EFT predictions in these partial
waves by imposing an energy independent boundary condition on the 
wavefunctions at $r=R$ in such a way that, for small enough $pr$, 
they match on to  
free wavefunctions at the cut-off radius (see Chapter \ref{pions}). 
In addition, we can explore 
deviations from the EFT prediction by allowing the boundary condition parameter $\alpha$
to be non-zero:
\be
R\frac{d}{dr}\ln{u}_l(p,r)\;\Big|_{r=R}\;=l+1+\alpha
\ee
in uncoupled waves, and the $2\times 2$ boundary condition matrix ${\cal A}$,
\be
R\frac{d}{dr}\vec{u_j}(p,r)\;\Big|_{r=R}\;
=\left[\mbox{Diag}(j,j+2)+{\cal A}\right]\vec{u_j}(p,r)
\ee
in the coupled waves.  We shall want to vary the boundary condition parameters
for two reasons.
Firstly, letting  $\alpha$ and ${\cal A}$ vary over a fixed range allows 
us to demonstrate explicitly how dependence on the boundary condition 
changes as angular momentum $l$ and laboratory energy $T_{lab}$ vary.
Secondly, it is interesting to test whether the addition of a single 
free parameter allows us to compensate for omitted higher order 
physics.  By doing this, we make contact with the new PWA of Rentmeester
{\it et al.} \cite{nijfriar} who allow several free parameters in each 
partial wave.

In a contrasting approach, Kaiser {\it et al.} \cite{kbw} calculated
the graphs of Figure \ref{fig:tpe} (including iterated OPE) at 
next to leading order in the expansion of the inverse nucleon mass
and obtained phase shifts by projecting the amplitude into the appropriate
partial wave.  Although not intended for use
in a particular Schr\"odinger equation, the irreducible part of their
interaction
transformed into coordinate space is the same as the potential used in the 
present work and also by the Nijmegen group \cite{nijfriar}.  

This perturbative calculation leads to finite predictions at finite
momenta and contains no free parameters.  It is not, however, a complete
calculation of the interaction to order $\nu=3$ because of the 
enhancement of reducible loop graphs.  
A full perturbative calculation to this order
would require the tedious evaluation of the reducible two and three-loop 
diagrams which contribute at order $\nu=2$ and $\nu=3$ respectively.
We shall sum these diagrams automatically by solving the Schr\"odinger 
equations.

In a further paper, Kaiser {\it et al.} \cite{kgw} add vector meson exchange
to the $\nu=3$ potential.  While the addition of $\rho$ and $\omega$ exchange
clearly permits a better description of the data, the inclusion of 
heavy mesons cannot be
reconciled with chiral power counting.  In particular,  the low energy
constants $c_i$ are no longer related in any simple way to those 
obtained from analyses of $\pi N$ scattering, and they must be refitted
to $NN$ scattering data.

More recently, Mei\ss ner has considered iterating the interaction calculated
to order $\nu=3$
in momentum space \cite{meissnew}.  This work is performed
in a similar spirit to the calculation
presented here.  The potential is regularised with
exponential and sharp cut-offs.  He finds evidence for $\nu=3$ TPE in
several partial waves.  

To test the full TPE predictions, we attempt to identify 
the partial waves, energies and cut-off radii for which phase shifts 
are simultaneously insensitive to non-perturbative effects and 
variation of $R$.  In these
regions, the TPE potential can be studied as unambiguously as possible.

In one partial wave in which TPE seems to be too strong, we shall 
demonstrate that adjusting the single parameter $\alpha$ to allow
for neglected higher order effects allows reasonably good global fits over 
similar energy ranges to those considered in \cite{kgw} when the full
TPE potential is included.  

\begin{table}
\begin{center}
\begin{tabular}{|c|c|c|c|c|c|c|c|c|c|c|} \hline
& $V_C$ & $V_S$ & $V_T$ & $V_{SO}$ & $V_{Q}$ & 
      $W_C$ & $W_S$ & $W_T$ & $W_{SO}$ & $W_{Q}$   \\ \hline
OPE 
&       &       &       &          &         &
        &       $\surd$ & $\surd$  &         &         \\ \hline
TPE ($\nu=2$) 
&  & $\surd$    & $\surd$ &          &         &
 $\surd$    &       &         &          &         \\ \hline

TPE ($\nu=3$)
&$\surd$&       &       &          &         & 
            & $\surd$      & $\surd$  &     &         \\ \hline
TPE ($1/M$)
&$\surd$&       & $\surd$ & $\surd$  &         &  
$\surd$     &       & $\surd$ & $\surd$  &         \\ \hline
$\sigma$ exchange 
& $\surd$   &       &       &          &  $\surd$ &
            &       &         &          &         \\ \hline
$\rho$ exchange
&       &       &       &          &         &
& $\surd$   & $\surd$  & $\surd$ & $\surd$ \\ \hline
$\omega$ exchange
&$\surd$&     &      & $\surd$  &      &
            &       &         &          &         \\ \hline
\end{tabular}
\end{center}
\caption[Contributions from the EFT potentials up to 
order $\nu=3$]{\label{q0sig} Contributions from the EFT potentials up 
to order $\nu=3$ compared 
with the most important OBE contributions \cite{pionsnuclei}.
$V_Q$ and $W_Q$ are isoscalar and isovector quadratic spin-orbit
potentials.  The remaining notation was defined in
Chapter \ref{pions}.}
\end{table}

\section{The Nijmegen Phase Shift Analyses}
We do not attempt to compare our results directly with NN scattering data.
Instead we consider the three most up-to-date potentials of the Nijmegen 
group \cite{nijpot}, and the 1993 
partial wave analysis PWA93 \cite{nijpwa}.  
Since the three potentials were determined in fits
to the same data set as the PWA93  \cite{nnonline}, and all have 
a $\chi^2$ per datum of around 1, they may be considered as 
independent PWA's and we shall refer to them as such.

Since these analyses have different characteristics, the spread 
in the corresponding phase shifts gives an indication of the 
systematic errors involved in choosing a particular PWA.
All four PWA's include the OPE interaction given in the Appendix.
To aid comparison, we also use this form which
takes account of pion mass differences. 
In our results (Figure \ref{fig:1d2pnp} to \ref{fig:3g5})
we show the full range of phase shifts produced by the PWA's
as a striped band. A short description of each follows.

\begin{itemize}
\item Reid93 is an update of the well known Reid68 potential ~\cite{reid68}.
Apart from the explicit appearance of the OPE interaction, 
the potential consists of a sum of regularised
Yukawa functions with arbitrarily chosen masses and coefficients
adjusted separately in each partial wave (up to $j=4$) to 
optimise the total $\chi^2$.  It should be noted that the potentials 
for $j\ge 5$ are set equal to potentials from lower partial waves, and 
are not allowed to 
vary independently.  A total of 50 parameters are used in the fit.
\item The Nijm I potential was constructed using an updated version of the
Nijmegen soft core potential \cite{nijsc} as a starting point. 
$\chi^2$ was then minimised by unfixing certain parameters in the 
potential and fitting them separately in each partial wave.
A total of 41 adjustable parameters were required.
\item The Nijm II potential was obtained in the same manner as Nijm I,
but with all non-local contributions removed. In this case, the fit 
made use of 47 parameters.
\item The PWA93 parameterisation of the NN interaction consists of a 
potential tail including OPE and heavy meson exchange contributions.
The remaining energy dependence of phase shifts is reproduced using an 
energy dependent boundary condition at $R=1.4\;$fm. 
\end{itemize}
\section{General Discussion}

The OPE phase shifts are obtained by solving the Schr\"odinger equations
(\ref{se1} - \ref{se3}) in the region $r>R$.  Four values of the 
cut-off radius $R$ are used: $R=0.8\;$fm, $1.0\;$fm, $1.4\;$fm and
$1.8\;$ fm.  In uncoupled partial waves, the TPE phase shifts are 
calculated both perturbatively using the DWBA (\ref{eqn:dwba}) with 
OPE wavefunctions as input, and non-perturbatively by solving the 
Schr\"odinger equation for the sum 
of OPE and TPE 
interactions.  In coupled partial waves, only the non-perturbative
calculation is performed. 

In general, we find that as $R$ is increased the agreement between 
perturbative and non-perturbative treatments of the TPE parts of the 
potential improves.  This can be understood from an RG point of view as 
discussed in Chapter \ref{pions}.  The inverse of the cut-off 
radius $1/R$ plays the role of the low 
cut-off scale $\Lambda$.  As $\Lambda\rightarrow 0$ the expansion parameter 
of the EFT, $\Lambda/\Lambda_0$, becomes very small and the convergence
properties of the theory improve.  
In terms of the potential, the large non-perturbative effects 
when $R$ is small are caused by the strong singularities 
encountered in the 
short distance part of potential.  When $R$ is made large enough,
these singularities are removed.

On the other hand, sensitivity to the boundary condition parameter 
$\alpha$ increases as $R$ becomes larger, and eventually the phase shifts
acquire a rapid energy dependence which is an artefact of the sharp cut-off.
An example of how such energy dependence can arise is given by the
non-trivial fixed point potential (\ref{eq:potfp.ere}) found
in the RG analysis of Chapter \ref{calc}. The cut-off dependent part of 
this potential contains a logarithm which blows up for a fixed 
momentum $p$ as $\Lambda$ approaches $p$ from above, causing
rapid variation of the potential with energy.

In the RG treatment, the energy dependence compensates for 
the effects of a sharp momentum space cut-off.  Since we 
have not solved an exact RG equation for $\alpha$, the energy
dependence caused by the boundary condition is still present in 
phase shifts and becomes important for high enough energies.

Sensitivity to the boundary condition $\alpha$
decreases for small $R$ and low energies where its effect
becomes small compared to the potential (or centrifugal barrier
in regions where the potential is sufficiently weak).
As would be expected,  sensitivity to non-perturbative effects and 
the regularisation
procedure is delayed to higher energies as $l$ becomes large because
of the centrifugal barrier.

In each partial wave we look for a region
of $R$ and $T_{lab}$ where non-perturbative and perturbative
predictions agree and are insensitive to the cut-off.  If the 
phase shift in this window is sufficiently different from the OPE
prediction, then we can interpret the difference as a
prediction of chiral TPE.

For each partial wave, we first show the full TPE prediction
as $R$ is varied (Figure \ref{fig:1d2pnp}, \ref{fig:1f3pnp} etc.).
In uncoupled waves both perturbative and non-perturbative
results are shown.  In general, 
we find that near $R=1.8\;$fm  phase shifts are very sensitive
to $R$ in the regions of energy where TPE contributions are
important and insensitive to $\alpha$.  For $R\sim 1.4\;$fm inwards,
insensitivity to $R$ persists to higher energies.  
$R=1.4\;$fm is one of the cut-off radii used by Rentmeester
{\it et al.} \cite{nijfriar} in their analysis of the $\nu=2$ and $\nu=3$
potentials.  For this reason we show the results obtained using this 
value of $R$ in each
partial wave considered here (Figure \ref{fig:1d2}, \ref{fig:1f3} etc.).
In these plots, the effect of adding first leading order TPE contributions,
and then the full TPE potential is shown.  Sensitivity (or
insensitivity) to the boundary condition parameter is also shown on 
this plot by allowing $-0.5<\alpha<0.5$ for the full TPE result in uncoupled
waves, and $-1/2<A_{ij}<1/2$ in coupled waves.

The effects of the $1/M$ potential in peripheral waves 
are quite small compared with rest of the $\nu=3$ potential. 
This suggests that the $1/M$ expansion is under control.  
For example, in the $^3G_5$
wave at $150$MeV where we see good evidence for chiral TPE, the $1/M$
effects are responsible for a $4\%$ reduction in the phase shift.  In
lower partial waves, they are more important, causing an $8\%$ increase
in the $^1D_2$ phase shift at $100$MeV.

In addition, phase shifts are not overly sensitive to fairly 
generous variations of the $c_i$ around their central values.  
Allowing for twice the quoted uncertainties \cite{cidet}, we find a 
maximum variation of $4\%$ in the $^3G_5$ partial wave at $150$MeV.  
The size of this change is comparable with the size of the 
shaded areas on the plot corresponding to variation 
of ${\cal A}$.  Similar sized effects are found 
in other partial waves in the regions of energy in which TPE is
important.

\section{Spin Singlet Partial Waves.}

\subsection{$^1D_2$}

The first counterterm which contributes to $D$-wave scattering enters
the EFT at order $\nu=4$.  Since this is only one order higher than
the effects we are looking for, it is not surprising that
$R$ dependence and $\alpha$ dependence are significant for energies
above about $20$MeV.  At $R=0.8\;$fm, non-perturbative effects cannot 
be ignored at any reasonable energy.

\begin{figure} [h]
\begin{center}
\includegraphics[width=6.0in,bb=38 10 536 536]{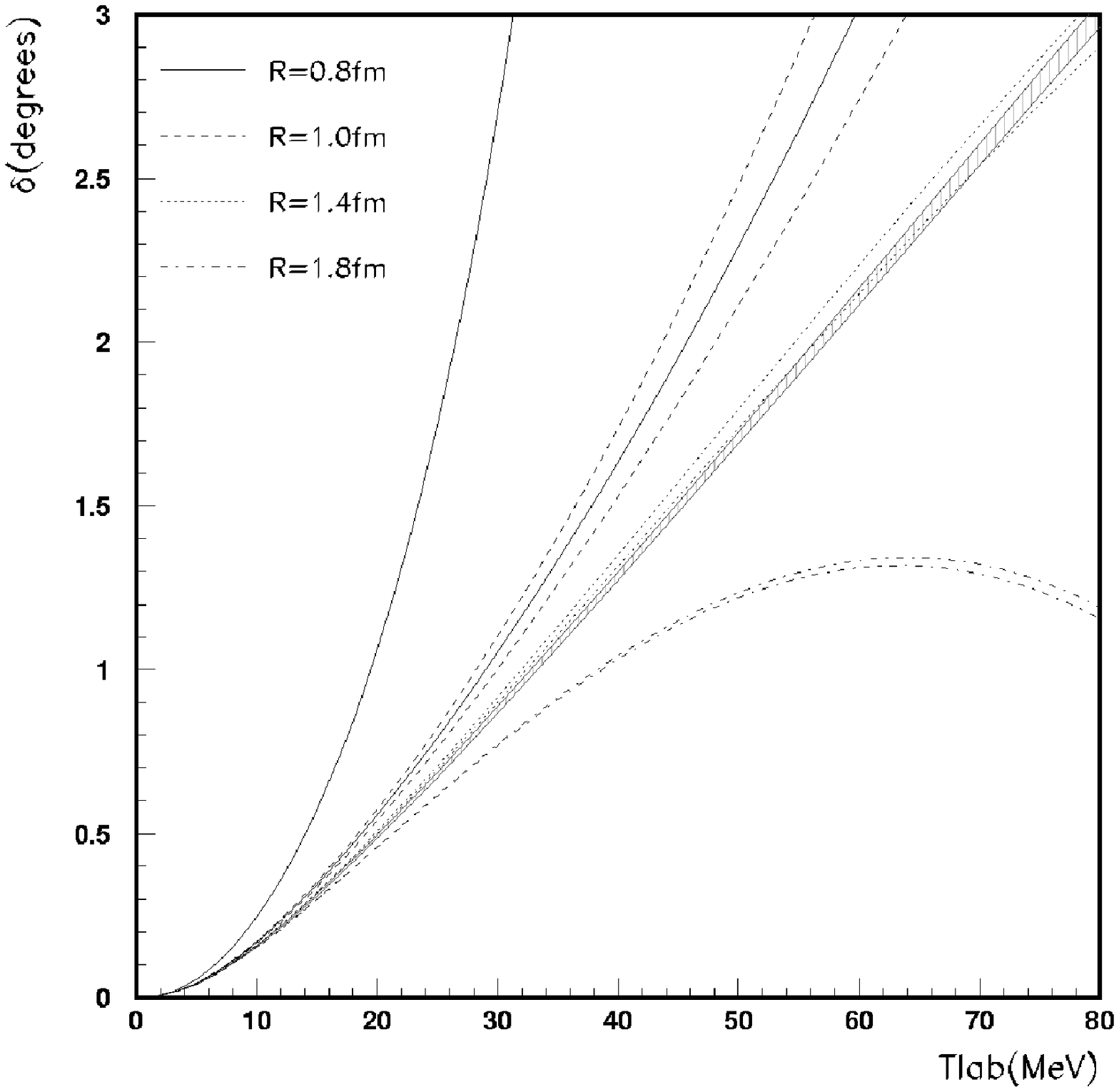}
\caption[The $^1D_2$ Phase Shift for four cut-off radii.]
{\label{fig:1d2pnp} The $^1D_2$ phase shift at $\alpha=0$ for four cut-off
radii.  At each radius, the upper curve is obtained by
iterating the full $\nu=3$ potential, while for the lower curve
the TPE contribution is included perturbatively.  The striped
band represents the Nijmegen Partial Wave Analyses (see text).} 
\end{center}
\end{figure}

The full TPE phase shifts are plotted in Figure \ref{fig:1d2pnp}
for four values of $R$ and for $\alpha=0$.  Two curves are shown
for each radius, corresponding to perturbative and non-perturbative
treatments of TPE contributions.  The full range of Nijmegen phase
shifts is also shown for comparison.

This figure illustrates quite nicely some of the general features 
discussed in the previous section.  At the smallest cut-off radius, 
non-perturbative effects are very large, while at $R=1.8\;$fm 
sensitivity to $R$ controls the phase shift.  Between $R=1.0\;$fm 
and $R=1.4\;$fm dependence on $R$ is 
less severe, and non-perturbative TPE effects do not overwhelm
the results below about $20$MeV.

At our chosen value of the cut-off radius ($R=1.4\;$fm), the full
TPE predictions lie almost on top of the Nijmegen values.  No 
conclusions should be drawn from this accident since in this partial
wave there is no region of $R$ and $T_{lab}$ in which TPE contributions 
can be isolated unambiguously.  

\begin{figure} [h]
\begin{center}
\includegraphics[width=6.0in,bb=38 48 536 536]{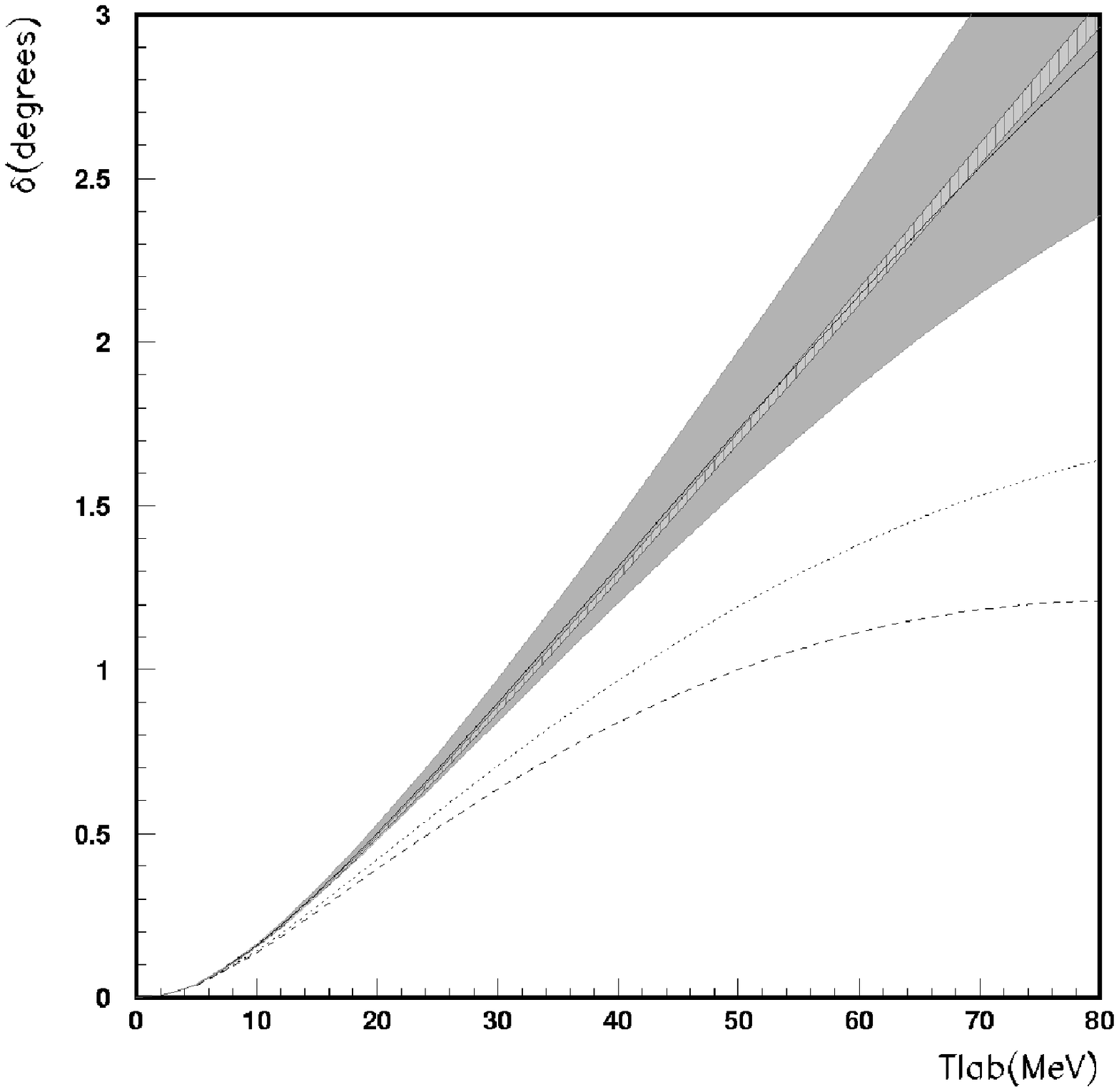}
\caption[The $^1D_2$ Phase Shift near $\alpha=0$ for $R=1.4\;$fm.]
{\label{fig:1d2} $^1D_2$ phase shift near $\alpha=0$ for cut-off 
radius $R=1.4\;$fm.  The dashed line is iterated OPE, the dotted line 
includes leading order TPE ($\nu=2$) perturbatively, and the solid
line is the full calculation to order $\nu=3$ with TPE treated 
perturbatively.   The shaded band shows variation of the $\nu=3$
phase shifts for $-0.5<\alpha<0.5$.} 
\end{center}
\end{figure}

In Figure \ref{fig:1d2}, the OPE, leading order TPE and full TPE
phase shifts are shown.  In this plot, TPE effects are
included perturbatively.  For all of these cases, the line $\alpha=0$ 
is shown, and for full TPE the shaded band indicates 
how phase shifts vary for $-0.5<\alpha<0.5$.

A pattern which is seen in most other partial waves is evident in Figure
\ref{fig:1d2}; the leading order ($\nu=2$) TPE contributions
are smaller than the next to leading order ($\nu=3$) contributions.
In this partial wave, despite problems with non-perturbative
effects and cut-off dependence we can, at least,
conclude that TPE contributes with the correct sign and roughly 
the correct strength.
  
\subsection{$^1F_3$}

As angular momentum $l$ increases, the $F$-wave phase shifts are the first 
for which we expect insensitivity to the contact interactions which 
are produced by the low energy effective theory 
up to order $\nu=3$ in the momentum expansion.  
The first contact interaction which contributes to $F$-wave
scattering would enter at order $\nu=6$.

\begin{figure} 
\begin{center}
\includegraphics[width=6in,bb=38 10 536 536]{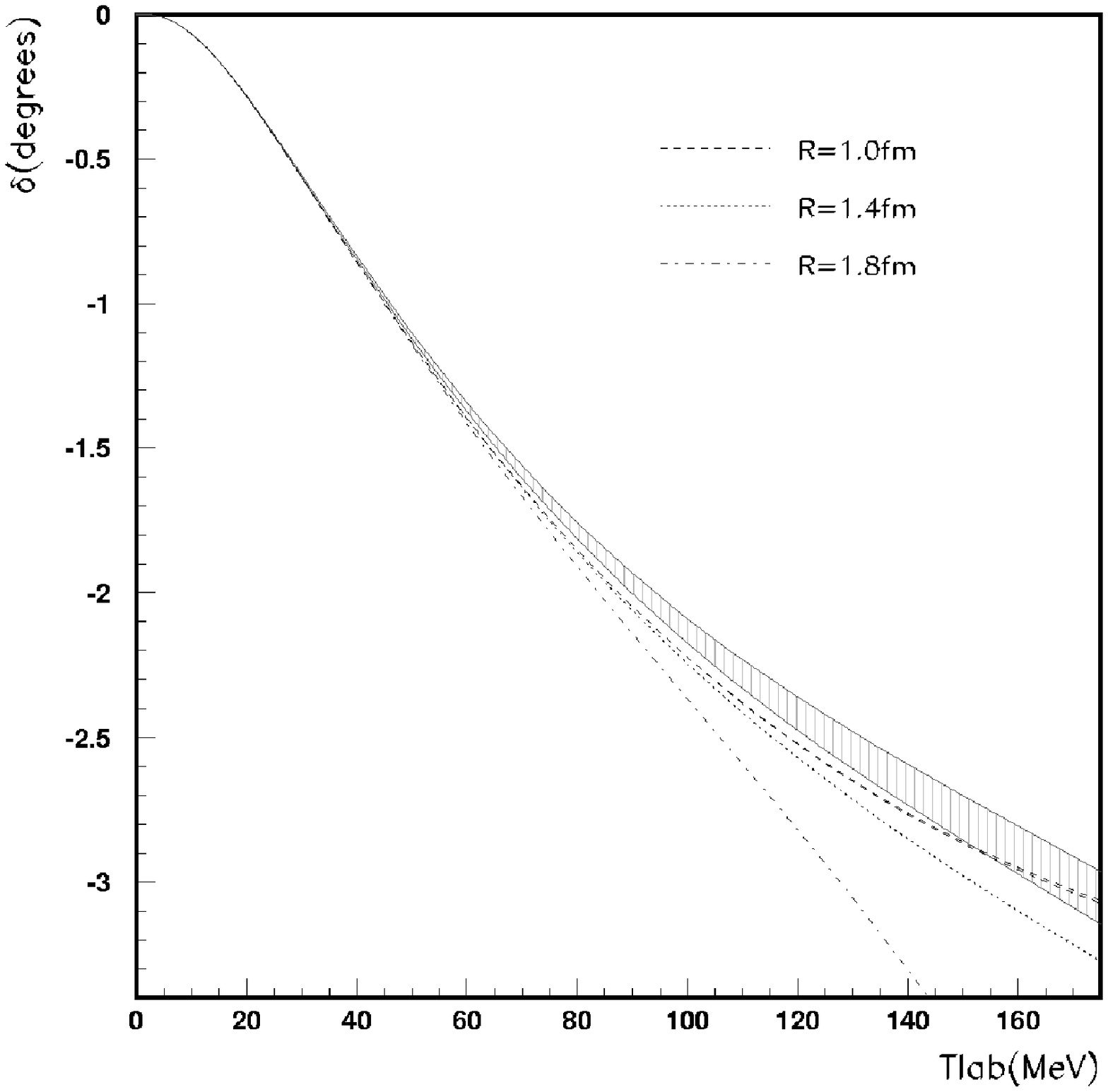}
\caption[The $^1F_3$ phase shifts for three cut-off radii.]
{\label{fig:1f3pnp} $^1F_3$ phase shift at $\alpha=0$ for three 
cut-off radii.  At each radius, the upper curve is obtained by
iterating the full $\nu=3$ potential, while for the lower curve
the TPE contribution is included perturbatively although the 
difference is only visible at $R=0.8\;$fm. 
The phase shifts at $R=0.8\;$fm are not shown,
as they lie almost exactly on top of the $R=1.0\;$fm results.
The striped band represents the Nijmegen phase shifts.}
\end{center}
\end{figure}

Figure \ref{fig:1f3pnp} shows the perturbative and non-perturbative
$\nu=3$ TPE predictions for three cut-off radii.  The difference
between iterated and DWBA curves is only visible at $R=1.0\;$fm.  The
curves for $R=0.8\;$fm are not shown because they lie nearly 
on top of the $R=1.0\;$fm curves.

Again, $R$ dependence is large near $R=1.8\;$fm and small for
$R$ below about $1.4\;$fm and $T_{lab}$ less than about 
$120$MeV.  Comparison with the $^1D_2$ case seems to 
indicate that our power counting expectations are justifed. 

More surprising is the almost exact agreement between perturbative
and non-perturbative calculations of TPE. Examination of the potential 
provides an explanation.  In spin singlet partial waves with
even angular momenta $l$, all contributions to the potential
are attractive. For odd angular momenta,
OPE and isovector parts of the asymptotic TPE potential 
(\ref{asym1},\ref{asym2}) are repulsive.  In the latter case, 
the cancellation between repulsive and attractive parts of the 
potential cuts its size significantly, allowing the cut-off
radius to be reduced without introducing parts of the 
potential which need to be treated non-perturbatively.

The ratio between $U_{even}(r)$ and $U_{odd}(r)$ for four values
of $r$ is shown in Table \ref{potcomp}.  The very small
ratio at $r=1.4\;$fm is caused by a change in sign of the 
odd $l$ potential near this radius from repulsive at large 
radii to attractive for small radii.  The even $l$ potential
is always attractive.

\begin{table} 
\begin{center}
\begin{tabular}{|c|c|c|c|c|} \hline
$r$ (fm)         & $0.8$ & $1.0$ & $1.4$ & $1.8$\\ \hline
$U_{odd}/U_{even}$ & $0.25$ & $0.21$ & $3.10^{-2}$ & $-0.32$\\ \hline 
\end{tabular}
\end{center}
\caption{\label{potcomp} Ratio of 
two-nucleon potentials in spin singlet 
partial waves for odd and even values of $l$.}
\end{table}

\begin{figure} [h]
\begin{center}
\includegraphics[width=6in,bb=38 48 536 536]{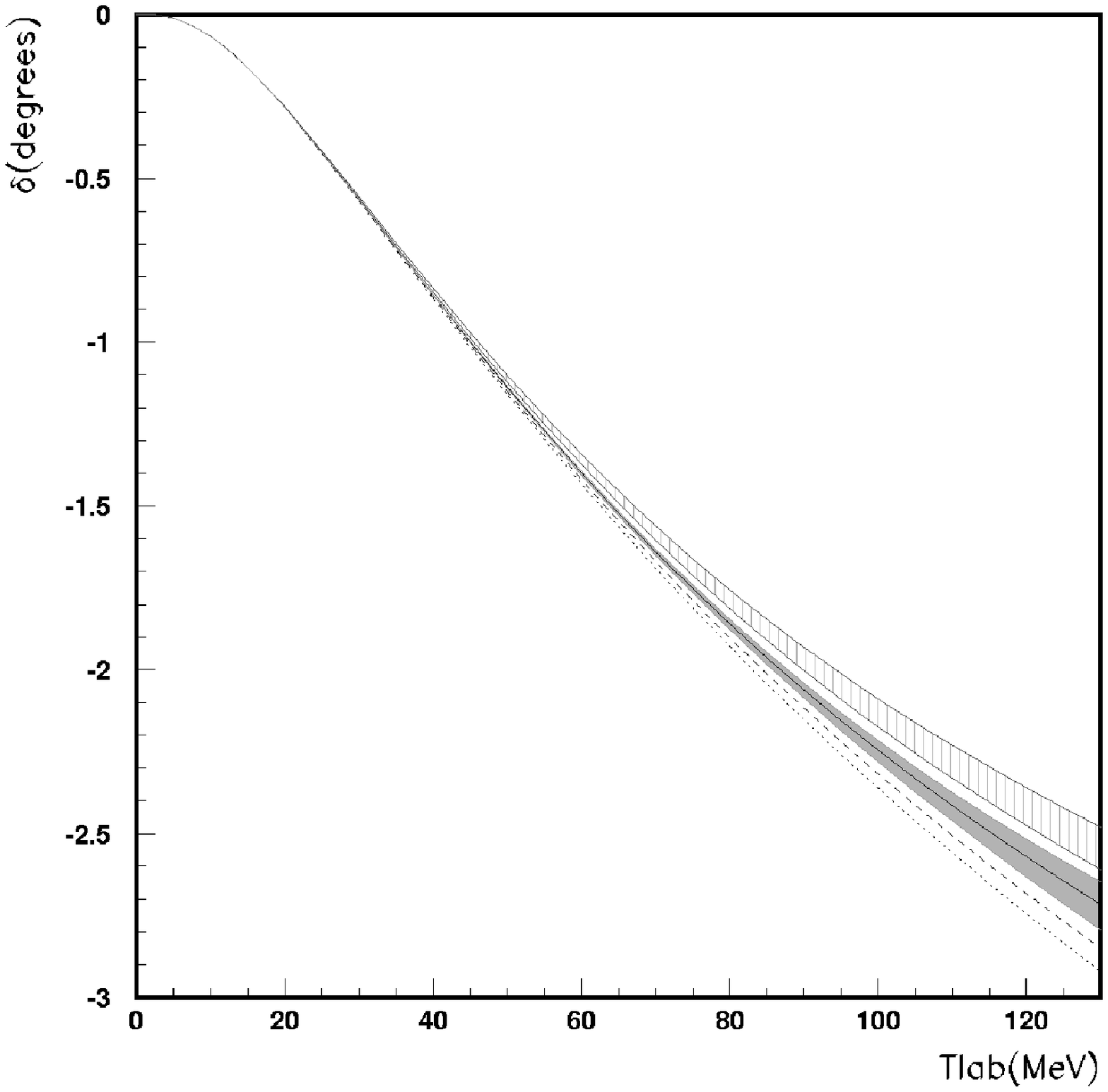}
\caption[The $^1F_3$ phase shifts near $\alpha=0$ for $R=1.4\;$fm.]
{\label{fig:1f3} $^1F_3$ phase shifts near $\alpha=0$ for cut-off 
radius
$R=1.4\;$fm.  The dashed line is iterated OPE, the dotted line 
includes leading order TPE ($\nu=2$) perturbatively, and the solid
line is the full calculation to order $\nu=3$ with TPE treated 
perturbatively.   The shaded band shows variation of the $\nu=3$
phase shifts for $-0.5<\alpha<0.5$.}
\end{center}
\end{figure}

In Figure \ref{fig:1f3}, the OPE, leading order TPE and full
TPE results are shown for $R=1.4\;$fm.  TPE is included
perturbatively.
Once again, the $\nu=3$ contributions are the dominant part of
the chiral TPE potential.  The leading order TPE potential is small and 
repulsive in the range of energy shown. The $\nu=3$ contribution is larger
and attractive, and closes the gap between OPE predictions and the
phase shifts of the Nijmegen potentials.  As expected from
power counting, the phase shifts are considerably less sensitive 
to the same variation of $\alpha$ at a given energy than in 
the $^1D_2$ partial wave.

\subsection{$^1G_4$}

\begin{figure} 
\begin{center}
\includegraphics[width=6in,bb=38 10 536 536]{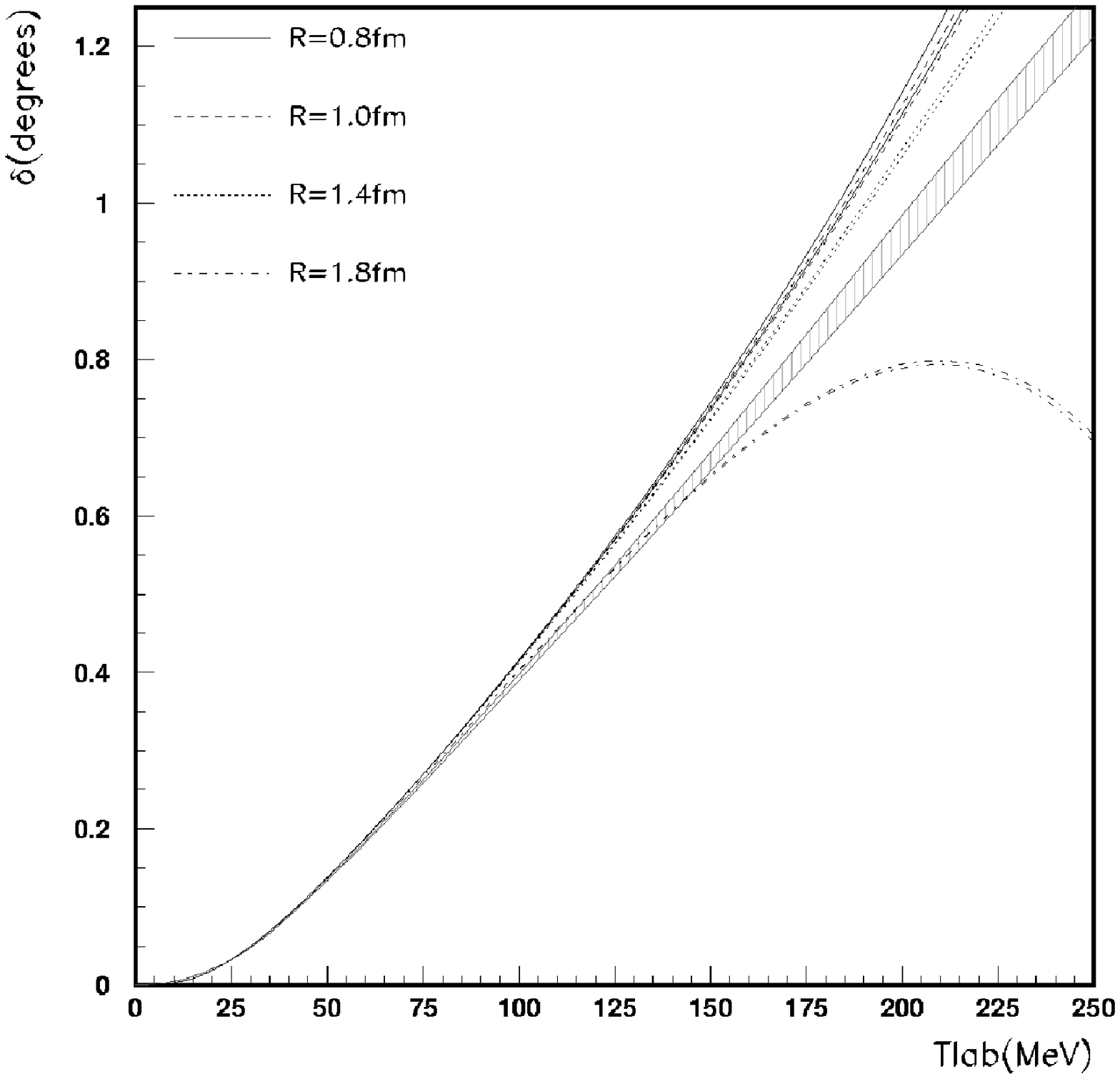}
\caption[The $^1G_4$ phase shift for four cut-off radii.]
{\label{fig:1g4pnp} $^1G_4$ phase shift at $\alpha=0$ for four 
cut-off radii.  At each radius, the upper curve is obtained by
iterating the full $\nu=3$ potential, while for the lower curve
the TPE contribution is included perturbatively.  The striped band
represents the Nijmegen phase shifts.
}
\end{center}
\end{figure}

\begin{figure} [h]
\begin{center}
\includegraphics[width=6in,bb=38 48 536 536]{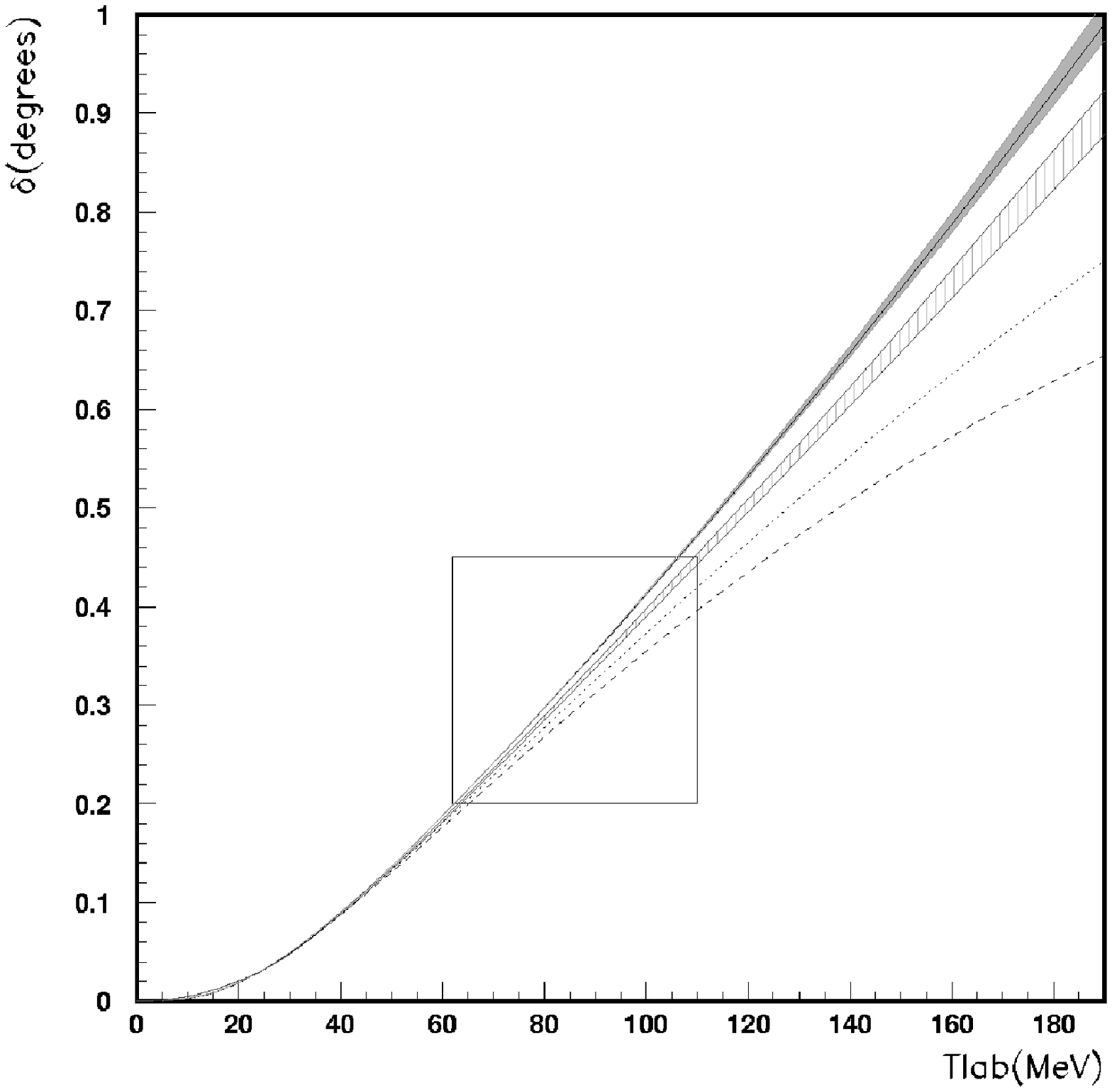}
\caption[The $^1G_4$ phase shift near $\alpha=0$ for $R=1.4\;$fm.]
{\label{fig:1g4} The $^1G_4$ phase shift near $\alpha=0$ for 
cut-off radius
$R=1.4\;$fm.  The dashed line is iterated OPE, the dotted line 
includes leading order TPE ($\nu=2$) perturbatively, and the solid
line is the full calculation to order $\nu=3$ with TPE treated 
perturbatively.   The shaded band shows variation of the $\nu=3$
phase shifts for $-0.5<\alpha<0.5$.  The boxed area is shown enlarged in
Figure \ref{fig:1g4small}.}
\end{center}
\end{figure}

The $^1G_4$ phase shift is an important place to look for the
effects of TPE.  In Figure \ref{fig:1g4}, it is clear that OPE
is too weak to account for the Nijmegen phase shifts.  
Figure \ref{fig:1g4pnp} shows the perturbative and non-perturbative TPE
results for four cut-off radii.  As usual, $R$ dependence is particularly
strong around $R=1.8\;$fm, but from $1.4\;$fm inwards the phase shifts 
are insensitive to $R$ below around $170$MeV (compared
with about $120$MeV in the $^1F_3$ wave).  We therefore have
a large window in which we can investigate TPE.  

\begin{figure} [h]
\begin{center}
\includegraphics[width=6in,bb=38 48 536 536]{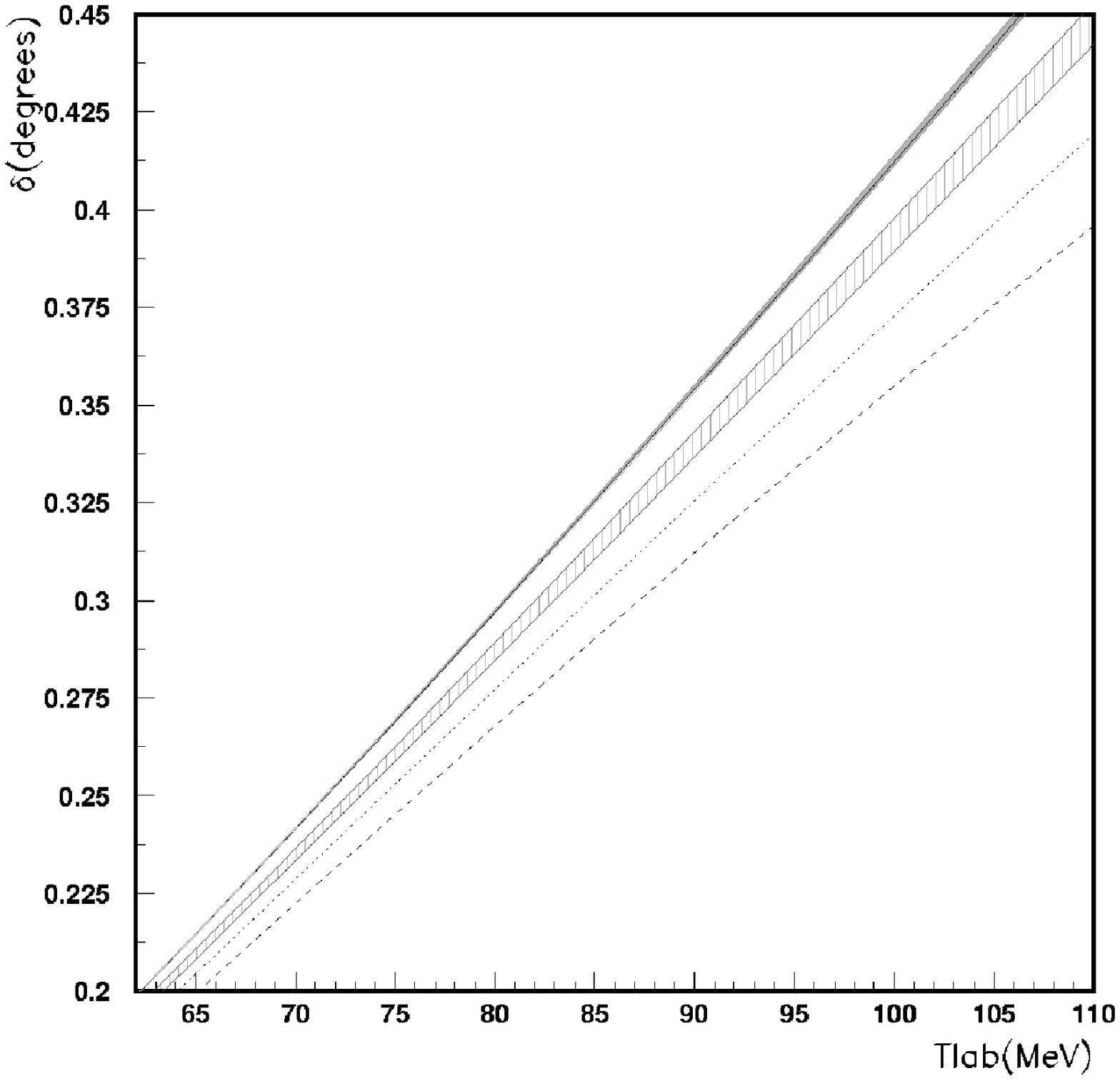}
\caption[An enlarged portion of the $^1G_4$ phase shifts 
near $\alpha=0$ for $R=1.4\;$fm.]
{\label{fig:1g4small} An enlarged portion of the $^1G_4$ phase shifts 
near $\alpha=0$ 
for cut-off radius
$R=1.4\;$fm.  The dashed line is iterated OPE, the dotted line 
includes leading order TPE ($\nu=2$) perturbatively, and the solid
line is the full calculation to order $\nu=3$ with TPE treated 
perturbatively.   The shaded band shows variation of the $\nu=3$
phase shifts for $-0.5<\alpha<0.5$.}
\end{center}

\end{figure}

In Figure \ref{fig:1g4} the OPE, leading order TPE and full TPE
results are presented.  Again, leading order TPE contributions
are smaller than $\nu=3$ contributions.  Over the chosen 
range, $\alpha$ dependence is very small.  The region from
$62$ to $110$MeV is enlarged in Figure \ref{fig:1g4small} and
shows the situation clearly in a region where TPE contributions are 
not sensitive to $\alpha$, $R$ or non-perturbative effects.

In this partial wave, TPE is clearly too attractive at all energies,
although it contributes with the correct sign.  This indicates that 
there are important higher order contributions missing.  It may be worth
noting that the PWA93 phase shift, (which has the lowest $\chi^2$ of all
of the Nijmegen potentials), lies at the top of the striped
band for $T_{lab}$ above about $150$MeV.

\begin{figure} [h]
\begin{center}
\includegraphics[width=6in,bb=38 48 536 536]{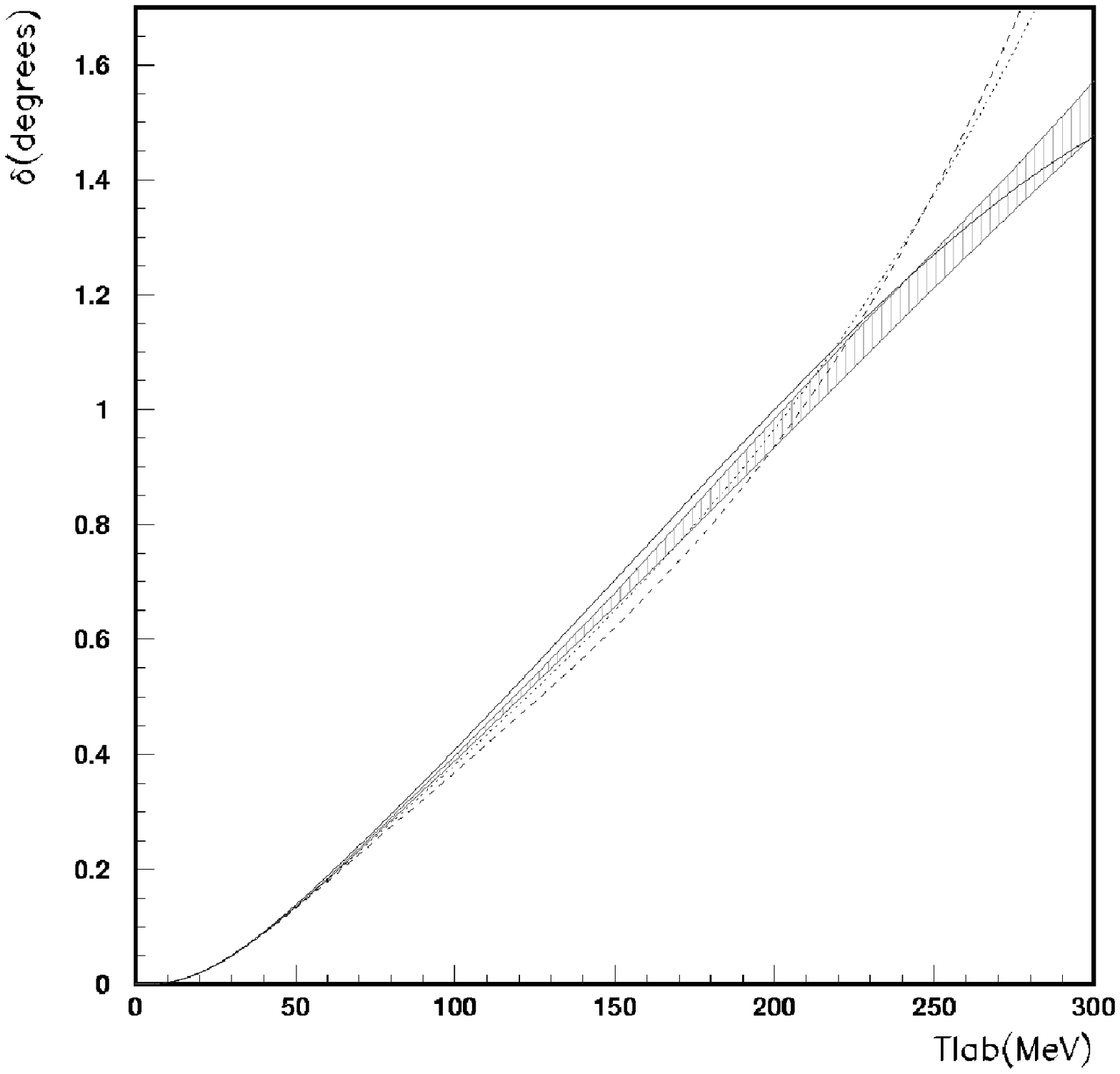}
\caption[The $^1G_4$ phase shift for adjusted values of 
$\alpha$ at $R=1.4\;$fm.]
{\label{fig:1g4best}$^1G_4$ phase shifts for adjusted values of 
$\alpha$ at cut-off radius $R=1.4\;$fm.  The dashed curve shows the OPE result
for $\alpha=-3.5$, the dotted curve is the leading order
TPE (DWBA) result at $\alpha=2.5$, and the solid line is the full
TPE (DWBA) result at $\alpha=1.5$}.
\end{center}
\end{figure}

It is interesting to ask whether a single parameter introduced to 
compensate for this discrepancy allows good agreement with the Nijmegen
results over a large energy range.  The results obtained when $\alpha$
is adjusted by hand to produce a rough global fit to the 
Nijmegen results up to $300$MeV are shown in Figure \ref{fig:1g4best}.
The OPE and $\nu=2$ curves can be brought into better agreement
at high energies at the cost of increasing the discrepancy at 
intermediate energies.  The `best' values of $\alpha$ were
as follows:
\be
\alpha_{OPE}=-3.5, \qquad \alpha_{TPE2} = 2.5 \qquad\mbox{and}\qquad
\alpha_{TPE3}=1.5.
\ee 
$\alpha_{TPE2}$ and $\alpha_{TPE3}$ are associated with $\nu=2$
and $\nu=3$ potentials respectively.  The full TPE calculation
gives the best fit overall, although it is still slightly
too attractive at intermediate energies.


\section{Spin Triplet Uncoupled Partial Waves}
\subsection{$^3D_2$}
In spin singlet partial waves, the tensor parts of the OPE
and TPE potentials do not contribute.  The uncoupled spin triplet waves
are therefore the first in which these parts of the potential
are tested.  The situation in the $^3D_2$ partial wave is similar
to the $^1D_2$ case, although OPE alone is much closer to the 
Nijmegen results in the triplet partial wave.  As in the $^1D_2$, the
$R=1.4\;$fm results are close to the PWA's but again, dependence
on $R$ and non-perturbative TPE effects prevents us from 
drawing too many conclusions from this.

\subsection{$^3F_3$}
\begin{figure} 
\begin{center}
\includegraphics[width=6in,bb=38 10 536 536]{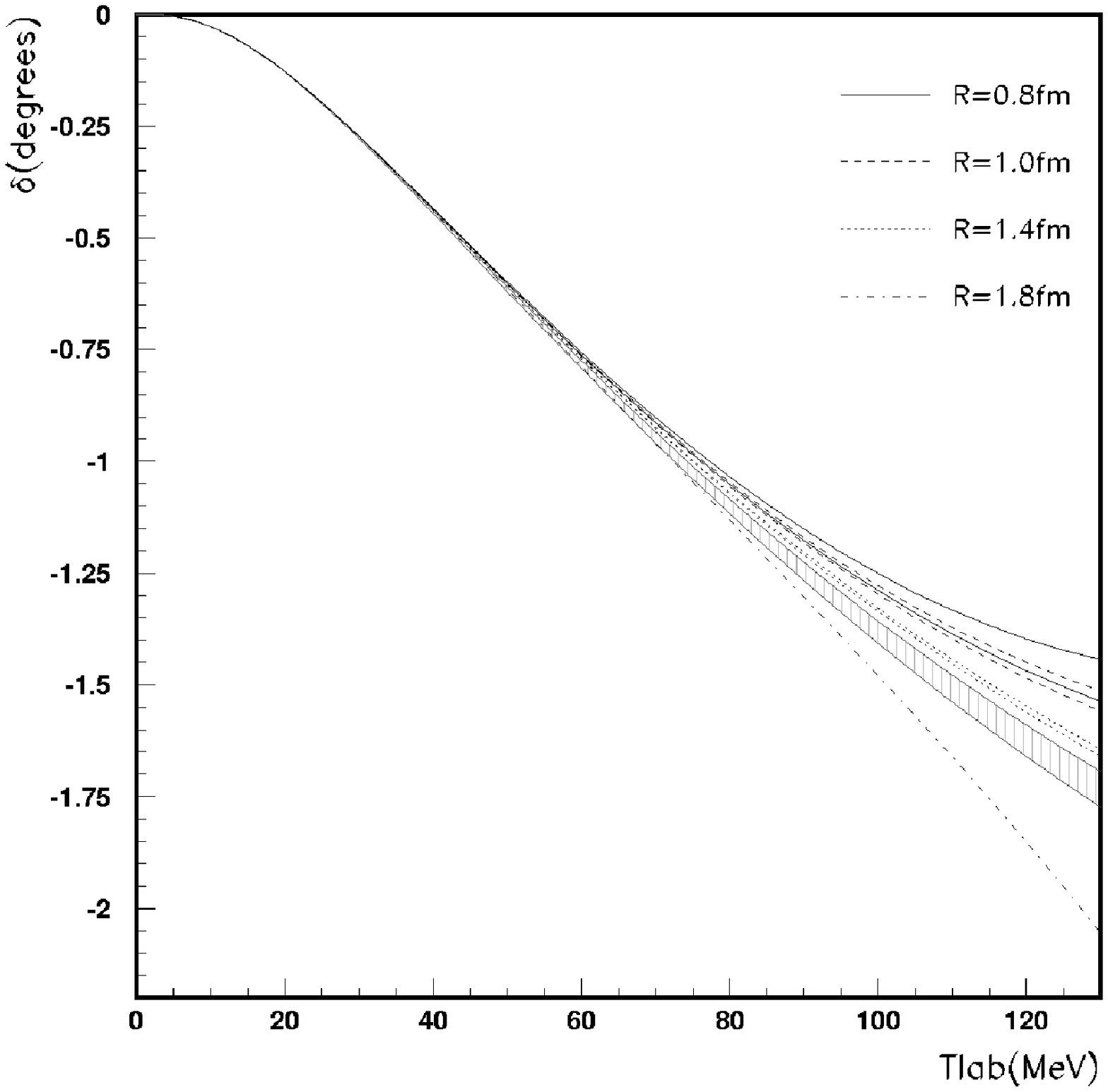}
\caption[The $^3F_3$ phase shift at $\alpha=0$ for 
four cut-off radii.]
{\label{fig:3f3pnp} $^3F_3$ phase shift at $\alpha=0$ for 
four cut-off radii.
For each value of $R$, the result of including full OPE + TPE results
perturbatively (DWBA) and non-perturbatively is shown.}
\end{center}
\end{figure}

\begin{figure} [h]
\begin{center}
\includegraphics[width=6in,bb=38 48 536 536]{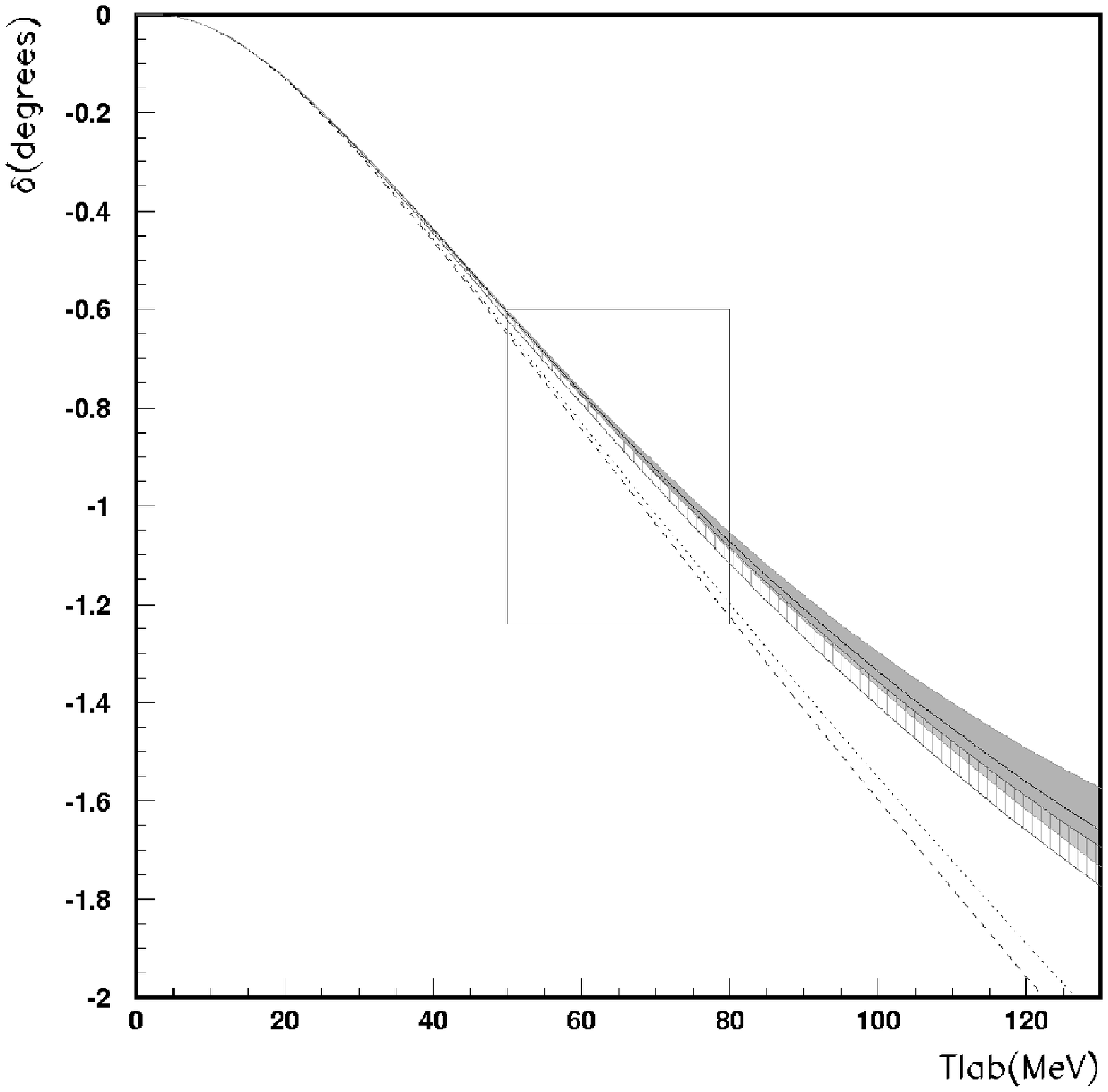}
\caption[The $^3F_3$ phase shifts near $\alpha=0$ for 
$R=1.4\;$fm.]
{\label{fig:3f3} $^3F_3$ phase shifts near $\alpha=0$ for 
cut-off radius $R=1.4\;$fm.  The dashed line is iterated OPE, the dotted line 
includes leading order TPE ($\nu=2$) perturbatively, and the solid
line is the full calculation to order $\nu=3$ with TPE treated 
perturbatively.   The shaded band shows variation of the $\nu=3$
phase shifts for $-0.5<\alpha<0.5$.  The boxed region is shown enlarged
in Figure \ref{fig:3f3small}}
\end{center}
\end{figure}

The $^3F_3$ partial wave, like the $^1G_4$, is a promising
place to look for TPE. Figure \ref{fig:3f3} shows that the 
discrepancy between OPE and the Nijmegen results is already
significant below $80$MeV.  Figure \ref{fig:3f3pnp} shows the full TPE
prediction for four values of cut-off radius $R$.  It is clear that
for $R$ between about $1.0\;$fm and $1.4\;$fm, sensitivity to
$R$ and non-perturbative TPE contributions is no larger than
the discrepancies among the Nijmegen phase shifts.

\begin{figure} [h]
\begin{center}
\includegraphics[width=6in,bb=38 48 536 536]{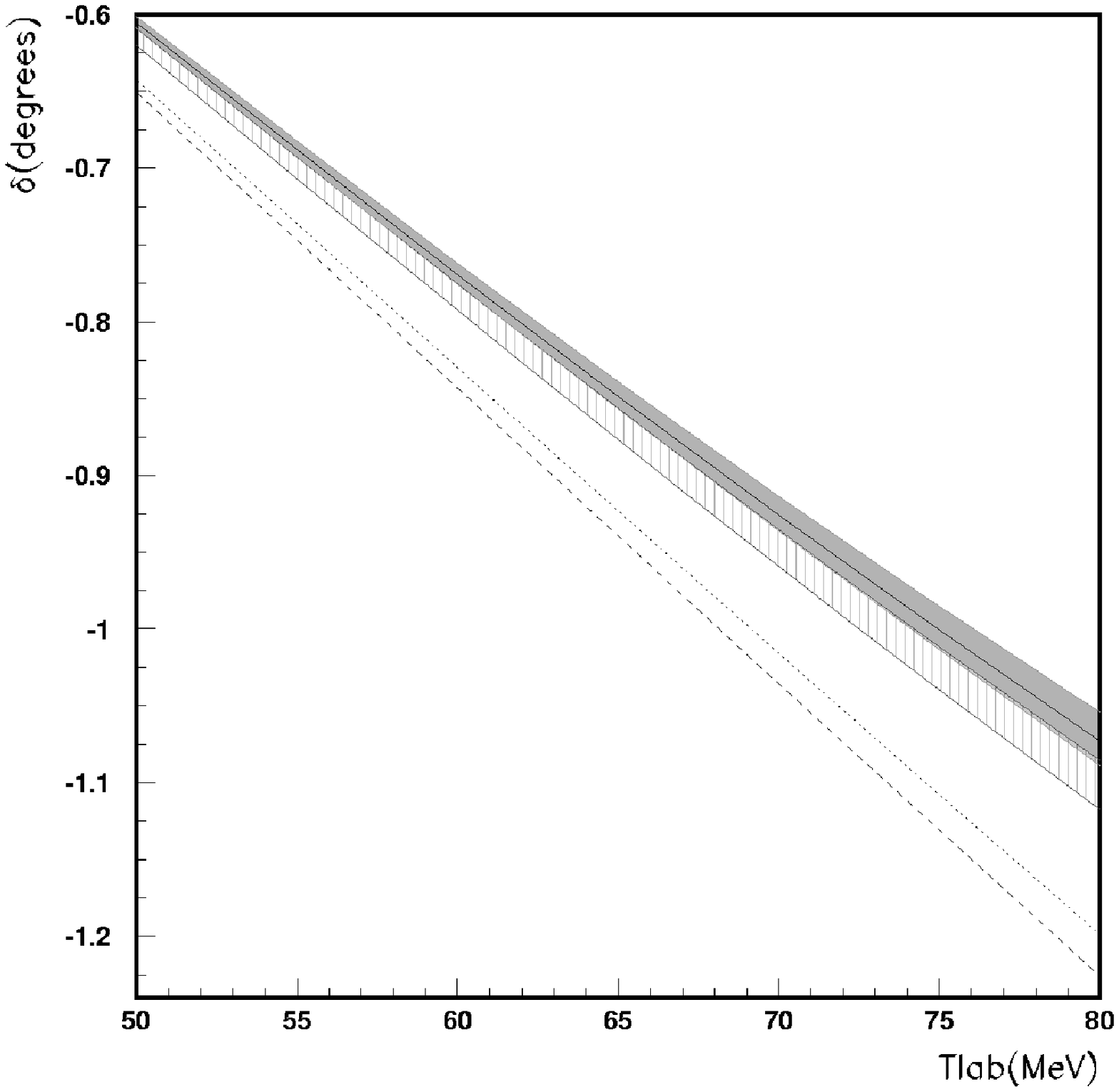}
\caption[An enlarged section of the $^3F_3$ phase 
shift near $\alpha=0$ for $R=1.4\;$fm.]
{\label{fig:3f3small} An enlarged section of the $^3F_3$ phase 
shift near $\alpha=0$ 
for $R=1.4\;$fm showing the effect of the full TPE calculation
in a region where the phase shift is not sensitive to the cut-off
or non-perturbative effects.
The dashed line is iterated OPE,
the dotted line is TPE to order $\nu=2$ (DWBA), and the 
shaded band is the full $\nu=3$ TPE (DWBA)calculation 
for $-0.5<\alpha<0.5$ with
the solid line showing $\alpha=0$.
The striped area shows the range of predictions of the
three Nijmegen potentials \cite{nijpot} and 1993 partial wave analysis 
\cite{nijpwa}.}
\end{center}
\end{figure}

Dependence on the cut-off radius is again large at $R=1.8\;$fm.
The comparison with OPE and leading order TPE is shown for $R=1.4\;$fm
in Figure \ref{fig:3f3}.  Over the entire energy range shown, the $\nu=3$
calculation represents a substantial improvement over OPE and 
the leading order TPE result.  The agreement with Nijmegen results
for $\alpha=0$ and $R=1.4\;$fm persists to about $250$MeV, although 
the prediction is strongly $R$ dependent at such a high energy. 
The region between $50$ and $80$MeV is  shown enlarged in 
Figure \ref{fig:3f3small}.
Once again the $\nu=3$ contribution is clearly dominant.  

\subsection{$^3G_4$}
The $^3G_4$ partial wave is described very well by OPE; even
at 250MeV the OPE prediction lies between the extremes 
of the Nijmegen PWA's.  In the sense that the TPE contributions do
not spoil this agreement, they have the correct magnitude.

\section{Spin Triplet Coupled Partial Waves}
\subsection{$^3P_2$ - $^3F_2$, ($\epsilon_2$)}
The $^3P_2$ phase shift is very sensitive to 
short distance effects, and we do not consider it here
except to verify that it has the correct behaviour at
very low energies when $\alpha=0$.  The mixing angle, $\epsilon_2$,
 is described well by OPE up to about $40$MeV where the results become 
dependent on $R$.  Up to about $100$MeV, both parts of TPE are
 very small corrections to the OPE prediction for $R=1.4\;$fm.

\begin{figure} 
\begin{center}
\includegraphics[width=6in,bb=38 10 536 536]{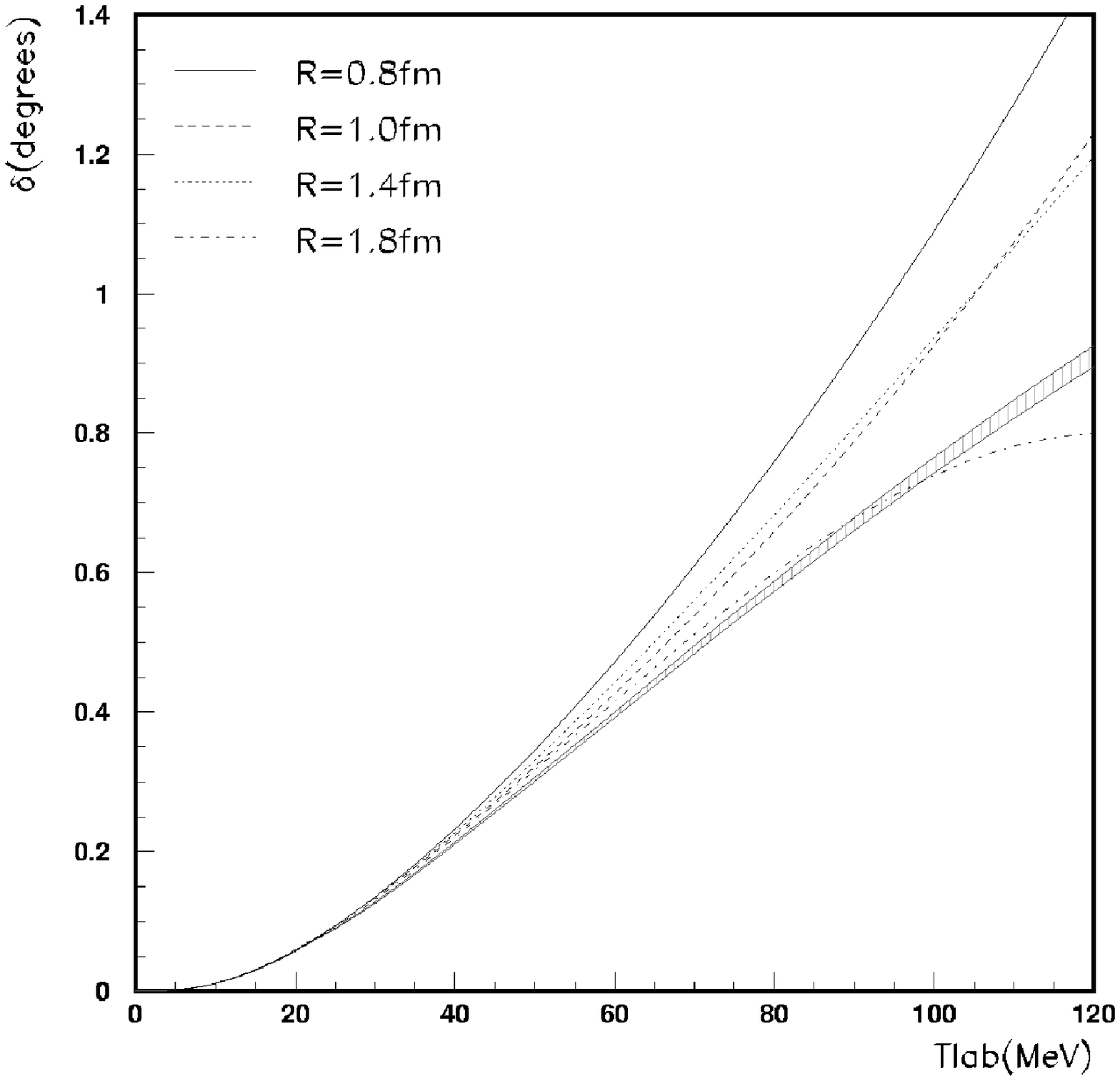}
\caption[The $^3F_2$ phase shift at ${\cal A}=0$ for 
four cut-off radii.]
{\label{fig:3f2pnp} The $^3F_2$ phase shift at ${\cal A}=0$ for 
four cut-off radii.
For each value of $R$, the result of including full OPE + TPE results
non-perturbatively is shown.}
\end{center}
\end{figure}

The $^3F_2$ phase shift is the first observable in a coupled
channel in which the opportunity to study TPE corrections
presents itself.  The full TPE phase shift is shown at
$\alpha=0$ in Figure \ref{fig:3f2pnp} for four cut-off radii.  
The rather unusual looking behaviour above $90$MeV 
as $R$ is varied between $1.0\;$fm and $1.4\;$fm is probably 
caused by interference with the $^3P_2$ phase shift.  In addition,
the difference between the $R=0.8\;$fm and $1.0\;$fm results
suggests that non-perturbative TPE effects may be important
at these radii.  The phase shift near $R=1.8\;$fm is sensitive
to the cut-off radius as usual. 

Despite the unusual features described above, the 
TPE phase shift lies consistently above the Nijmegen
PWA's in the region of energy where dependence on $R$ is small
(below about $60$MeV).  

\begin{figure} [h]
\begin{center}
\includegraphics[width=6in,bb=38 48 536 536]{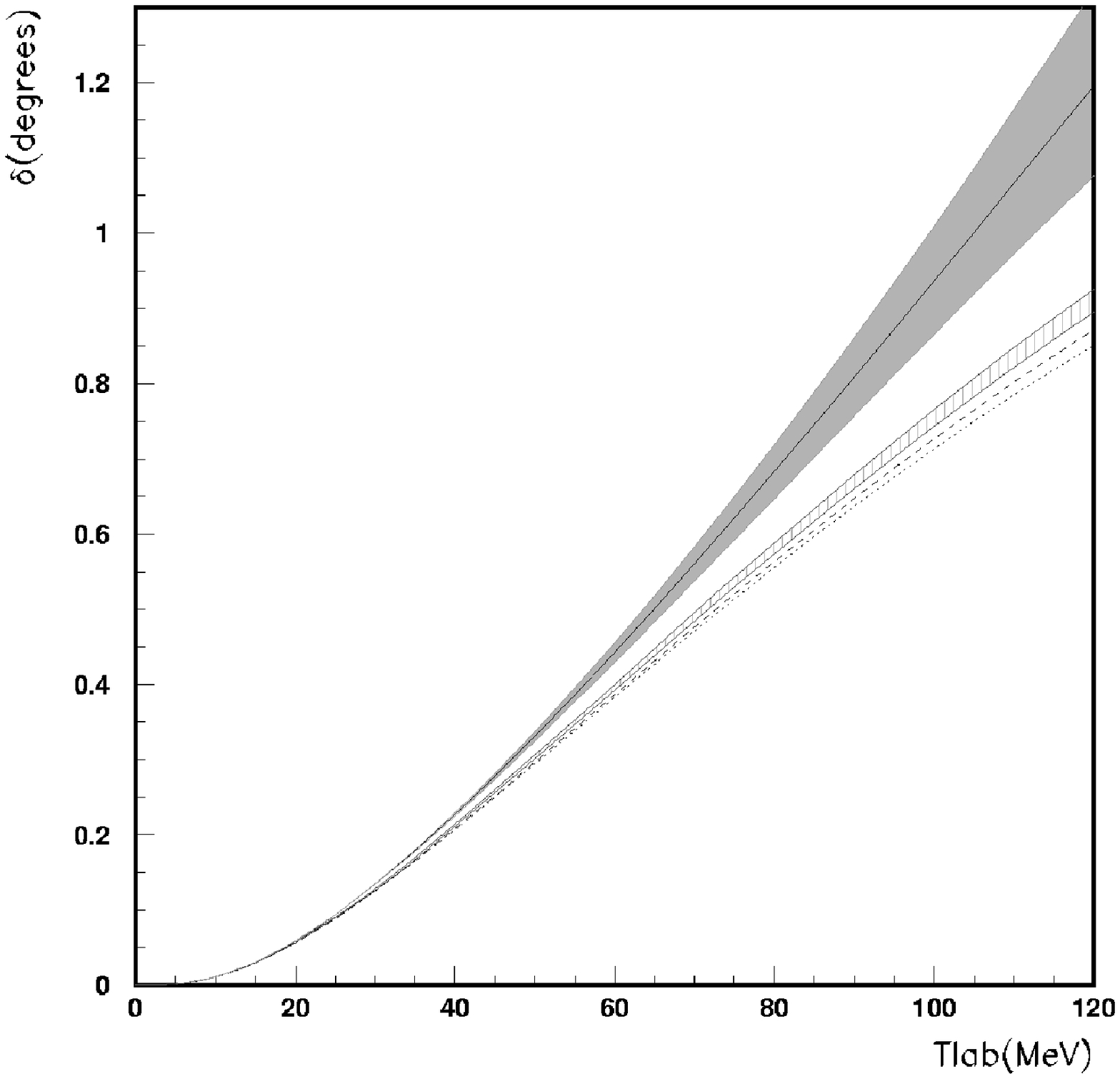}
\caption[The $^3F_2$ phase shifts near ${\cal A}=0$ for 
$R=1.4\;$fm.]
{\label{fig:3f2} $^3F_2$ phase shifts near ${\cal A}=0$ for 
cut-off radius
$R=1.4\;$fm.  The dashed line is iterated OPE, the dotted line 
includes leading order TPE ($\nu=2$) perturbatively, and the solid
line is the full calculation to order $\nu=3$ with TPE treated 
perturbatively.   The 
shaded band corresponds to the $\nu=3$
phase shifts, allowing each of the independent components of
${\cal A}$ to explore the range $-1/2<{\cal A}_{ij}<1/2$.} 
\end{center}
\end{figure}

The OPE, leading order TPE and full TPE results are shown 
for $R=1.4\;$fm in Figure \ref{fig:3f2}.  
In this, and other coupled waves, the shaded band shows the
maximum variation of the full TPE result as the three 
independent elements of the boundary condition matrix ${\cal A}$
are allowed to vary between $-1/2$ and $1/2$.

As is clear from the figure, 
the OPE results describe the Nijmegen PWA's quite well in the
region of energy shown.  The leading order TPE corrections
are small and have the correct sign, but the full TPE corrections
are much too large.  It is worth noting at this point that
Kaiser {\it et al.} \cite{kbw} also find too much attraction in this partial
wave.  They also note that the VPI \cite{vpi} PWA and the Nijmegen PWA93
$^3F_2$ phase shifts are very similar.  For
these reasons, it is unlikely that the over-attraction 
is an artefact of the way the calculation has been performed, or 
due to the choice of a particular PWA.

\subsection{$^3D_3$ - $^3G_3$, ($\epsilon_3$)}
The $^3D_3$ phase shift is very sensitive to $R$ at all
energies.  As in the other two $D$-waves,
we observe that if the cut-off is placed at $R=1.4\;$fm, the full 
TPE phase shift lies within the Nijmegen range over a reasonable
range of energies, in this case up to about $100$MeV.
The agreement may be of more significance in this partial wave than the 
other two $D$-waves because the OPE phase shift has the wrong sign.

$\epsilon_3$ is described extremely well by OPE up to around
$250$MeV.  As with $\epsilon_2$ the TPE corrections are both very
small for $R=1.4\;$ fm and do not spoil this agreement.  
Kaiser {\it et al.} \cite{kbw}
do not find quite such good agreement with OPE at high energies.  
This may be an indication that the two and three loop reducible 
iterated OPE graphs not calculated in that work are significant 
for this mixing angle.  

In the $^3G_3$ phase shift, OPE is too repulsive.  As in other
isoscalar partial waves, the leading order TPE potential 
is small and repulsive.  Adding $\nu=3$ TPE brings the phase
shift back into agreement with the Nijmegen PWA's, particularly
at $R=1.4\;$fm.  The differences among the Nijmegen phase shifts
are larger than the improvement however, so we do not show these
results here.

\subsection{$^3F_4$ - $^3H_4$, ($\epsilon_4$)}
\begin{figure} 
\begin{center}
\includegraphics[width=6in,bb=38 10 536 536]{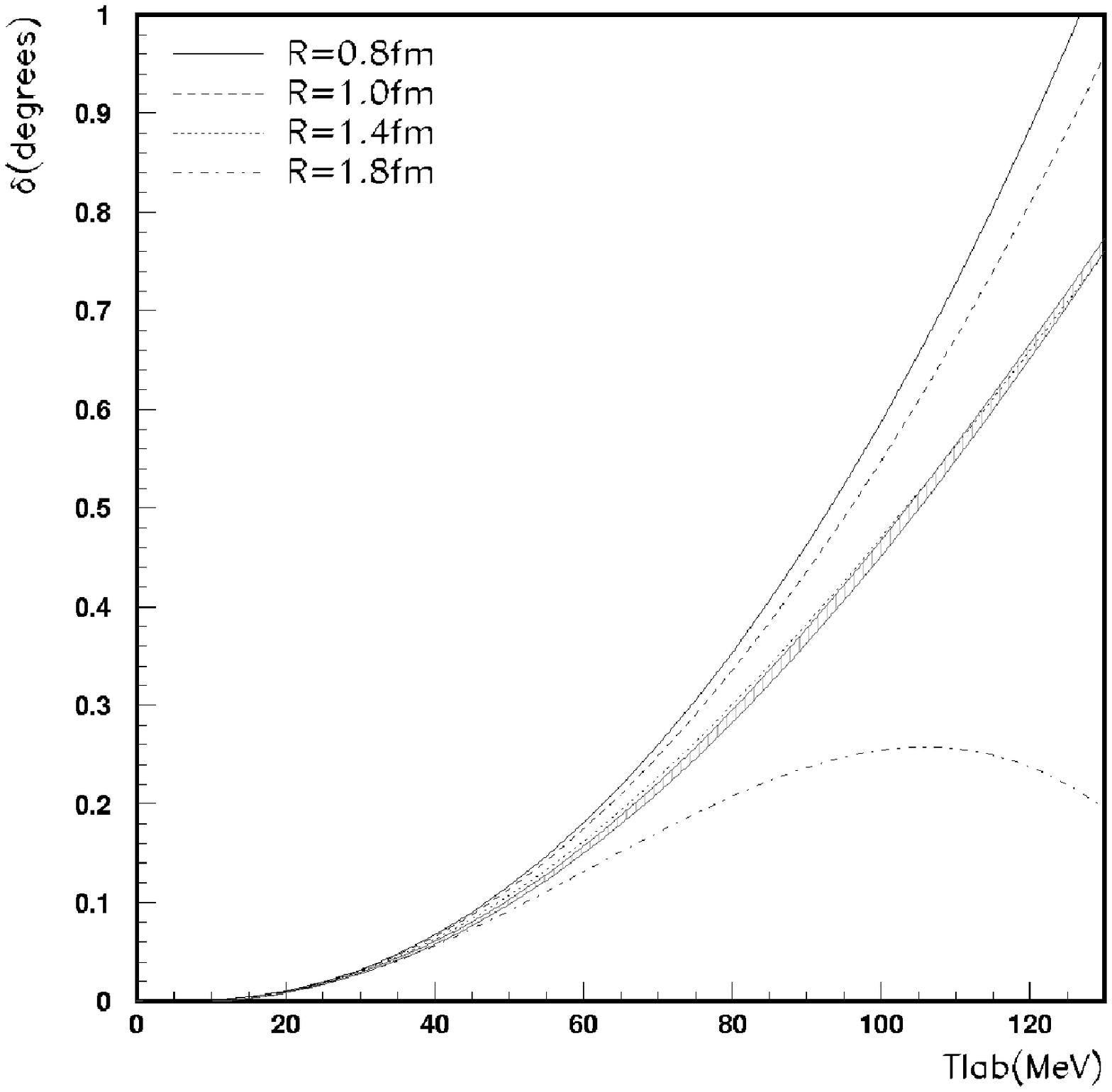}
\caption[The $^3F_4$ phase shift at ${\cal A}=0$ for 
four cut-off radii.]
{\label{fig:3f4pnp} $^3F_4$ phase shift at ${\cal A}=0$ for 
four cut-off radii.
For each value of $R$, the result of including full OPE + TPE results
non-perturbatively is shown.}
\end{center}
\end{figure}

\begin{figure} [h]
\begin{center}
\includegraphics[width=6in,bb=38 48 536 536]{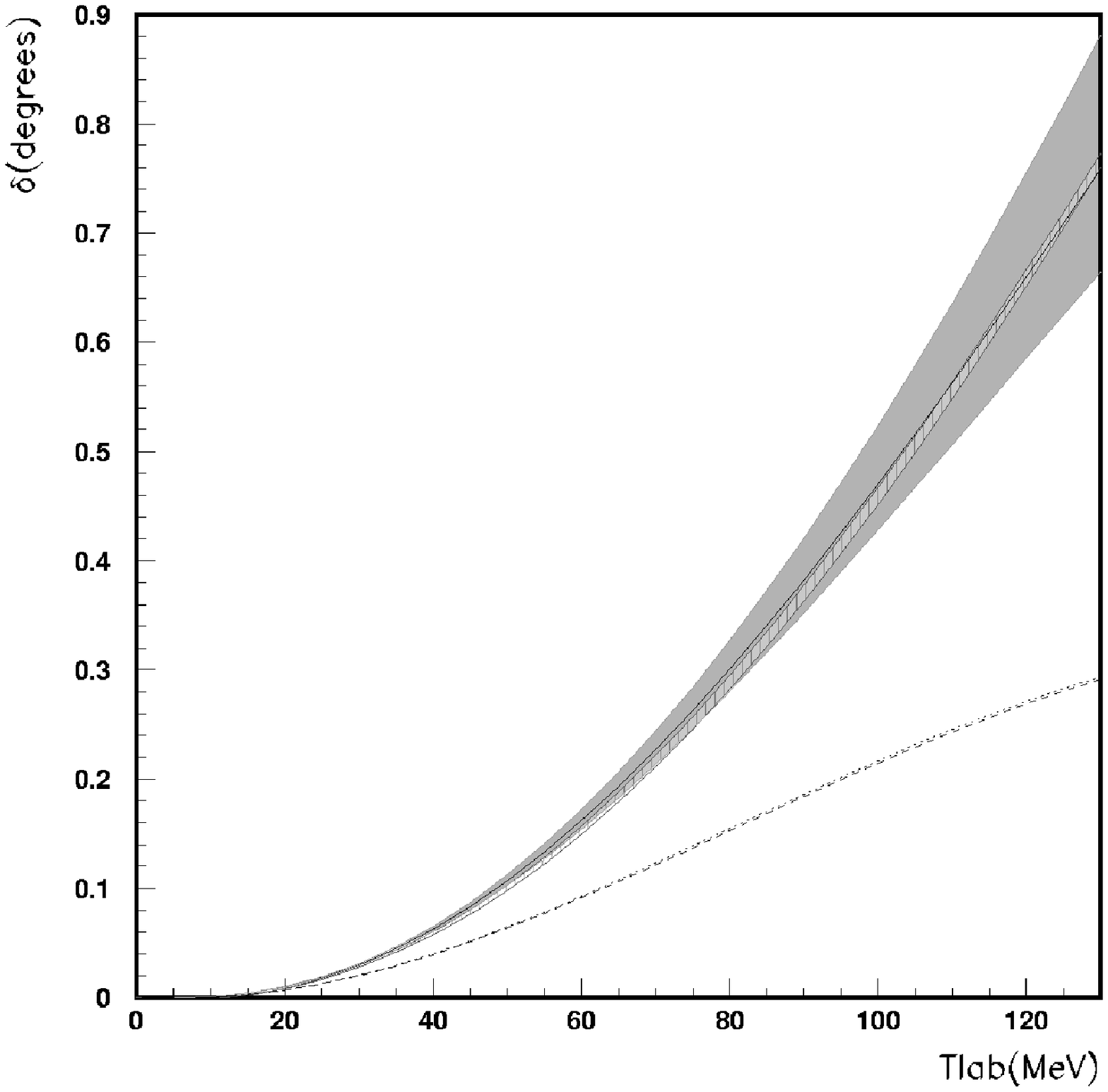}
\caption[The $^3F_4$ phase shifts near ${\cal A}=0$ for 
$R=1.4\;$fm.]
{\label{fig:3f4} $^3F_4$ phase shifts near ${\cal A}=0$ for 
cut-off radius
$R=1.4\;$fm.  The dashed line is iterated OPE, the dotted line 
includes leading order TPE ($\nu=2$) perturbatively, and the solid
line is the full calculation to order $\nu=3$ with TPE treated 
perturbatively.  The 
shaded band corresponds to the $\nu=3$
phase shifts, allowing each of the independent components of
${\cal A}$ to explore the range $-1/2<{\cal A}_{ij}<1/2$.}
\end{center}
\end{figure}

The $^3F_4$ phase shifts for TPE at the usual four cut-off radii are shown 
in Figure \ref{fig:3f4pnp}.  Dependence on $R$ is still surprisingly
large at $R=1.4\;$fm, and sets in at lower energies than in other $F$-waves.
$R=1.4\;$fm happens to be a particularly good choice in this partial wave.
This is shown alongside the  OPE and leading order TPE results in Figure 
\ref{fig:3f4}.
As usual, for the full TPE result, sensitivity to ${\cal A}$ is shown 
by a shaded band. 

The $\nu=2$ corrections have almost no effect on the leading order 
predictions. 
The result to order $\nu=3$, however, represents a marked improvement over
OPE.  Despite sensitivity to the cut-off radius, this is true
over a wide range of values of $R$.  

$\epsilon_4$, like $\epsilon_3$ and $\epsilon_2$, is dominated by OPE.
TPE corrections are small enough not to spoil the agreement.

The Nijmegen PWA's are not expected to provide a good description of
$H$-waves and higher, as the potentials used to produce the quoted phase
shifts are taken from a lower partial wave.  Nevertheless,
there is a tendency for full TPE to be over-attractive in the $^3H_4$
partial wave.

\subsection{$^3G_5$ - $^3I_5$, ($\epsilon_5$)}
\begin{figure} 
\begin{center}
\includegraphics[width=6in,bb=38 10 536 536]{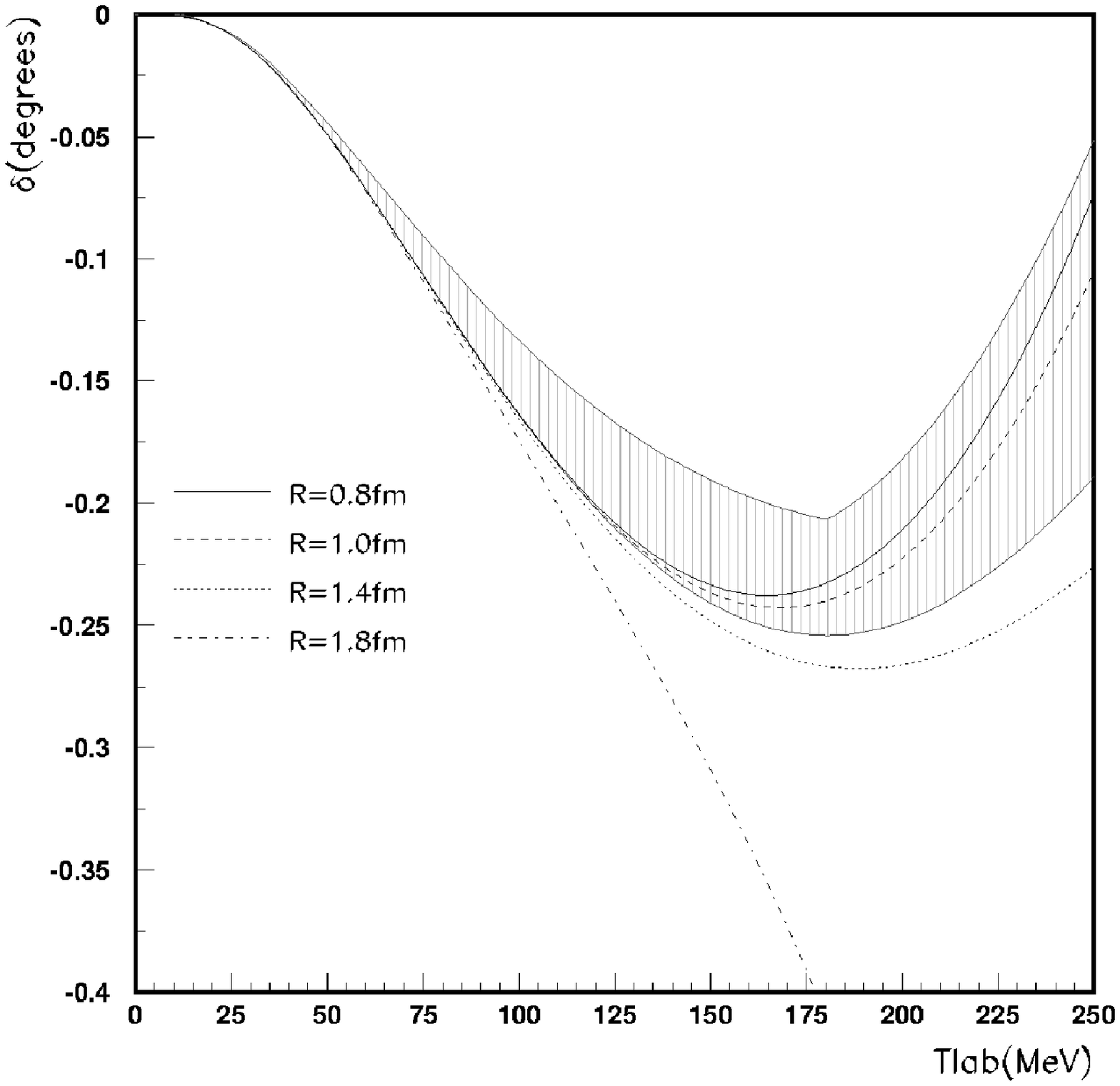}
\caption[The $^3G_5$ phase shift at ${\cal A}=0$ for 
four cut-off radii.]
{\label{fig:3g5pnp} The $^3G_5$ phase shift at ${\cal A}=0$ for 
four cut-off radii.
For each value of $R$, the result of including full OPE + TPE results
non-perturbatively is shown.}
\end{center}
\end{figure}

\begin{figure} [h]
\begin{center}
\includegraphics[width=6in,bb=38 48 536 536]{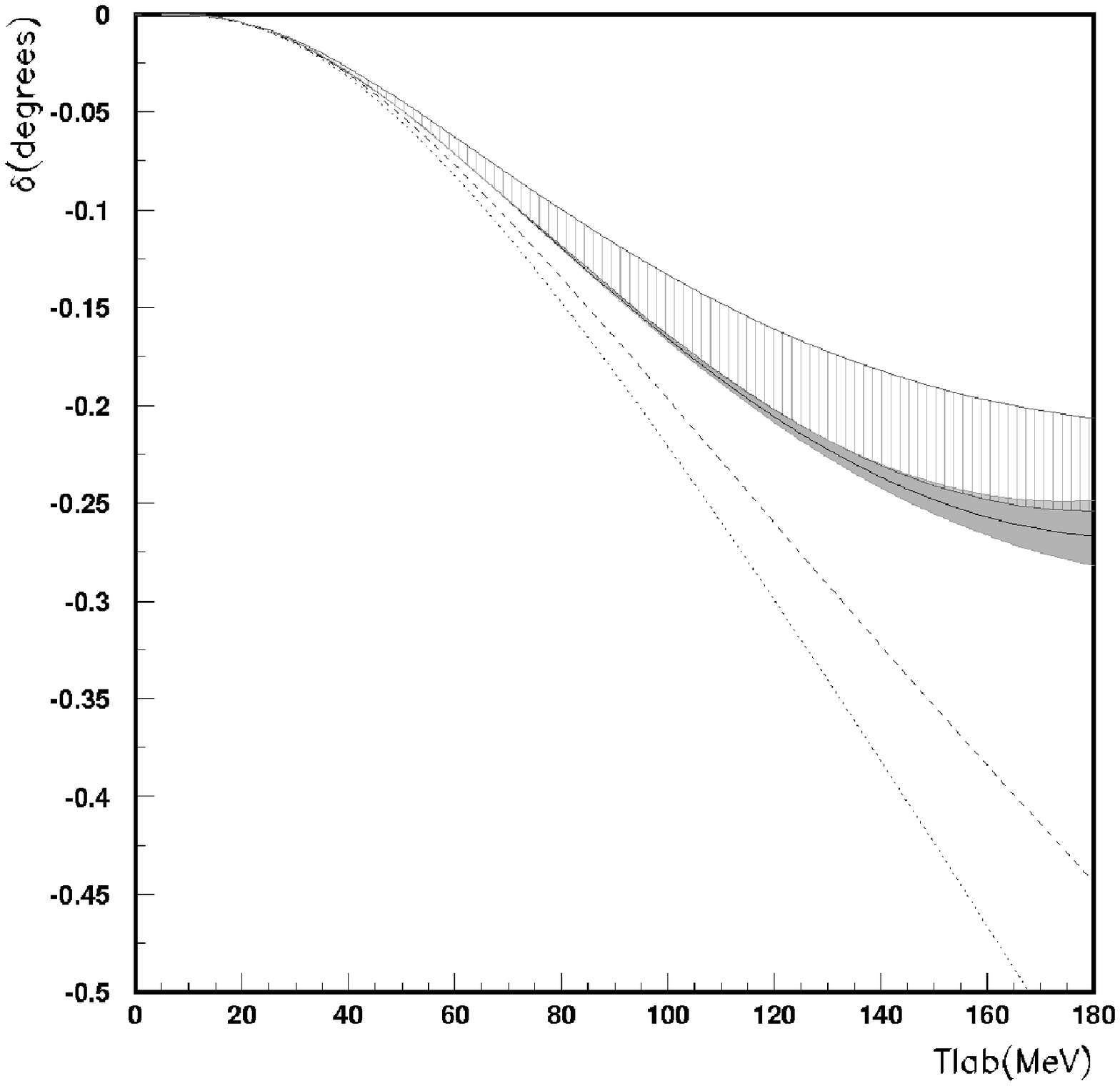}
\caption[The $^3G_5$ phase shift near ${\cal A}=0$ for $R=1.4\;$fm.]
{\label{fig:3g5} $^3G_5$ phase shifts near ${\cal A}=0$ for 
cut-off radius
$R=1.4\;$fm.  The dashed line is iterated OPE, the dotted line 
includes leading order TPE ($\nu=2$) perturbatively, and the solid
line is the full calculation to order $\nu=3$ with TPE treated 
perturbatively.  The 
shaded band corresponds to the $\nu=3$
phase shifts, allowing each of the independent components of
${\cal A}$ to explore the range $-1/2<{\cal A}_{ij}<1/2$.}
\end{center}
\end{figure}

Despite large differences among the PWA phase shifts, the $^3G_5$
partial wave shows very good evidence for the importance of 
chiral TPE.  Figure \ref{fig:3g5pnp} shows the full TPE for our
four choices of cut-off radius.  Disregarding the $R=1.8\;$fm
case as usual, $R$ dependence is small below around $130$MeV.
Figure \ref{fig:3g5} shows OPE, leading order TPE and full
TPE for ${\cal A}=0$.  The shaded band shows variation of
full TPE with ${\cal A}$ over the usual range.  The OPE potential
is substantially too repulsive above around $60$MeV and leading order
TPE has the wrong sign, but the agreement of the full TPE result with the 
PWA's is quite impressive, particularly as variation with ${\cal A}$
is small.

As would be expected, $\epsilon_5$ and the $^3I_5$ phase shift
are dominated by OPE.  

\section{Summary}
The $D$-wave phase shifts are very sensitive to the cut-off radius
and effects generated by iterating the TPE potential. We note that,
in all three cases, the choice $R=1.4\;$fm leads to surprisingly
good agreement with the Nijmegen PWA's.  

In all of the $F$-wave phase shifts, the results are much
less sensitive to the regularisation procedure.  In three
of these four partial waves, the TPE results are in good agreement with
the Nijmegen PWA's for $R=1.4\;$fm.  The full TPE contribution 
to the $^3F_2$ potential is too large and attractive at all energies.

The $G$-wave phase shifts are the least sensitive to the effects of
iteration of TPE and boundary condition parameters.  In the $^3G_3$ partial
wave uncertainties in the Nijmegen PWA's prevent any firm conclusion,
and the $^3G_4$ partial wave is very well described by OPE alone.
Clean signals of TPE are seen in the $^1G_4$ and $^3G_5$ phase
shifts.  In the $^1G_4$ case, TPE is too attractive, but adjusting
the boundary parameter $\alpha$ can correct for this.  Of any partial
wave, 
the most convincing evidence for TPE is seen in the $^3G_5$ phase 
shift where OPE is much too repulsive, but including both TPE 
contributions leads to very good agreement with the PWA's.

The mixing angles $\epsilon_2$,$\epsilon_3$ and $\epsilon_4$
are well described by iterated OPE, and the TPE corrections
to these are small.

In most partial waves considered here, $\nu=3$ TPE corrections 
are much larger than $\nu=2$ corrections. 
In general, the TPE potential is slightly underattractive
in isoscalar partial waves, and too attractive in isovector 
partial waves. 

\chapter{Conclusion} \label{conc}

In this thesis, we have considered the application of effective
field theory to the Nucleon-Nucleon interaction, with particular
emphasis on the constraints imposed by chiral symmetry.

Consideration of the Wilson renormalisation group in a nucleon-only 
theory, which is valid at very low momenta, led us to identify
two very different power counting schemes which are valid in
different physical regimes. 

A systematic power counting scheme for 
contributions to the reactance matrix $K$
can be obtained in the case of weak scattering by expanding around the 
unique trivial RG fixed point.  The power counting found here
is straightforward, and was originally proposed by Weinberg.

For strong scattering at low energies, a useful expansion of $1/K$
can be obtained by expanding the potential around an unstable
non-trivial fixed point corresponding to a bound state of two nucleons 
at threshold.  This expansion can be ordered in such a way that it 
reproduces the familiar effective range expansion order by order. 

The treatment of pions in the case of strong scattering remains 
problematic.  We compared modified effective range expansions
obtained by removing pions both perturbatively and non-perturbatively
using a distorted wave treatment. 
In the perturbative treatment, the effective range decreased while 
non-perturbatively we found an increase of similar magnitude.
The disagreement in these results suggests either that long-range
physics other than OPE makes important contributions to the effective
range, or that unknown short distance physics with a complicated 
structure causes very slow convergence of the small momentum expansion.

In partial waves with angular momentum greater than one, 
coefficients of four-nucleon contact interactions do not contribute 
to NN scattering.  Furthermore,
scattering in these partial waves is weak, allowing us to
use straightforward power counting corresponding to expansion
of the potential around the trivial fixed point.

Iterating the Nucleon-Nucleon potential we obtained from HBCHPT 
in Chapter \ref{pions} produces all contributions to the 
interaction up to order $\nu=3$.

We have found several partial waves in which the EFT
predictions can be isolated to order $\nu=3$ in a parameter-free way. 
Of these, the $^1F_3$ and $^3G_5$ are isoscalar, and the $^1G_4$, $^3F_2$,
$^3F_3$ and $^3F_4$ are isovector.  In these phase shifts, there
is a range of energies and cut-off radii for which sensitivity
to the regularisation procedure and higher order effects
generated by iterating the potential are small. 
Since the above  list includes spin 
singlet and triplet waves, we have tested both the central and
tensor parts of the potential. 

In the partial waves mentioned above, with the exception
of $^1G_4$ and $^3F_2$, the $\nu=3$ potential is an improvement
over both OPE and leading order TPE.  In the $^1G_4$ partial 
wave, adjusting the single number $\alpha$ allows a satisfactory global 
fit to the Nijmegen PWA's up to about $300$MeV.  This is not
possible for OPE or leading order TPE.

The highest order,
$\nu=3$, contribution tends to be much larger than the leading
order, $\nu=2$, correction, so convergence of the small momentum
expansion has not been established.
On the other hand, the $\nu=3$ potential is the first 
in which important contributions, such as the $\sigma$ - like
contribution can appear.  The calculation to order $\nu=4$ involves
no new counterterms for $F$-waves and above, and may offer a fairer 
test of convergence.  

A trend which seems to emerge from the results is a tendency
for the full TPE potential to provide slightly too little attraction
in isoscalar partial waves, and too much attraction in isovector
partial waves (especially for $R<1.4\;$fm).  Isovector contributions
which occur at next order may improve
this situation.  One example is the two loop graph with a 
three-pion nucleon vertex from $\Lcal_{\pi N}^{(1)}$ 
in which the three-pion lines terminate on the other nucleon.    

\appendix
\chapter{The Nijmegen OPE Potential.}
\label{nijope}
In the Nijmegen PWA's \cite{nijpwa,nijpot}
described in Chapter \ref{chiral},  the distinction 
between neutral-pion and charged-pion 
exchange is explicitly taken into account.  Since we 
are looking for signals of shorter range physics, it is
helpful to use their form of the OPE potential.
Defining,
\be
V(m)=\left(\frac{m}{m_{\pi^\pm}}\right)^2m\left[
\vec{\sigma_1}.\vec{\sigma_2}+\frac{3+3x+x^2}{x^2}S_{12}
\right]\frac{e^{-x}}{3x}
\ee
where $x=mr$,  the Nijmegen OPE in the case of 
neutron-proton scattering may then be written as:
\be
V(r)=-f_\pi^2V(m_{\pi^0})\pm 2f_\pi^2V(m_{\pi^\pm}), 
\ee
where $m_{\pi^0}$ and $m_{\pi^\pm}$ are the neutral and charged
pion masses.  The plus sign applies in the case of total isospin
$I=1$, and the minus sign is used when $I=0$.

\references{References}{}{ref.;}

\begin{thebibliography}{999}
\bibitem{polemics} R. Machleidt and G. Q. Li, Phys.\ Rept.\ {\bf 242} 5 (1994).
\bibitem{mikesol} M. C. Birse, Prog.\ Part.\ Nucl.\ Phys.\ {\bf 25} 1 (1990). 
\bibitem{modofnuc} R. K. Bhaduri {\it Models of the Nucleon}, 
(Addison-Wesley, 1988). 
\bibitem{pich98} A. Pich, hep-ph/9806303.
\bibitem{mikerev} M. C. Birse, \jpg{20}{94}{1537}.
\bibitem{chpt}V. Bernard, N. Kaiser and Ulf-G. Mei{\ss}ner
{{Int. J. Mod. Phys. E}} {\bf{4}}, No.2 193-344 (1995). 
\bibitem{pdg} C. Caso et al, The European Physical Journal {\bf{C3}},1 (1998)  
     and 1999 off-year partial update for the 2000 edition available on 
     the PDG WWW pages (URL: http://pdg.lbl.gov/). 
\bibitem{pcac} M. K. Banerjee, {{Prog. Part. Nucl. Phys}} {\bf{31}}, 77 (1993).
\bibitem{weinbook2} S. Weinberg, {\it The Quantum Theory of Fields},
(Cambridge University Press, 1996).
\bibitem{Gold} J. Goldstone, {{Nuovo Cimento}} {\bf{19}}, 154 (1961).
\bibitem{gor}M. Gell-Mann, R. Oakes and B. Renner, Phys. Rev. {\bf 175}, 2195 (1968).
\bibitem{wein79} S. Weinberg {{Physica}} {\bf{96A}}, 327-340 (1979).
\bibitem{dotsm} J. F. Donoghue, E. Golowich and B. R. Holstein,
{\it Dynamics of the Standard Model}, (Cambridge University Press, Cambridge,
1992).
\bibitem{hqet} E. Jenkins and A. V. Manohar, Phys. Lett. {\bf 255} 558 (1991). 
\bibitem{wein2} S. Weinberg, \pl{B251}{90}{288}; S. Weinberg, \np{363}{91}{3}.
\bibitem{orvk} C. Ord\'o\~nez, L. Ray and U. van Kolck \prc {53}{96}{2086}.
\bibitem{curpin} S. Weinberg {{Phys. Rev. Lett.}} {\bf{17}}, (1966).
\bibitem{pionsnuclei} T. Ericson and W. Weise, {\it Pions and Nuclei}
(Clarendon Press, Oxford, 1988).
\bibitem{meisstriangle} P. B\"uttiker and Ulf-G. Mei\ss ner, hep-ph/9908247.
\bibitem{cidet} V. Bernard, N. Kaiser and Ulf-G. Mei\ss ner 
{{Nucl.Phys.}} {\bf{A615}}, 483 (1997).
\bibitem{mpi4} V. Bernard, N. Kaiser and Ulf-G. Mei\ss ner,
\prc{52}{95}{2185}.
\bibitem{afg}S. K. Adhikari and T. Frederico,
Phys.\ Rev.\ Lett.\ {\bf 74}, 4572 (1995); S. K. Adhikari and A. Ghosh,
J. Phys.\ A: Math.\ Gen.\ {\bf 30}, 6553 (1997). 
\bibitem{ksw}D. B. Kaplan, M. J. Savage and M. B. Wise, 
Nucl.\ Phys.\ {\bf B478}, 629 (1996).
\bibitem{md1}T. D. Cohen, Phys.\ Rev.\ {\bf C55}, 67 (1997); 
D. R. Phillips and T. D. Cohen, Phys.\ Lett.\ {\bf B390}, 7 (1997).
\bibitem{md2}S. R. Beane, T. D. Cohen and D. R. Phillips, Nucl.\ Phys.\ {\bf
A632}, 445 (1998).
\bibitem{lep}G. P. Lepage, nucl-th/9706029.
\bibitem{rbm}K. G. Richardson, M. C. Birse and J. A. McGovern, hep-ph/9708435.
\bibitem{bvk}P. F. Bedaque and U. van Kolck, Phys.\ Lett.\ {\bf B428}, 221
(1998); P. F. Bedaque, H.-W. Hammer and U. van Kolck, \prc {58}{98}{R641}.
\bibitem{pkmr}T.-S. Park, K. Kubodera, D.-P. Min and M. Rho, \prc {58}{98}{637}.
\bibitem{ksw2}D. B. Kaplan, M. J. Savage, and M. B. Wise, Phys.\ Lett.\ {\bf
B424}, 390 (1998); \np {534}{98}{329}; \prc {59}{99}{617}.
\bibitem{geg}J. Gegelia, nucl-th/9802038; \jpg {25}{99}{1}.
\bibitem{uvk}U. van Kolck, Talk given at the Joint Caltech/INT Workshop
on  {\it Nuclear Physics with Effective Field Theory},
KRL preprint MAP-228 (1998); U. van Kolck, \np {645}{99}{273}. 
\bibitem{ch}T. D. Cohen and J. M. Hansen, \prl{440}{98}{233}.
\bibitem{drp}D. R. Phillips, nucl-th/9804040.
\bibitem{mcb}M. C. Birse, nucl-th/9804028.
\bibitem{wrg}K. G. Wilson and J. G. Kogut, Phys.\ Rep.\ {\bf 12},
75 (1974); J. Polchinski, Nucl.\ Phys.\ {\bf B231}, 269 (1984);
R. D. Ball and R. S. Thorne, Ann.\ Phys.\ {\bf 236}, 117 (1994); 
T. R. Morris, Prog.\ Theor.\ Phys.\ Suppl.\ {\bf{131}} 395 (1998).
\bibitem{bmr} M. C. Birse, J. A. McGovern and K. G. Richardson, hep-ph/9807302.
\bibitem{bbp} D. R. Phillips, S. R. Beane and M. C. Birse, 
J.\ Phys.\ {\bf{A32}} 3397 (1999).  
\bibitem{ere}J. M. Blatt and J. D. Jackson, Phys.\ Rev.\ {\bf 76}, 18 (1949); 
H. A. Bethe, Phys.\ Rev.\ {\bf 76}, 38 (1949).
\bibitem{bl}H. A. Bethe and C. Longmire, Phys.\ Rev.\ {\bf 77}, 647 (1950).
\bibitem{paris} M. Lacombe, B. Loiseau, J.M. Richard, R. Vinh Mau,
J. C\^ot\'e, P. Pir\`es and R. de Tourreil, \prc {21}{80}{861}
\bibitem{stony} G. E. Brown and A. D. Jackson, {\it The Nucleon-Nucleon 
Interaction} (North-Holland, Amsterdam 1976).
\bibitem{kbw} N. Kaiser, R. Brockmann and W. Weise, \np{624}{97}{527}.
\bibitem{fl} E. L. Lomon and H. Feshbach, {{Ann. Phys.}} {\bf{48}}, 
94-172 (1968). 
\bibitem{newton} R. G. Newton, {\it Scattering Theory of Waves and 
Particles}, (Springer-Verlag, New York, 1982).
\bibitem{nniso} E. Epelbaum, Ulf-G. Mei\ss ner, nucl-th/9903046.
\bibitem{egm} E. Epelbaum, W. Glockle, Ulf-G. Mei\ss ner
{{Nucl.Phys.}} {\bf{A637}}, 107 (1998).
\bibitem{tmo} M. Taketani, S. Machida and S. Ohnuma, Prog.\ Theor.\
Phys.\ (Kyoto) {\bf{7}}, 45 (1952).
\bibitem{bw} K. A. Brueckner and K. M. Watson, Phys.\ Rev.\ {\bf 92},
1023 (1953).
\bibitem{coonfriar} S. A. Coon and J. L. Friar, \prc {34}{86}{1060}.
\bibitem{friarcoon} J. L. Friar and S. A. Coon, \prc {49}{94}{1272}. 
\bibitem{friar97} J. L. Friar, nucl-th/9901082.
\bibitem{nijfriar} M. C. M. Rentmeester, R. G. E. Timmermans, J. L. Friar, 
J. J. de Swart, nucl-th/9901054.
\bibitem{steefur} J. V. Steele and R. J. Furnstahl, \np{645}{99}{439}.
\bibitem{lands} D. B. Kaplan and J. V. Steele, nucl-th/9905027.
\bibitem{dwere} H. van Haeringen and L. P. Kok, \pra {26}{82}{1218}.
\bibitem{vpi} R. A. Arndt, J. S. Hyslop and L. D. Roper \prd {35}{87}{121}.
\bibitem{nnonline} http://NN-OnLine.sci.kun.nl/.
\bibitem{kgw} N. Kaiser, S. Gerstend\"orfer and W. Weise, \np {637}{98}{395}.
\bibitem{meissnew} Ulf-G. Mei\ss ner, nucl-th/9909011.
\bibitem{nijpot} V. G. J. Stoks, R. A. M. Klomp, C. P. F. Terheggen 
and J. J. de Swart, \prc {49} {94} {2950}.
\bibitem{nijpwa} V. G. J. Stoks, R. A. M. Klomp, M. C. M. Rentmeester
and J. J. de Swart, \prc {48} {93} {792}.
\bibitem{reid68} R. V. Reid, Jr., Ann Phys. (NY) {\bf 50}, 411 (1968).
\bibitem{nijsc} M. M. Nagels, T. A. Rijken and J. J. de Swart, Phys. Rev. D 
{\bf 17}, 768 (1978).




\end{thebibliography}
\end{document}